      [1999/12/01 v1.4c Il Nuovo Cimento]
\documentclass{cimento}

\newcommand{\eps}{\epsilon}

\newcommand{\oo}{{\omega_\oplus}}
\newcommand{\om}{\omega}
\newcommand{\be}{\begin{equation}}
\newcommand{\ee}{\end{equation}}
\newcommand{\bea}{\begin{eqnarray}}
\newcommand{\eea}{\end{eqnarray}}

\newcommand{\al}{\alpha}
\newcommand{\bet}{\beta}
\newcommand{\gam}{\gamma}

\newcommand{\bal}{\bar\alpha}

\newcommand{\bg}{\bar g}

\newcommand{\barm}{\bar m}

\newcommand{\Om}{\Omega}

\newcommand{\Oo}{\Omega_\oplus}

\newcommand{\is}{i_4\bar s}
\newcommand{\kemin}{\tilde\kappa_{e-}}
\newcommand{\kopl}{\tilde\kappa_{o+}}

\usepackage{epsfig}
\usepackage{graphicx}
\usepackage{amsfonts,amsmath}

\title{Quantum mechanics, matter waves, and moving clocks}
\author{Holger M\"uller\from{ins:x}}
\instlist{\inst{ins:x} Department of Physics, 366 Le Conte Hall, University of California, Berkeley, CA 94720, USA}
\PACSes{\PACSit{03.75.Dg}
\PACSit{37.25.+k}
\PACSit{03.65.Pm}
\PACSit{04.62.+v}
\PACSit{31.15.xk}\PACSit{03.65.Ta}}

\begin{document}

\maketitle

\begin{abstract}
This paper is divided into three parts. In the first (section 1), we demonstrate that all of quantum mechanics can be derived from the fundamental property that the propagation of a matter wave packet is described by the same gravitational and kinematic time dilation that applies to a clock. We will do so in several steps, first deriving the Schr\"odinger equation for a nonrelativistic particle without spin in a weak gravitational potential, and eventually the Dirac equation in curved space-time describing the propagation of a relativistic particle with spin in strong gravity.

In the second part (sections 2-4), we present interesting consequences of the above quantum mechanics: that it is possible to use wave packets as a reference for a clock, to test general relativity, and to realize a mass standard based on a proposed redefinition of the international system of units, wherein the Planck constant would be assigned a fixed value. The clock achieved an absolute accuracy of 4 parts per billion (ppb). The experiment yields the fine structure constant $\alpha=7.297\,352\,589(15)\times 10^{-3}$ with 2.0\,ppb accuracy. We  present improvements that have reduced the leading systematic error about 8-fold and improved the statistical uncertainty to 0.33\,ppb in 6 hours of integration time, referred to $\alpha$.

In the third part (sections 5-7), we present possible future experiments with atom interferometry: A gravitational Aharonov-Bohm experiment and its application as a measurement of Newton's gravitational constant, antimatter interferometry, interferometry with charged particles, and interferometry in space.

We will give a review of previously published material when appropriate, but will focus on new aspects that haven't been published before.
\end{abstract}
\newpage

\tableofcontents
\newpage

%

\section{Quantum mechanics as a theory of waves oscillating at the Compton frequency}

We will show that all of quantum mechanics can be derived from a picture of matter waves as clocks together with simple assumptions such as the principle of superposition. This picture assumes that a quantum mechanical wave packet has an oscillation frequency of $\om_C=mc^2/\hbar$, where $m$ is the particle's mass, $c$ the velocity of light, and $\hbar$ the reduced Planck constant. The oscillation frequency is shifted by the gravitational redshift and time dilation as the particle moves through space and time. The propagation of arbitary quantum states can be decomposed into such wave-packets (``matter-wave clocks") taking all possible paths through phase-space. We will show that this path integral formalism will yield the quantum mechanical wave equations, starting with the Schr\"odinger equation for nonrelativistic, spinless particles, then for relativistic particles with spin, first without gravity, then in curved space-time. This shows that the picture of matter wave packets as Compton frequency clocks is not just exact. It can even be used to re-derive all of quantum mechanics.

The description of matter waves as matter-wave clocks has been the basis of de Broglie's invention of matter waves \cite{deBroglie}. 
It has recently been applied to tests of general relativity \cite{redshift,redshiftPRL,Nobili,Bonder,Klein,Lindesay,Hacyan,Ridgley,RidgleyForces,Carloni,Chou}, matter-wave experiments \cite{Poli,Berrett,ZhuoLin,JMO,Hakimov,Sorrentino,Kajari,Alberti,Abele,GravAB}, the foundations of quantum mechanics \cite{Dolce,Bassi}, quantum space-time decoherence \cite{Blencowe}, the matter wave clock/mass standard \cite{CCC,Bouchendira2011,Bouchendira2013}, and led to a discussion on the role of the proper time in quantum mechanics  \cite{ZychCQG,ZychNatureComm}. It is generally covariant and thus well-suited for use in curved space-time, e.g., gravitational waves \cite{Dimopoulos,AGIS,Hogan,Graham}. It has also given rise to a fair amount of controversy \cite{Naturecomment,CQGComment,WolfCQG,JoeSam,Giulini,SchleichNJP,SchleichPRL,Greenberger,Jaeckel}.
Within the broader context of quantum mechanics, however, this description has been abandoned, in part because it could not be used to derive a relativistic quantum theory, or explain spin.

The descriptions that replaced the clock picture achieve these goals, but do not motivate the concepts used. For example, the Dirac equation can be derived from a Lagrangian {\em density}, where $\psi$ takes the role of the coordinates: $\mathcal L_D=i\hbar c \bar \psi \gamma^\mu \partial_\mu\psi-mc^2\bar\psi\psi$, where the $\gamma^\mu$ are the Dirac matrices, the operator $\psi$ annihilates, and $\bar \psi$ creates, a particle, and $\partial_\mu\equiv\partial/\partial x^\mu$. This Lagrangian density is quadratic in $\psi$ and thereby allows to construct a path integral in Hilbert space. It, however, 
takes the existence of spinors and Dirac matrices for granted rather than explaining or motivating the need for them.

We shall construct a path integral directly from a Lagrangian that is a function of the space-time coordinates $L=-mc^2 d\tau/dt$, where $t$ is the coordinate time, without making a nonrelativistic approximation or introducing additional fields. This will require us to introduce the Dirac matrices and spinors, and will thus explain their use. Since the phase accumulated by a wave packet is given by $\phi=-L/\hbar$, it corresponds to a description of matter waves as clocks. We will thus arrive at a space-time path integral \cite{Feynman1948} in which $\phi=-\om_C\tau$ is maintained exactly, that is equivalent to the Dirac equation. 

This derivation shows that De Broglie's matter wave theory naturally leads to particles with spin-1/2. It relates to Feynman's search for a formula for the amplitude of a path in 3+1 space and time dimensions which is equivalent to the Dirac equation \cite{Feynman,Jacobson}. It yields a new intuitive interpretation of the propagation of a Dirac particle and reproduces all results of standard quantum mechanics, including those supposedly at odds with it. Thus, it illuminates the role of the gravitational redshift and the proper time in quantum mechanics. Finally, we hope it offers an intuitive way to think about quantum mechanics and its possible generalizations. 


\subsection{Notation}
We use letters from the second half of the Greek alphabet $\kappa, \lambda, \mu, \nu,\ldots =0,1,2,3$ to denote the space-time coordinates. Letters from the second half of the Latin alphabet $j,k,l,m\ldots$ denote the spatial coordinates. In curved space-time, we shall employ both a coordinate frame with a metric $g^{\mu\nu}$ and a local Lorentz frame with a Minkowski metric $\eta^{\alpha\beta}$. The determinant of $g^{\mu\nu}$ is denoted $g$. Greek letters from the start of the alphabet $\alpha, \beta,\ldots$ will denote coordinates in the local Lorentz frame, the letters $a,b,c,\ldots $ denote the spatial coordinates in the local Lorentz frame. The two frames are connected by the vierbein $g^{\mu\nu}=e^\mu_\al e^\nu_\bet \eta^{\al\bet}$. Our Minkowski metric has a signature $-+++$. The conventional Dirac matrices in the coordinate frame are $\alpha^k$ and $\beta$ as well as $\gamma^0=\beta, \gamma^k =\gamma^0\alpha^k$ and $\sigma^{\alpha\beta}=\frac12[\gamma^\alpha, \gamma^\beta]$, where $[a,b]=ab-ba$ is the commutator. In weak gravitational fields, we write the metric as $g_{\mu\nu}=\eta_{\mu\nu}+h_{\mu\nu}$, where  $|h_{\mu\nu}|\ll 1$.

\subsection{De Broglie's relations}

De Broglie started with Einstein's equation $E=mc^2$ and Planck's $E=h\nu$, where $E$ is an energy, $m$ the mass of a particle, $c$ the velocity of light, $h$ the Planck constant, and $\nu$ a frequency \cite{deBroglie}. The first relation implies that a massive particle has energy, and the second implies that a process having an energy is associated with an oscillation. The two relations together determine a frequency $\nu_C=mc^2/h$. That leads us to guess that maybe a particle is associated with an oscillation at that frequency. Since $\nu_C$ is related to the Compton wavelength by $\nu_C=c/\lambda_C$, we will call it the particle's Compton frequency.

Na\"{\i}vely, a particle moving at a velocity of $v$ could be described in two ways: The proper time $\tau$ measured by a co-moving clock for a moving reference frame is related to the coordinate time by $\tau=t/\gamma$, where $\gamma=1/\sqrt{1-v^2/c^2}$. Consequently, the moving particle should accumulate fewer oscillations, as $\om_Ct$ is replaced by $\omega_C \tau=(\omega_C/\gamma) t$. As measured by a clock at rest, we thus expect to observe a frequency
\be\label{gamma}
\omega_C'=\omega_C\frac{d\tau}{dt}=\omega_C\gamma^{-1}.
\ee
However, one can make the converse argument: The energy of a moving particle is given by $mc^2\gamma$ and should thus correspond to a frequency of
\be\label{En}
\omega_C''=\omega_C\gamma.
\ee

These seemingly contradictory results can be reconciled. For a wave, there are two velocities, phase velocity $v_p$ and group velocity $v_g$. We assume the group velocity is identical to the classical velocity of the particle, $v_g=v$. Thus, $v_g$ will determine the time dilation factor $\gamma$. The phase accumulated by the particle in its rest frame is $\omega_C\tau=\omega_C't$. If a wave originates at $x=0, t=0$ then the same wave has the phase $-\omega t+kx$ at a different location, where $k=\omega/v_p$ (by definition of $v_p$). We will try to determine $v_p$ such that this wave has the phase $\omega_C' t$ everywhere. In other words, we require
\be\label{k}
\omega_C''t-k'' x=\omega_C't, \quad k''= \frac{\omega_C''}{v_p}.
\ee
We substitute $x=vt$ and find
\be
\om_C\gamma\left(1-\frac{v}{v_p}\right)=\frac{\om_C}{\gamma},
\ee
which is solved by $v_p=c^2/v$ or $v_gv_p=c^2$. We have thus been able to overcome the first hurdle. {\em A particle corresponds to an oscillation of frequency $\om_C$ in its rest frame. Seen in the lab frame, it is a wave of frequency $E=\hbar \omega_C$ where $E$ is the total energy, group velocity $v$, and phase velocity $v_p=c^2/v$.}

Let us denote the oscillation $\psi(x,t)$. Obviously, with hindsight we could identify it with the wave function, but we want to adopt a perspective that we do not know what it means just now. For example, we do not know whether it has to be a complex number, or how its amplitude is determined. We hope that these things will become clear when we know more about the wave's behavior, and the theory will eventually be justified if it makes correct predictions for observable quantities. For now, we will speculate that, if the amplitude is high at a certain location, we will find a large number of particles there. We will adopt the latter point of view and defer the details for later study.) What we do know is that the phase of the wave is given by either the left or the right hand side of Eq. (\ref{k}), e.g.,
\be
\psi\propto e^{-i\om_C\tau}.
\ee


A first experimentally observable effects can be deduced by studying the momentum $p=m\gamma v$ of a particle. According to Eq. (\ref{k}), \be
k=\frac{\om_C''}{v_p}=\frac{mc^2}{\hbar}\gamma\frac{v}{c^2}=\frac1\hbar m\gamma v
\ee
or
\be
\boxed{p=\hbar k.}
\ee
This is de Broglie's famous relation. It can be used to analyze, e.g., Young's double slit experiment (using the principle of superposition). 


\subsection{Construction of a path integral}
\begin{figure}
\centering
\epsfig{file=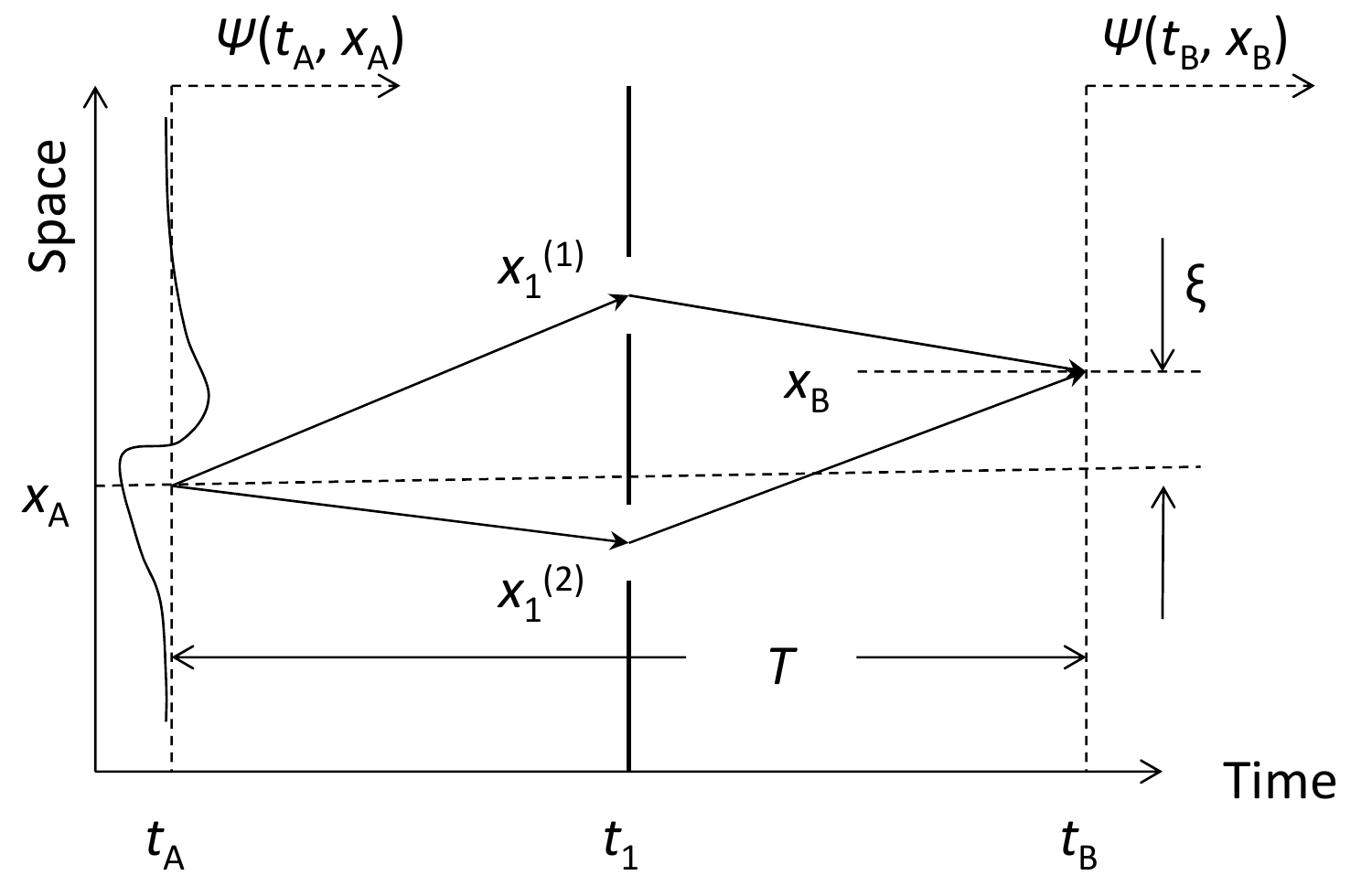,width=0.45\textwidth}\quad
\epsfig{file=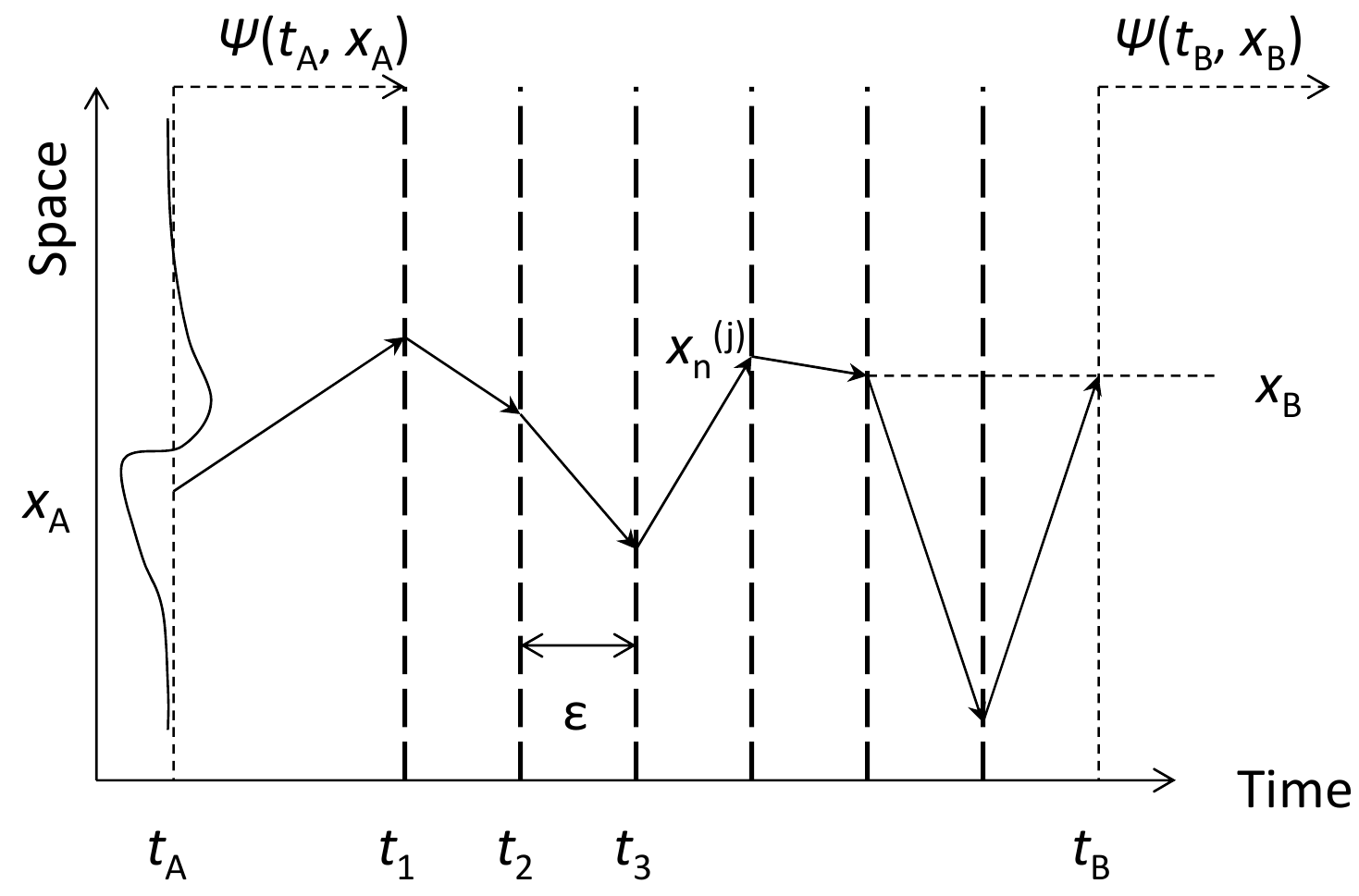,width=0.45\textwidth}
\caption{\label{slit} Left: A particle going through a double slit on the way from $x_A, t_A$ to $x_B, t_B$. Right: Continuous space-time can be approximated by putting infinitely many slits, with infinitely many holes each, in the particle's way.}
\end{figure}

So far, we can only analyze non-interacting particles, traveling on a straight line at constant velocity. We will gradually extend our formalism to study a particle in a potential and general trajectories. We assume we know $\psi(x_A,t_A)$ and want to know $\psi(x_B,t_B)$, where $t_B=t_A+T$ and $x_B=x_A+\xi$. Take a look at the double-slit experiment shown in Fig. \ref{slit}, left). At some time $t_1$ between $t_A$ and $t_B$, the particle has to pass through holes located at $x_1^{(1,2)}$. Clearly, the contribution of $\psi(\vec x_A, t_A)$ to $\psi(x_B,t_B)$ is given by the sum
\be
\psi(x_B,t_B)\propto \psi(x_A,t_A)(e^{-i\om_C\tau(A, 1, B)}+e^{-i\om_C\tau(A, 2, B)})
\ee
where $\tau(A,1,B)$ is the proper time elapsed on the path from $A$ via $1$ to $B$. The exact form of it is unimportant for now. If the screen has, say, $n$ holes located at $x_1^{(1,2,\ldots n)}$, we obtain
\be
\psi(x_B,t_B)\propto \sum_{n_1=1}^n \psi(x_A,t_A)e^{-i\om_C\tau(A, n_1, B)}.
\ee
What about many screens, each with many holes at $x_1^{(1,2,\ldots n)}, x_2^{(1,2,\ldots n)},\ldots x_N^{(1,2,\ldots n)}$, as shown in Fig. \ref{slit}, right? Well,
\be
\psi(x_B,t_B)\propto \sum_{n_1=1}^n \sum_{n_2=1}^n\ldots \sum_{n_N=1}^n\psi(x_A,t_A) e^{-i\om_C\tau(A, n_1, n_2, \ldots, n_N, B)}.
\ee
If each screen has an infinite number of holes
and there are infinitely many screens, we obtain\footnote{With hindsight, by going from the sum without to the integral and thereby introducing the line elements $dx$, the interpretation of $|\psi|^2$ changed from a probability to a probability density.}
\be
\psi(x_B,t_B)\propto \lim_{N\rightarrow \infty} \int dx_1 \int dx_2\ldots \int dx_N \psi(x_A,t_A)e^{-i\om_C\tau(A, x_1, x_2, \ldots, x_N, B)}.
\ee

To evaluate the proper time $\tau(A, x_1, x_2, \ldots, x_N, B)$, we split it up in sections $\tau(A, x_1, x_2, \ldots, x_N, B)=\tau(A,x_1)+\tau(x_1,x_2)+\ldots \tau(x_N,B)$. For each section,
\be
\tau(x_j,x_{j+1})=\epsilon\sqrt{1-v_j^2/c^2},
\ee
where we used that $T=\epsilon N$ is split into $N$ sections and $v_j$ is the velocity of the particle within that section, and $v_j = (x_{j+1} - x_j) / \epsilon$. So,
\be
\psi(x_B,t_B)\propto 
\lim_{N\rightarrow \infty} \int dx_1 \int dx_2\ldots \int dx_N \psi(x_A,t_A) e^{-i\epsilon \om_C\left(\sqrt{1-\frac{v_A^2}{c^2}}+\sqrt{1-\frac{v_1^2}{c^2}}+\ldots \sqrt{1-\frac{v_N^2}{c^2}}\right)}.
\ee
In the exponent, we recognize the Riemannian sum and replace it by its limit, the integral 
%
\[
-\frac{mc^2}{\hbar}\int dt \sqrt{1-\frac{v^2(t)}{c^2}}=\frac 1\hbar \int L dt=\frac 1\hbar S,
\]
where $L$ is the Lagrangian of a point particle in special relativity and $S$ the action. 
So we can write
\be
\psi(x_B,t_B) \propto \lim_{N\rightarrow \infty} \int dx_1 \int dx_2\ldots \int dx_N \psi(x_A,t_A)\exp\left[\frac{i}{\hbar} \int dt L(x,\dot x)\right]
\ee
or
\be\label{pathint}
\boxed{\psi(x_B,t_B) = \int \mathcal D x \psi(x_A,t_A)\exp\left[\frac{i}{\hbar} \int dt L(x,\dot x)\right].}
\ee
The factor of $\sqrt{1-v^2/c^2}$ in the Lagrangian is nothing but the relationship between proper time and coordinate time, $L= -mc^2 d\tau/(dt)$.
To include an interaction, we may use general relativity (GR), a description of gravity. The relationship between proper time and coordinate time in GR is
\be
d\tau=\sqrt{-g_{\mu\nu}dx^\mu dx^\nu}/c.
\ee
The Lagrangian of a point particle is still $L= -mc^2 d\tau/(dt)$.

\subsection{Derivation of the Schr\"odinger equation}

We shall follow the approach of Feynman \cite{Feynman1948}. We start by using the action
\be
S=-\int mc^2 \sqrt{-g_{\mu\nu}u^\mu u^\nu}dt \approx -\int mc^2\left(1-\tfrac12h_{00}+h_{0j}\frac{u^j}{c}
-\tfrac12(\delta_{jk}-h_{jk})\frac{u^j}{c}\frac{u^k}{c}\right)dt
\ee
where we have expanded the square-root to leading order, choosing as a laboratory frame one in which the particle is moving slowly and the gravitational potential is weak.\footnote{The minus sign of $h_{00}$ comes from $\eta_{00}=-1$}  In this frame, $u^j$ is the usual 3-velocity. We now compute the path integral for an infinitesimal time interval $t\rightarrow t+ \epsilon$ and an infinitesimal distance $q^\mu=(x_B)^\mu-(x_A)^\mu$. For an infinitesimal $\epsilon,$ we have $v^j=q^j/\epsilon$, so
\be
\psi(t+\epsilon,(x_A)^j)=N\int d^3q\, \psi(t,(x_A)^j-q^j)e^{-i \frac{mc^2\epsilon}{\hbar}\left(1-\tfrac12 h_{00}\right)} e^{-\frac12 A_{jk}q^jq^k+B_jq^j}
\ee
where $N$ is a normalization factor and
\be
A_{jk}\equiv -\frac{im}{\hbar \eps} (\delta_{jk}-h_{jk}), \quad B_j\equiv  \frac{im c }{\hbar}h_{0j}.
\ee
We can expand in powers of $\eps, q^\mu$:
\bea
\psi+\epsilon\partial_t \psi \\ =N\int d^3q \left(\psi-q^j \partial_j \psi+\tfrac12 q^jq^k\partial_j\partial_k \psi\right) \left(1-i \frac{mc^2\epsilon}{\hbar}\left(1-\tfrac12 h_{00}\right)\right) \exp\left[\frac{1}{2} A_{jk}q^jq^k+B_jq^j \right]\nonumber
\eea
where $\psi\equiv \psi(t,\vec x_A)$. We compute
\be
\int
e^{ -\frac{1}{2} A_{jk}q^jq^k + B_jq^j}d^3q =\frac{(2\pi)^{3/2}}{\sqrt{\det A}}e^{-\frac12 B_j(A^{-1})_{jk}B_k},
\ee
where $\det A$ is the determinant of $A$ and $A^{-1}$
is the inverse matrix. We obtain
\bea
\psi+\epsilon\partial_t \psi&=&N\frac{(2\pi)^{3/2}}{\sqrt{\det A}} \left[\left(1-i \frac{mc^2\eps}{\hbar}(1-\tfrac12 h_{00})\right)\psi \right. \nonumber \\ && \left.
-(\partial_j\psi)\frac{\partial}{\partial B_j}
+\frac 12 (\partial_j\partial_k\psi)\frac{\partial}{\partial B_j} \frac{\partial}{\partial B_k}  \right] \exp\left(\frac12 B_jB_k(A^{-1})_{jk}\right).
\eea
The normalization factor is determined from the fact that $\psi(t+\epsilon,\vec x_A)$ must approach $\psi(t,\vec x_A)$ for $\eps\rightarrow 0$. 
We carry out the derivatives. 
We now neglect all terms that are suppressed by two powers of $1/c$ or more, including the $h_{jk}$ terms, and terms proportional to $\eps^2$.
This leads to a Schr\"odinger equation
\be\label{Schrodinger}
\boxed{i\hbar \frac{d}{dt} \psi =- mc^2\tfrac12 h_{00}\psi
- \frac{\hbar^2}{2m} \left(\vec \nabla-m\vec H \right)^2 \psi,}
\ee
where we have substituted $\psi\rightarrow e^{-i \om_C t}\psi$. The 3-vector $\vec H$ is defined by $H_j\equiv (i c/\hbar)h_{0j}$. 

To see that this is the familiar Schr\"odinger equation, we note that $U=-h_{00}c^2/2$ is the scalar gravitational potential. The significance of $\vec H$ is a gravitational vector potential that describes ``frame dragging" for a rotating source mass. This post-Newtonian effect of GR is extremely small on Earth. 

From here on, we may derive the entire program of quantum mechanics, e.g., derive the conservation of the probability current to arrive at a interpretation of the wave function, the uncertainty relationship or commutation relations, and generalize the theory to describe multiple particles. This shows that quantum mechanics is a description of waves oscillating at the Compton frequency that explore all possible paths through curved spacetime.


\subsection{Derivation of the Dirac equation without gravity}

The theory still has important gaps. We do not know about spin yet, and while we started relativistically, the Schr\"odinger equation we obtained is only nonrelativistic. It is not straightforward to obtain a relativistic theory in analogy to Eq. (\ref{pathint}). The difficulties are substantial, so we will tackle them for a special relativistic framework, without gravity.

The difficulties arose when integrating the exponential $\exp(-imc^2\sqrt{1-v^2/c^2})$ over all of space, because there is no limit on the velocity $v$. 
In particular, the integrand is not well behaved when $v\rightarrow c$ and beyond. One might attempt to cut the integral before $v=c$ or anywhere else, but this would not lead to a Lorentz-invariant theory. The reason is that any speed below $v=c$ is the rest frame of a physically possible  observer, and can thus not be excluded from the theory. Cutting at $v=c$, on the other hand, doesn't avoid divergence. Our luck in the previous chapter was that paths at and outside the light cone were suppressed by gaussian functions in the nonrelativistic framework. But now that we want to develop the relativistic theory, this is no longer possible. We are led to accept that the divergence is not a computational problem, but an indication that the model that we have used so far needs to be refined.

\subsubsection{Re-writing the proper time}

Since the difficulty arises from the square-root in the exponential, we shall try to avoid the square root. Using the momentum $\vec p =\nabla_{\dot {\vec  q}} L=m \vec v \gamma$ we shall re-write $L= \vec p \cdot \dot{\vec q}-H$. The function $H$, the Hamiltonian, turns out to be $H=mc^2\gamma=\sqrt{p^2c^2+m^2c^4}$. We then use Dirac's trick of replacing
\be
\sqrt{p^2c^2+m^2c^4}\equiv c(-\vec \alpha)\cdot \vec p+\beta mc^2.
\ee
In order for this to work, we must require $(-\vec \alpha)^2=1, \beta^2=1$, and $(-\vec \alpha)\beta+\beta(-\vec \alpha)=0$. (The sign of $\alpha$ is arbitrary. We choose it to be negative, so that our end result has the familiar form.) It is clear that $\vec \alpha$ and $\beta$ cannot be ordinary numbers, but they may be $4\times 4$ matrices, e.g.,
\be
\vec \alpha=\left(\begin{array}{cc} 0 & \vec \sigma \\ \vec \sigma & 0 \end{array}\right),\quad \beta=\left(\begin{array}{cc} 1 & 0 \\ 0 & -1 \end{array}\right),
\ee
where $\vec \sigma$ are the Pauli matrices. We now have
\be
L_\Box=\vec p\cdot \dot{\vec q}+c\vec \alpha \cdot \vec p-mc^2\beta.
\ee
Note that this Lagrangian is a matrix. For now, we shall continue our calculation and interpret this fact if and when we obtain a result.

We could now try inserting the new Lagrangian into the path integral, Eq. (\ref{pathint}) and use $\vec p=m\vec v/\sqrt{1-v^2/c^2}$. This, however, brings back the square-root and thus an integrand which is not well-behaved at the light cone. We can, however, generalize the path integral by treating $\vec p, \vec q$ as independent variables and integrate over all trajectories in phase-space, not just all trajectories in real space. We thus write
\be\label{Diracpath}
\boxed{\psi(\vec x_B,t_B) \int \frac{\mathcal D^3 p_1}{(2\pi)^3} \int \mathcal D^3x\, \exp\left[\frac{-i}{\hbar} \int dt \left(\vec p\cdot \dot{\vec q}+c\vec \alpha \cdot \vec p-mc^2\beta\right)\right]\psi(\vec x_A,t_A).}
\ee

\subsubsection{Derivation of the Dirac equation}
As before, consider an infinitesimal interval $t\rightarrow t+\epsilon, \vec x\rightarrow \vec x+\vec q$. We may use just one integration each. Noting that $\dot{\vec q}=\vec q/\epsilon$, we obtain
\be
\psi(t+\epsilon, x) =N \int \frac{d^3 p}{(2\pi)^3} \int d^3q\exp\left[\frac{-i}{\hbar} \vec p\cdot \vec q+\frac{i\epsilon}{\hbar}\left(-c\vec \alpha \cdot \vec p+mc^2\beta\right)\right]\psi(t,\vec x-\vec q).
\ee
We note that $\int d^3 q e^{-i \vec p\cdot \vec q/\hbar}\psi(\vec x-\vec q) =-e^{i\vec p\cdot \vec x/\hbar} \Phi(-\vec p,t)$ is given by the momentum-space wave function $\Phi(\vec p,t)$. Inserting this into the path integral gives
\be
\psi(t+\epsilon, x) =-N \int \frac{d^3 p}{(2\pi)^3} \exp\left[\frac{i\epsilon}{\hbar}\left(-c\vec \alpha \cdot \vec p+mc^2\beta\right)\right]e^{-i\vec p\cdot \vec x/\hbar} \Phi(-\vec p,t).
\ee
Since $\epsilon$ is an infinitesimal quantity, we may expand to first order on both sides of the equation:
\bea
\psi(t,\vec x)+\epsilon \dot \psi(t,\vec x)
\\ =-N \int \frac{d^3 p}{(2\pi)^3} e^{-i\vec p\cdot \vec x/\hbar} \Phi(-\vec p,t)
-N \int \frac{d^3 p}{(2\pi)^3} \frac{i\epsilon}{\hbar}\left(-c\vec \alpha \cdot \vec p+mc^2\beta\right)e^{-i\vec p\cdot \vec x/\hbar} \Phi(-\vec p,t).\nonumber
\eea
The first term is the reverse Fourier transform and yields the position-space wave function. We determine the normalization factor by noting that if $\epsilon=0$, the right hand side must equal the left hand side, i.e., $N=-1$. The remaining terms are
\be
\dot \psi(t,\vec x)= \int \frac{d^3 p}{(2\pi)^3} \frac{i}{\hbar}\left(-c\vec \alpha \cdot \vec p+mc^2\beta\right)e^{-i\vec p\cdot \vec x/\hbar} \Phi(-\vec p).
\ee
We can replace the $\vec p$ in the parenthesis by the derivative $-(\hbar/i)\vec \nabla$ acting on the exponential,
\be
\boxed{i\hbar \dot \psi= \left[\frac{\hbar}{i}c\vec \alpha \cdot \vec \nabla+mc^2\beta\right]\psi,}
\ee
the Dirac equation!\footnote{I derived this on board the train to Varenna on July 14, 2013.} We have thus arrived at a relativistic wave equation, and discovered spin. Our need to introduce the $4\times 4$ matrices $\vec \alpha$ and $\beta$ means the wave function is a vector having 4 components. We could now derive conserved quantities, find solutions to the Dirac equation, and recover the Schr\"odinger equation in the nonrelativistic limit. This would show us that the 4 components of $\psi$ are the particle and antiparticle with spin up and spin down, respectively.

Our notion of an elementary particle as a single clock turned out to be incompatible with relativity. Rather, a particle is a set of four clocks, two of which tick forward, two backward. The $4\times 4$ langrangian gives the time lags in an experiment comparing any of the four to another one.

\subsubsection{Interpretation}

We now come back to the interpretation: Let us label the spinor components of $\psi$ by an index $s=1\ldots 4$. If a particle is found at four-position $x_A^\mu$ in a spin state $s$, we may call this a spinor event $(A,s)$. The components of the Lagrangian $(L_\Box)^r_s dt$ then represent the phase accumulated by the state between two infinitesimally separated spinor events $(A,s)$ and $(B,r)$. The phase is, e.g., $\phi=(L_\Box)^1_1\eps= (pv+mc^2)\eps$ for $r=s=1$, $(pv-mc^2)\eps$ for $r=s=3$, and $-cp\eps$ if $r=3, s=1$, where $\eps$ is an infinitesimal coordinate time interval. To calculate the phases between two events, the events have to be amended by a discrete coordinate $s$.

\begin{figure}
\centering
\epsfig{file=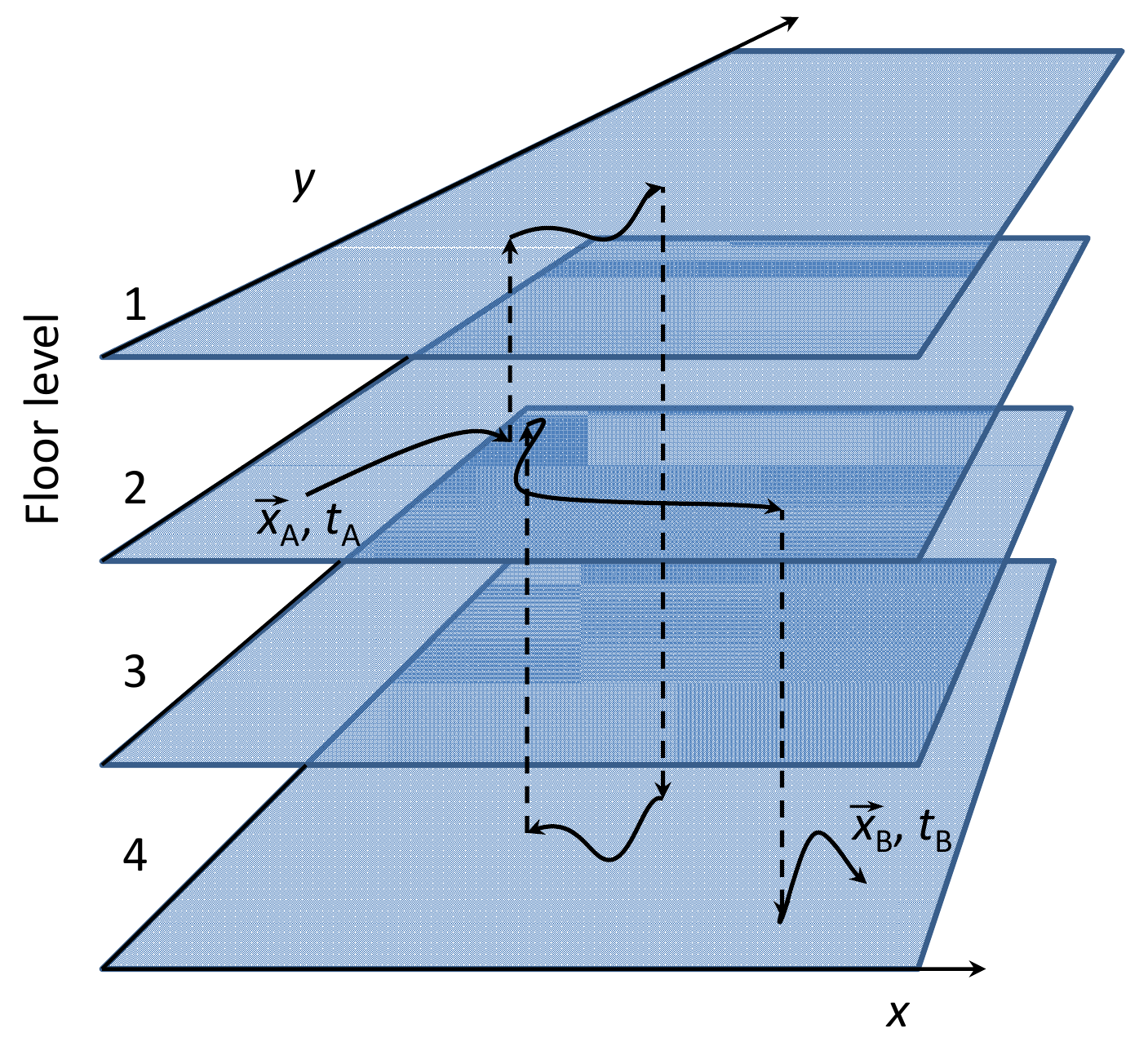,width=0.3\textwidth}
\caption{\label{building} One example for a path included in the path integral Eq. (\ref{Diracpath}). The start and end points need to be specified by location, time and floor level; the particle randomly jumps between floor levels on its way from A to B.}
\end{figure}

The path integral Eq. (\ref{Diracpath}) is over all of phase space, $\int \mathcal D p \mathcal D q$. Thus, there are arbitrary combinations of matrices $\alpha_x,\alpha_y,\alpha_z$ in the exponential of one path, e.g., $\ldots \times e^{-\frac i\hbar c \alpha_x p_x}  e^{-\frac i\hbar c \alpha_z p_z} e^{-\frac i\hbar c \alpha_y p_y}\times \ldots \psi$. Since each term with a matrix may change the spin $s$, the particle not only takes all possible paths through phase space, but thereby also goes through all possible paths through spin space (Fig. \ref{building}). Loosely, we may draw an analogy between the propagation of a Dirac particle and observers carrying clocks on random paths through a building having four floors in which proper time passes at different rates - forward and backward. In such a building, time, geographical latitude and longitude $A$ as well as the floor level $s$ constitute a full description of an event $(A,s)$.

We consider two special cases: (i) Eigenstates of $L_\Box$,
\be
e^{iL_\Box/\hbar dt}\psi=e^{-i\om_C d\tau}\psi=e^{-ip_\mu d x^\mu/\hbar}\psi,
\ee
are characterized by a definite momentum $\vec p$ and do not change spin while propagating. The accumulated phase is equal to the proper time times the Compton frequency, i.e., the picture of matter waves as clocks applies exactly - not just in the nonrelativistic limit as before. (ii) A particle on a classical path extremizes its action. It will thus keep its spin state constant, as switching between such states (floor levels in the analogy) reduces the absolute value of the phase. Such particles can be treated without regard to spin and the phase accumulated along the path is $\phi=-\om_C\tau$.

\subsubsection{Derivation of the matter-waves-as-clocks picture from the Dirac equation}

To complete the demonstration that the clock picture and standard quantum mechanics follow from each other we outline how the clock picture can be derived from the Dirac equation. With $H=c\vec \alpha \cdot \vec p+mc^2\beta$, we see that \be
|\psi(t+T)\rangle=e^{iH(t_1) \epsilon/\hbar}\ldots e^{iH(t_N) \epsilon/\hbar}|\psi(t)\rangle,
\ee
where $t_1\equiv 0, t_N\equiv t+T$ divide the interval $T$ in $N-1$ parts. Using position and momentum eigenstates $|\vec x, s\rangle, |\vec p, s\rangle$ with spin $s$, we insert one each of the unity operators
\be
\int dx \sum_s |\vec x, s\rangle\langle \vec x, s|,\quad \sum _s \int \frac{d^3p}{(2\pi)^3} |\vec p, s\rangle\langle \vec p, s|
\ee
between the exponentials. Noting that $\langle \vec p, s|\vec x, t\rangle =\langle \vec p|\vec x\rangle \delta_{st}=e^{i\vec p\vec x/\hbar}\delta_{st}$ leads to Eq. (\ref{Diracpath}).

\subsection{Derivation of a Dirac equation with electromagnetic potentials}

The generalization to a particle in an electromagnetic field is straightforward by starting with the classical Lagrangian of a charged particle
\be
L=-mc^2\sqrt{1-v^2/c^2}+\frac ec \vec A\vec v-e\Phi,
\ee
where the vector and scalar potential $\vec A, \Phi$ are differentiable but otherwise arbitrary functions of $\vec x, t$ (there is no restriction to potentials that are at most quadratic in the coordinates as in nonrelativistic path integrals). Proceeding as above, we obtain
\bea
\vec p&=&\gamma m \vec v+\frac ec \vec A, \quad H=\sqrt{(c\vec p-e\vec A)^2+m^2c^4}+e\Phi, \nonumber \\
L_\Box&=&\vec p \cdot \vec{\dot q}-\vec \alpha\cdot(c\vec p-e\vec A)-mc^2\beta-e\Phi,
\eea
and again calculate a path integral over an infinitesimal interval $t\rightarrow t+\epsilon, \vec x\rightarrow \vec x+\vec q$, as in Eq. (\ref{Diracpath}). This leads to the Dirac equation
\be
i\hbar \dot\psi=\left[ c \vec \alpha  \cdot \left(\frac i\hbar \vec \nabla -\frac ec \vec A(\vec x)\right)-mc^2\beta-e\Phi(\vec x)\right]\psi.
\ee
From the basic equations of motion, we could now proceed to construct the theory of interacting Fermions, i.e., quantum electrodynamics. Of course, this is a huge undertaking, requiring second quantization as a way of dealing with multi-particle systems. We will not consider this.

\subsection{Derivation of the Dirac equation with gravity, in curved space-time}
\subsubsection{Derivation}
The proper time is expressed by the Lagrangian
\be
L=\frac{d\tau}{dt} = -mc\sqrt{-g_{\mu\nu}\dot x^\mu\dot x^\nu}.
\ee
The momentum is
\be
p_\mu=\frac{\partial L}{\partial \dot x^\mu}=mc \frac{g_{\mu\nu}\dot x^\nu}{\sqrt{-g_{\mu\nu}\dot x^\mu\dot x^\nu}}
\ee
and satisfies
\be\label{em2c2}
g^{\kappa\lambda}p_\kappa p_\lambda=m^2c^2\frac{g^{\kappa\lambda}g_{\kappa\nu}\dot x^\nu g_{\lambda\mu}\dot x^\mu}{-g_{\mu\nu}\dot x^\mu\dot x^\nu}=-m^2c^2.
\ee
We note that
\be
p_\mu \dot x^\mu-L=mc \frac{g_{\mu\nu}\dot x^\nu \dot x^\mu}{\sqrt{-g_{\mu\nu}\dot x^\mu\dot x^\nu}}-L=0.
\ee
Now we work in a specific frame and use
\be
H=p_k\dot x^k-L= p_\mu\dot x^\mu-p_0\dot x^0-L = -p_0\dot x^0=-p_0c= -c\sqrt{p_0^2}.
\ee
From Eq. (\ref{em2c2}), we obtain
\be
-m^2c^2=g^{\mu \nu}p_\mu p_\nu=g^{00}p_0p_0+2g^{0j}p_0p_j+g^{jk}p_j p_k,
\ee
which we may solve for $p_0^2$ and insert:
\be
H=c\sqrt{\frac{1}{-g^{00}}\left(m^2c^2+2g^{0j}p_0p_j+g^{jk}p_jp_k\right)} \ee
At this point, let us define
\be
\bg^{\mu\nu}=\frac{g^{\mu\nu}}{-g^{00}}, \quad \barm^2=\frac{m^2}{-g^{00}}.
\ee
So that
\be
H=c\sqrt{\barm^2c^2+2\bg^{0j}p_0p_j+\bg^{jk}p_j p_k}.
\ee
In flat spacetime, this reduces to $\sqrt{p^2c^2+m^2c^4}$ as it should. We note that $p_0=-H/c$ under the square-root, so we have $H$ on the right hand side and the left hand side,
\be
H^2=\bar m^2c^4+2c\bg^{0j}p_j H+c^2\bg^{jk}p_j p_k. 
\ee
We obtain
\be
H=c\bg^{0j}p_j \pm c\sqrt{(\bg^{0j}\bg^{0k} +\bg^{jk})p_j p_k+\barm^2c^2}.
\ee
We pick the plus sign so the Hamiltonian reduces to the usual one in flat space-time. We now introduce a dreibein $d^a_j$ so that
\be
\bg^{0j}\bg^{0k} +\bg^{jk}=d^j_ad^k_b\eta^{ab}=d^j_ad^k_b\delta^{ab}.
\ee
We define
\be
\bal^j=d^j_a\alpha^a,
\ee
where $\alpha^{1,2,3}$ are the familiar Dirac matrices. It is easy to check that
\bea
\{\bal^j,\bal^k\}&=&d^j_ad^k_b(\alpha^a\alpha^b+\alpha^b\alpha^a)
=2d^j_ad^k_b\delta^{ab} =2(\bg^{0j}\bg^{0k} +\bg^{jk})\nonumber \\
\{\bal^j,\beta\} &=&d^j_a(\alpha^a\beta+\beta\alpha^a)=0.
\eea
where $\{a,b\}=ab+ba$ denotes the anticommutator. Thus,
\bea
\left(\bal^jp_j+\beta \barm c\right)^2&=&\bal^j\bal^kp_jp_k+(\bal^j\beta+\beta\bal^j)p_j\barm c+\beta^2\barm^2c^2\nonumber \\
&=&\tfrac12(\bal^j\bal^k+\bal^k\bal^j)p_jp_k+\beta^2\barm^2c^2\nonumber \\
&=&(\bg^{0j}\bg^{0k} +\bg^{jk})p_jp_k+\barm^2c^2.
\eea
So we define
\be
L_\diamond=p_k\dot q^k-c\bg^{0j}(x^k,t)p_j-c\left[(-\bal^j(x^k,t))p_j+\beta \barm(x^k,t)c\right]
\ee
where we have explicitly denoted that the $\bal$ and $\barm$ depend on the coordinate and the time. (As before, the sign before $\bar \alpha^j$ is arbitrary and chosen such that the end result will reduce to the familiar Dirac equation in flat space time.) If all that works, our path integral will be
\bea
\boxed{
\begin{aligned}
\psi(t+T)&=\int\frac{\mathcal D^3 p}{(2\pi)^3\sqrt{-g}}\int \mathcal D^3 x\sqrt{-g} \\ &\times \exp\left\{\int \left[-\frac i\hbar p_kq^k + \frac{ic}\hbar\bg^{0j}p_j -\frac{ic}{\hbar}\left(-\bal^jp_j+\beta \barm c\right)\right] dt\right\}\psi(t,\vec x).
\end{aligned}
}
\eea
As before, we calculate an infinitesimal step
\be
\psi(\vec x, t+\eps)=\int\frac{d^3p}{(2\pi)^3\sqrt{-g}}\int d^3 q\sqrt{-g} e^{-\frac i\hbar p_kq^k}e^{\frac{ic}{\hbar}\eps\bg^{0j}p_j} e^{-\frac{ic}{\hbar}\eps\left[-\bal^jp_j+\beta \barm c\right]}\psi(\vec x-\vec q,t).
\ee
Just as in the case without gravity, we are allowed to evaluate $\sqrt{-g}, \bg, \barm$ at $\vec x$ instead of $\vec x-\vec q$. That leaves us with
\bea
\psi(\vec x, t+\eps)&=&\int\frac{d^3p}{(2\pi)^3}e^{\frac{ic}{\hbar}\eps\bg^{0j}p_j}
e^{-\frac{ic}{\hbar}\eps\left[-\bal^jp_j+\beta \barm c\right]}\int d^3 q e^{-\frac i\hbar p_kq^k} \psi(\vec x-\vec q,t) \nonumber \\
&=&\int\frac{d^3p}{(2\pi)^3}e^{\frac{ic}{\hbar}\eps\bg^{0j}p_j}
e^{-\frac{ic}{\hbar}\eps\left[-\bal^jp_j+\beta \barm c\right]}  e^{-\frac i\hbar p_k x^k}\Phi(-\vec p,t)
\eea
We use $\psi(\vec x,t+\eps)=\psi(\vec x ,t)+\eps\dot\psi(\vec x,t)$ on the left hand side and obtain
\bea
i\hbar\dot\psi&=&-\int\frac{d^3p}{(2\pi)^3} c \left(\bg^{0j}p_j -\left[-\bal^jp_j+\beta \barm c\right]\right)  e^{-\frac i\hbar p_k x^k}\Phi(-\vec p,t) \nonumber \\ &=&
 \left[\frac{\hbar}{i}\left(\bal^j-\bg^{0j} \right)\partial_j+\beta \barm c^2\right]\int\frac{d^3p}{(2\pi)^3}   e^{-\frac i\hbar p_k x^k}\Phi(-\vec p,t).
\eea
We are now able to write the Dirac equation in curved space-time in compact form
\be\label{Diracgrav}
\boxed{i\hbar\dot\psi=
 \left[\frac{\hbar}{i}\left(\bal^j-\bg^{0j} \right)\partial_j+\beta \barm c^2\right]\psi,}
\ee
where the barred symbols are defined by
\be\label{diamonddefs}
\{\bal^j,\bal^k\}=2(\bg^{\mu\nu}+\bg^{0j}\bg^{0k}), \quad \{\bal^j,\beta\}=0, \quad \bar g^{\mu\nu}=\frac{g^{\mu\nu}}{-g^{00}}, \quad \barm=\frac{m}{\sqrt{-g^{00}}}.
\ee
The $\bal$ can be constructed from the standard Dirac matrices using the dreibein, as explained above. This Dirac equation describes the propagation of relativistic particles with spin through gravitational fields, which may be arbitrarily strong. Note that it has been derived from the picture of the matter wave as a clock, the way we derived the flat-space time Dirac eqution before.

\subsubsection{A simple limiting case}
In the weak-gravity limit, we have $g_{\mu\nu}=\eta_{\mu\nu}+h_{\mu\nu}$ with $|h_{00}|\ll 1$ and $h_{0j}=h_{jk}=0$. Thus, the dreibein satisfies
\be
\frac{1}{-g^{00}}\delta^{jk}=d^j_ad^k_b\delta^{ab}
\ee
so we may choose $d^j_a=\delta^j_a/\sqrt{-g^{00}}$. Thus, our Dirac equation reduces to
\be
i\hbar \dot\psi=
 \left[\frac{\hbar}{i(1-\tfrac 12 h^{00})}\alpha^j\partial_j+\beta m c^2+\tfrac 12\beta h^{00}mc^2\right]\psi
\ee
For a particle with low momentum, $\frac12 \beta h_{00} mc^2$ appears like a scalar potential. Newtonian mechanics, here we come. The $\beta$ in that potential makes sure that antimatter falls downward, another nice feat. An alternative way of writing this
\be
i (1-\tfrac 12 h^{00})\dot\psi=
 \left[\frac{1}{i}\alpha^j\partial_j+\beta \om_C\right]\psi
\ee
reveals once more that gravity in quantum mechanics is described by the gravitational redshift to the Compton frequency.


\subsubsection{Comparison to the usual form}
The Dirac equation in curved space time found in the literature \cite{BirrellDavies} is sometimes called the tensor representation of the Dirac equation (TRD) \cite{Arminion}. It reads
\be\label{standirac}
[i\hbar e^\mu_\alpha \gamma^\alpha (\partial_\mu-\Gamma_\mu)-mc]\psi=0
\ee
where
\be
\Gamma_\mu=\frac i4 \sigma_{\alpha\beta}[e^a_\nu\partial_\mu e^{\nu b}+e_\nu^a e^{\sigma b}\Gamma^\nu_{\sigma \mu}]
\ee
is the spin connection, which is not a tensor. Our Dirac equation, on the other hand, does not have a spin connection and thus belongs to the Quadruplet Representation of the Dirac theory ($QRD-0$) in which $e^\mu_\al \gamma^\al \Gamma_\mu =0$. It was recently shown that in an open neighborhood of each spacetime point, every TRD equation is in fact equivalent to a QRD equation and vice versa. This holds under ``mild assumptions" on the metric, the G\"odel universe being a notable exception \cite{Arminion}. We can use $e^\mu_\al \gamma^\alpha \Gamma_\mu=0$ in Eq. (\ref{standirac}) and re-write is as
\be
i\hbar e^0_\alpha \gamma^\alpha \dot \psi =-i\hbar e^k_\alpha \gam^\al \partial_k\psi +mc\psi.
\ee
We multiply both sides with $e^0_\bet\gam^\bet$
\be
-i \hbar g^{00} \dot \psi =-i\hbar e^0_\bet e^k_\alpha \gam^\al\gam^\bet \partial_k\psi +e^0_\bet \gam^\bet mc\psi
\ee
Let's consider the first term on the right hand side:
\be
e^0_\bet e^k_\al \gamma^\al \gamma^\bet=\frac12 (e^0_\bet e^k_\al\gamma^\al \gam^\bet+e^0_\al e^k_\bet \gam^\bet\gam^\al) =\frac12 (e^0_\bet e^k_\al\gamma^\al \gam^\bet+2e^0_\al e^k_\bet \eta^{\al\bet}-e^0_\al e^k_\bet \gam^\al\gam^\bet).
\ee
Note that the definition of the vierbein involves six unphysical degrees of freedom. They are three Lorentz boosts and three rotations. If we use the three Lorentz boosts to set
\be
e^0_a=0 \quad ({\rm for\,} a\neq 0),
\ee
we obtain
\be
e^0_\bet e^k_\al \gamma^\al \gamma^\bet
=\frac12 (e^0_0 e^k_\al\gamma^\al \gam^0-2e^0_0 e^k_0-e^0_0 e^k_\bet \gam^0\gam^\bet)=-e^0_0 e^k_a \al^a-e^0_0 e^k_0,
\ee
where we used
\be
\gam^0\equiv \beta, \quad \gam^k\equiv \gam^0\alpha^k, \quad \al^a\bet=-\bet \al^a, \quad \gam^a\gam^0=\gam^a\bet=\bet \al^a\bet=-\al^a.
\ee
We also use
\be
e^0_\bet\gam^\bet=e^0_0\bet+e^0_b\gam^b=e^0_0\bet
\ee
to bring the Dirac equation into the form
\be
-i \hbar g^{00} \dot \psi =i\hbar (e^0_0 e^k_a \al^a+e^0_0 e^k_0) \partial_k\psi +e^0_0 \bet mc\psi.
\ee
We can replace $e^0_0 e^k_0$ by the metric, since
\be
g^{0k}=e^0_\al e^k_\bet \eta^{\al\bet}=e^0_0 e^k_\bet \eta^{0\bet}=e^0_0 e^k_0.
\ee
Therefore,
\be
i \hbar \dot \psi =\left[\frac \hbar i \left(\breve a^k -\bar g^{0k}\right) \partial_k -\frac{e^0_0 \bet mc}{g^{00}}\right]\psi
\ee
where
\be
\breve{a}=\frac{e^0_0 e^k_a \al^a}{g^{00}}.
\ee
It remains to show that the $\breve \al$ satisfy the anticommutator Eq. (\ref{diamonddefs}). This can be done by calculating
\bea
(g^{00})^2\{\breve a^j,\breve a^k\} &=& (e^0_0)^2 e^k_a e^j_b( \al^a \al^b + \al^b\al^a) =2(e^0_0)^2 e^k_a e^j_b\eta^{ab} \nonumber \\
&=&2(e^0_0)^2 [e^k_\al e^j_\bet \eta^{\al\bet} - e^k_0 e^j_0\eta^{00} -e^k_0 e^j_b\eta^{0b} +e^k_a e^j_0\eta^{a0} ]
=2(e^0_0)^2 [g^{kj} + e^k_0 e^j_0]
\eea
and
\bea
g^{0j}g^{0k}=e^0_\al e^j_\bet e^0_\gam e^k_\delta \eta^{\al\bet}\eta^{\gam\delta}=(e^0_0)^2 e^j_0 e^k_0.
\eea
Finally, inserting $e^0_0=\sqrt{-g^{00}}$ brings the standard form of the Dirac equation into the form that we derived from the path integral, Eq. (\ref{Diracgrav}). Our equation and the standard form are equivalent.

\subsection{Discussion}

Assuming that the phase accumulated by a matter wave packet is always proportional to the Compton frequency times the proper time measured along the path taken by the wave packet, we have derived the equations of motion of quantum mechanics. Our results hold for gravitational fields of any strength, wave packets of any speed, and with or without spin (the case of a spinless particle can be derived by iterating the Dirac equation). Note that all Lagrangians we have used are more or less complicated restatements of the Compton frequency times the proper time, for eigenfunctions of the Lagrangian. There is no exception to the rule that ``rocks" (massive wave packets) are clocks.




\subsection{Review of some counterarguments}
Having completed our demonstration, we briefly revisit some arguments that have been raised against the ``clock picture."  In particular, we examine those arguments that reject the notion that wave-packets in matter-wave interferometers can be treated like two clocks that measure the proper time difference along two trajectories.  Those who make these arguments find support in the fact that the phase of a matter-wave interferometer can be determined in a representation-free (with respect to the wave-packet's position or momentum) formalism \cite{SchleichNJP,SchleichPRL}, without explicit reference to the gravitational redshift, Compton frequency, or the proper time in the non-relativistic limit \cite{WolfCQG}, and that for some interferometer geometries, the free evolution phase difference accumulated by wave-packets traveling along different arms of the interferometer is zero \cite{JoeSam,WolfCQG,Jaeckel}. While these points are technically correct, they do not refute the clock picture, as they are all based on the Schr\"odinger/Dirac formulation of quantum mechanics, which we have shown can be derived from the clock picture. 

\subsection{Conclusion}

In general relativity, the trajectory of a freely falling test particle is the one that leads to extremal proper time $\tau$. The phase accumulated by a wave packet traveling between events $A$ and $B$ is given by the proper time elapsed along its path
\be
\phi_{\rm clock}=\omega_{\rm clock}\tau_{AB},
\ee
where $\om_{\rm clock}$ is the frequency of the clock in its own rest frame. The path of a matter wave packet is determined from the same principle of least action, and its phase given by
\be
\phi=-\om_C \tau_{AB},
\ee
and hence identical (equal and opposite) to the one of a clock ticking at the particle's Compton frequency. For free Dirac particles, these statements apply exactly to semiclassical states as well as to eigenspinors of $L_\Box$ (or $L_\diamond$ in curved space-time). We derived a path integral for the Dirac equation in which particles explore all paths in real space, momentum space, and spin space by starting only from a simple and easily motivated Lagrangian, $-mc^2d\tau/dt$, and the requirement that the theory be Lorentz invariant.

Dirac's trick is used as one of several mathematical devices to avoid the square-root in the action without changing the action, requiring the addition of unphysical degrees of freedom, or simply squaring the action. Note this led naturally to fermions, whereas we have not found a way to treat bosons directly (it is possible to find a Klein-Gordon equation by iterating the Dirac equation). The restriction of path integrals to potentials that are at most quadratic in the coordinates is lifted and thus found to be an artifact of nonrelativistic physics.  We also found an intuitive analogy between Dirac particles and paths in a building.
We may conclude that matter waves can be exactly treated as clocks.
Standard quantum mechanics is, in fact, predictated on the validity of general relativistic time dilation; in particular, if the gravitational redshift of the Compton frequency of matter waves \cite{redshift} was any different from the redshift of conventional clocks, the standard description of gravity by quantum mechanics would be incorrect.  

We note that in the theory of bosonic strings, the action is proportional to the area of the world-sheet swept out by the string, in generalization of the proper time. Perhaps a generalization of our methods will help find an alternative method introducing fermions to string theory \cite{Strings,Lust}.


\section{Brief summary of basics of atom interferometers}

\begin{figure} \centering
\epsfig{file=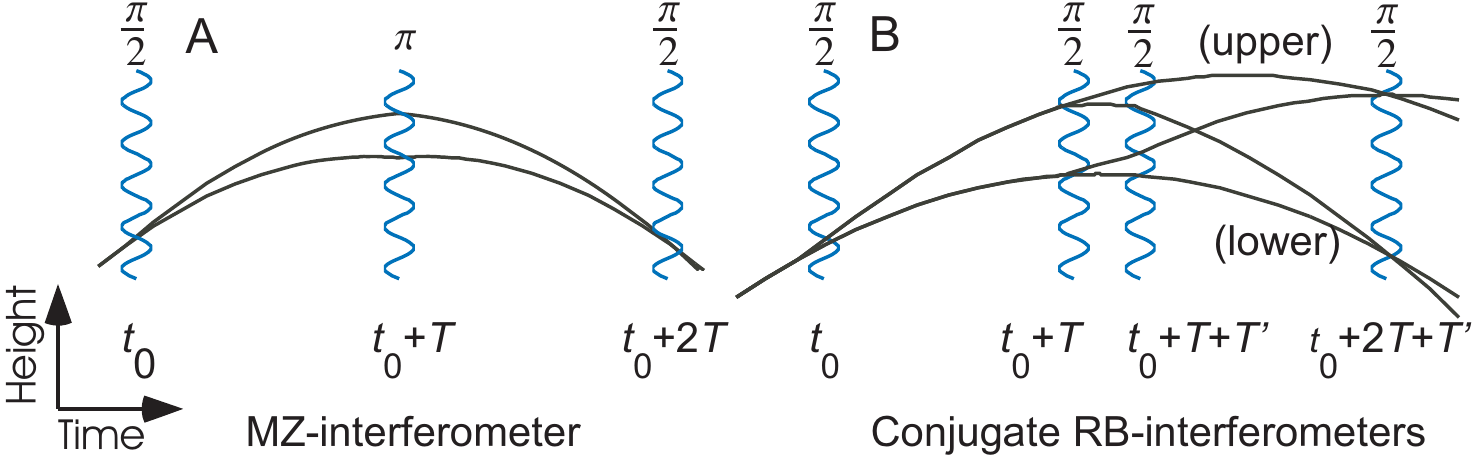,width=0.6\textwidth}
\caption{\label{schematic} A: MZI. ``$\pi/2$" pulses transfer
momentum with a probability of 1/2. They thus act as beam
splitters; ``$\pi$" pulses act as mirrors. B: Conjugate RBIs;
either is selected by the last $\pi/2$ pulse pair as described in
the text. Not shown are outputs of the third beam splitter, which
do not interfere.}
\end{figure}

We assume that the reader is already familiar with atom interferometry. Here, we give a brief description of the interferometer relevant in this article. The phase difference $\phi=\phi_F+\phi_I$ measured in an atom interferometer contains a contribution of the atom's evolution between the beam splitters $\phi_F$, and one of their interaction $\phi_I$. To discuss specifically the effects of large momentum transfer beam splitters, it is useful to consider Mach-Zehnder and Ramsey-Bord\'e interferometers (MZI and RBI) separately. In MZIs (Fig. \ref{schematic} A), $\phi_F$ vanishes for constant $g$, but gravity causes a $\phi_I$ by lowering the height at which the arms interact with the beam splitters. If the momentum transferred by the beam splitter is $2n\hbar k$, where $n$ is an integer, a MZI thus has a phase difference of \cite{PritchardReview,Peters,Petersmetrologia}
\be
\phi_{\rm MZ}=n(2kgT^2-\phi_L),
\ee
where $\phi_L=\phi_1-2\phi_2+\phi_3$ are
the phases $\phi_{1-3}$ of the laser fields at some reference
point. Here, multiphoton beam splitters lead to a linear increase
in phase. In RBIs, only one arm receives momentum from the beam
splitters (Fig. \ref{schematic} B). Thus, $\phi_F=2E_{\rm
kin}T/\hbar $ is nonzero due to the difference in kinetic energy
$E_{\rm kin}$. The same term, times minus two, enters $\phi_I$ due
to the modified locations at which the atoms interact. Summing up,
\begin{equation}\label{RBIphase} \phi_{\rm RB}=\pm 8n^2\omega_r T+2n k
g(T+T')T+n\phi_L.
\end{equation}
The plus and minus signs are for the upper and lower
interferometer, respectively, and
$\phi_L=\phi_2-\phi_1-\phi_4+\phi_3$ is given by the phases
$\phi_{1-4}$ of the laser pulses. The recoil term in RBIs scales
quadratically with the momentum splitting.

\subsection{Mach-Zehnder atom interferometers as redshift measurements}

It is amusing how closely a Mach-Zehnder interferometer resembles a classical measurement of the gravitational redshift with moving clocks. For simplicity, we assume a constant gravitational acceleration $g$ everywhere, i.e., we neglect the gravity gradient.

\subsubsection{Conventional redshift measurements with clocks}
\begin{figure}
\centering
\epsfig{file=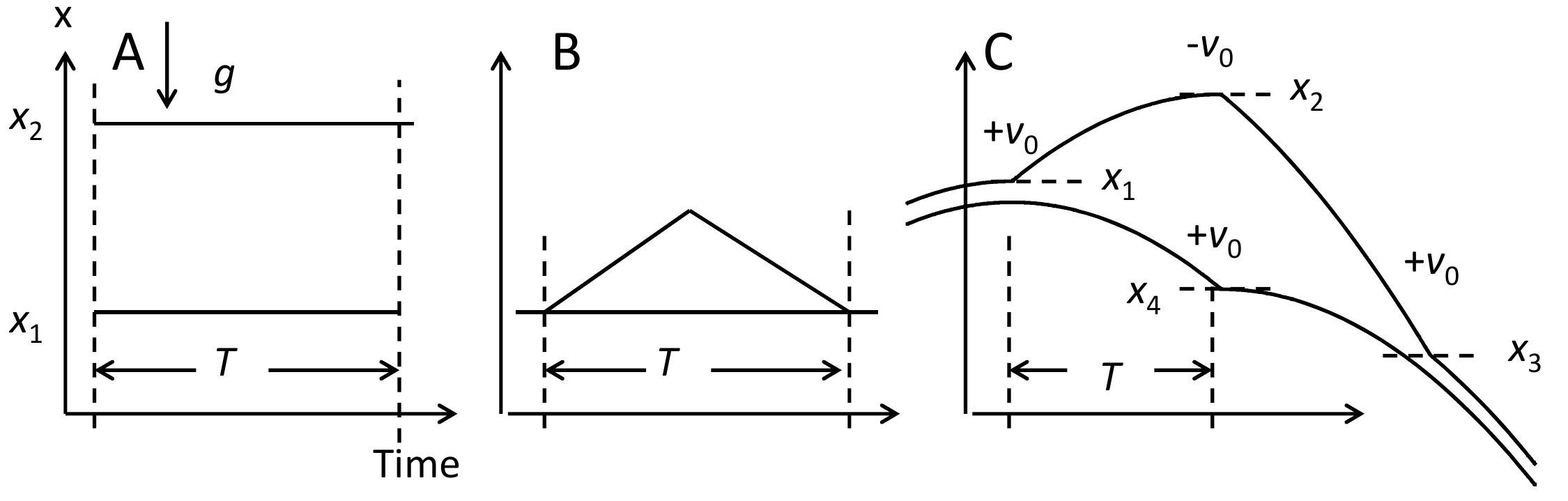,width=0.65\textwidth}
\caption{\label{ClockInterf} (A): World-lines of two clocks at constant locations $x_{1,2}$ in a particular coordinate frame. (B): One clock at a constant location is compared to a clock that is slowly transported away from the first clock and back. (C): A clock-comparison experiment with two clocks starting out at location $x_1$. One is then kicked upwards with a velocity change of $v_0$ and travels in free-fall to $x_2$, where it experiences a velocity change of $-v_0$, to arrive at $x_3$. The other moves from $x_1$ to $x_3$ on two free-fall trajectories, with a velocity change of $+v_0$ at $x_4$. The small distance between the clocks at the beginning and end is for purposes of this drawing only, and assumed to be negligibly small.}
\end{figure}

Consider the experiment shown in Fig. \ref{ClockInterf}, A. A pair of similar clocks having a proper frequency $\om$ each are held at constant positions, having a height difference $h$ that gives rise to a gravitational potential difference $\Delta U$. They will exhibit a frequency ratio $\om_1/\om_2=1+\Delta U/c^2$ due to the gravitational redshift.\footnote{Note that the absolute frequency of the clocks $\om_1, \om_2$ drops out of this expression} While running for a coordinate time interval $T$, they will accumulate a phase shift. The phase shift could be measured, e.g., by comparing the clocks via light signals or by the experiment shown in Fig. \ref{ClockInterf}, B: Two clocks are synchronized when they are at a common location, then moved apart and brought back together. The gravitational potential difference is now time-dependent, and so
\be
\phi_U=\omega \int_0^T \frac{\Delta U}{c^2}dt .
\ee
If the velocity of the clock's motion is not negligible, the special relativistic time dilation reduces the proper time. To leading order,
\be\label{vessotequation}
\phi=\phi_U+\phi_{\rm TD}=\omega \int_0^T \left(\frac{\Delta U}{c^2}-\frac12 \frac{v^2}{c^2}\right)dt.
\ee

\subsubsection{Correcting for time dilation}
To measure the gravitational redshift with moving clocks,\footnote{In such experiments, the linear Doppler effect has to be compensated for. Two-way radio links are available for this purpose. We will not consider this.} one may measure the clock's velocity $v$ as function of time, for example by radar. The time dilation term $\phi_{\rm TD}$ can be calculated and subtracted from Eq. (\ref{vessotequation}), so that a measurement of the gravitational redshift $\phi_U$ is obtained as $\phi_U=\phi-\phi_{\rm TD}$. This is the basic principle of, e.g., gravity-probe B and other experiments with spaceborne clocks \cite{Vessot}.

\subsubsection{Experiment with piecewise freely falling clocks}
Consider the slightly more complicated clock-comparison experiment shown in Fig. \ref{ClockInterf}, C. The clocks are initially synchronized at a common location and made to take two different paths by kicking (sit venia verbo) in intervals $T$. Each kick provides a velocity change by $v_0$. The clocks are compared after their paths merge at $t=2T$.\footnote{Several other versions of this experiment are possible, for example one in which both clocks are kicked two times each, or one in which the lower clock is kicked three times. The reader is invited to verify that the results of this chapter apply to any of these configurations, as well as to different initial locations and velocities of the two clocks, so long as the clocks are in the same position and same velocity as each other initially and finally.} We can easily generalize Eq. (\ref{vessotequation}) to calculate the phase difference shown by the clocks:\footnote{We assume that the velocity change does not perturb the operation of the clock so that the clocks are perfect realizations of proper-time measurements.}
\be
\phi=\omega \int_0^{2T} \left(\frac{\Delta U}{c^2}-\frac12 \frac{v_1^2-v_2^2}{c^2}\right)dt.
\ee
To subtract the time dilation term, we can monitor the trajectories as before. Under our assumptions of free fall with a constant gravitational acceleration, there is, however, a simpler method. The time dilation phase equals
\be
\phi_{\rm TD}=-\frac{\omega}{2c^2} \int_0^{2T} \left(v_1^2-v_2^2\right)dt =
-\om\frac{v_0}{c^2} gT^2=\om\frac{v_0}{c^2}(x_1-x_2+x_3-x_4),
\ee
where we labeled the coordinates of the turning points as in Fig. \ref{ClockInterf} C. It is thus sufficient to measure the coordinates of the turning points. We have only assumed that Newtonian mechanics is valid and that the clocks are falling with a constant acceleration of free fall that is identical for both clocks. We did not make any assumptions about the origin or magnitude of $g$.\footnote{The reader is invited to verify that the above results hold for arbitary initial positions and initial velocities.} We now have a strategy for our redshift experiment with clocks: send the clocks on the trajectories given in Fig. \ref{ClockInterf} C and measure the total phase shift $\phi$ accumulated between them. Also measure $x_{1-4}$ and recover the redshift phase as
\be
\phi_U=\phi-\om\frac{v_0}{c^2}(x_1-x_2+x_3-x_4).
\ee

\subsubsection{Comparison to atom interferometer}
The clock-comparison experiment has an exact correspondence to a Mach-Zehnder atom interferometer. The free evolution of the wave packets yields a phase shift in analogy to the one between the clocks in the above experiment, if the clock frequency is replaced by the Compton frequency:
\be
\phi_{\rm F}=-\om_C \int_0^{2T} \left(\frac{\Delta U}{c^2}-\frac12 \frac{v_1^2-v_2^2}{c^2}\right)dt
\ee
As before, $\phi_{\rm F}=\phi_U+\phi_{\rm TD}$ can be decomposed into the redshift part $\phi_U$ and the time dilation $\phi_{\rm TD}$ which can be expressed as
\be
\phi_{\rm TD}=-\om_C\frac{v_0}{c^2}(x_1-x_2+x_3-x_4).
\ee
In an atom interferometer, the velocity changes by $v_0$ is provided by laser-atom interactions. For a laser with wavenumber $k$, the recoil velocity is $v_r=n\hbar k/m$, where $n$ is the number of photons that the atom interacts with. Inserting $\om_C=mc^2/\hbar$ and $v_0=n\hbar k/m$, we obtain
\be
\phi_{\rm TD}=-n k(x_1-x_2+x_3-x_4).
\ee
The laser-atom interaction also imparts a phase to the matter wave, $\phi_{\rm I}$: Whenever a photon is absorbed, its phase is added to the matter wave. When a photon is emitted, its phase is subtracted. As the photons propagate by a distance $x$, they accumulate a phase $kx$.\footnote{For the following calculation, we shall refer all photon phases to the location $x=0$, though other conventions would lead to the same result.} Referring to Fig. \ref{ClockInterf} C, phase is imparted on the upper wave packet three times: A phase $+nk x_1$ at $t=0$; a phase $-nkx_2$ at $t=T$ (the negative sign arises because the atom is kicked down at this point), and a phase $+nkx_3$ at $t=2T$. The lower atom received a phase shift of $+nkx_4$ at $t=T$. Taking the difference between the total phases imparted by the laser on the upper and lower path, respectively, the laser phase evaluates to
\be\label{cancellation}
\phi_{\rm I}=nk(x_1-x_2+x_3-x_4).
\ee
So we see that $\phi_{\rm TD}+\phi_{\rm I}=0$, i.e., the laser phase acts like a laser-based tracker for the atoms position that automatically adds a counterterm that cancels the time dilation phase. This means, the atom interferometer is in every respect analogous to a redshift measurement using a pair of clocks on the trajectories shown in Fig. \ref{ClockInterf} C. As before, our only assumptions were freely falling motion with a constant acceleration of arbitrary magnitude or origin, and an arbitrary initial velocity.

\subsubsection{Examples where interpretations as force measurements fail}

As is well-known, the free evolution phase $\phi_{\rm free}=0$ for the above situation of freely falling wave packets. It is tempting to generalize this notion and assert that it is always true, ignoring the fact that atom interferometers fundamentally measures potentials. This would mean the atom interferometer measures nothing but the physical acceleration of the trajectory of the atoms relative to the reference plane used in defining the laser phase \cite{WolfCQG}. However, if the physical acceleration is modified without changing the potential difference between the paths, the interferometer will not register the change; if the potential difference is changed without changing the acceleration, the interferometer will. These observations are inconsistent with an interpretation of the interferometer as a pure accelerometer, but consistent with an interpretation as a redshift measurement.

\begin{figure}
\centering
\epsfig{file=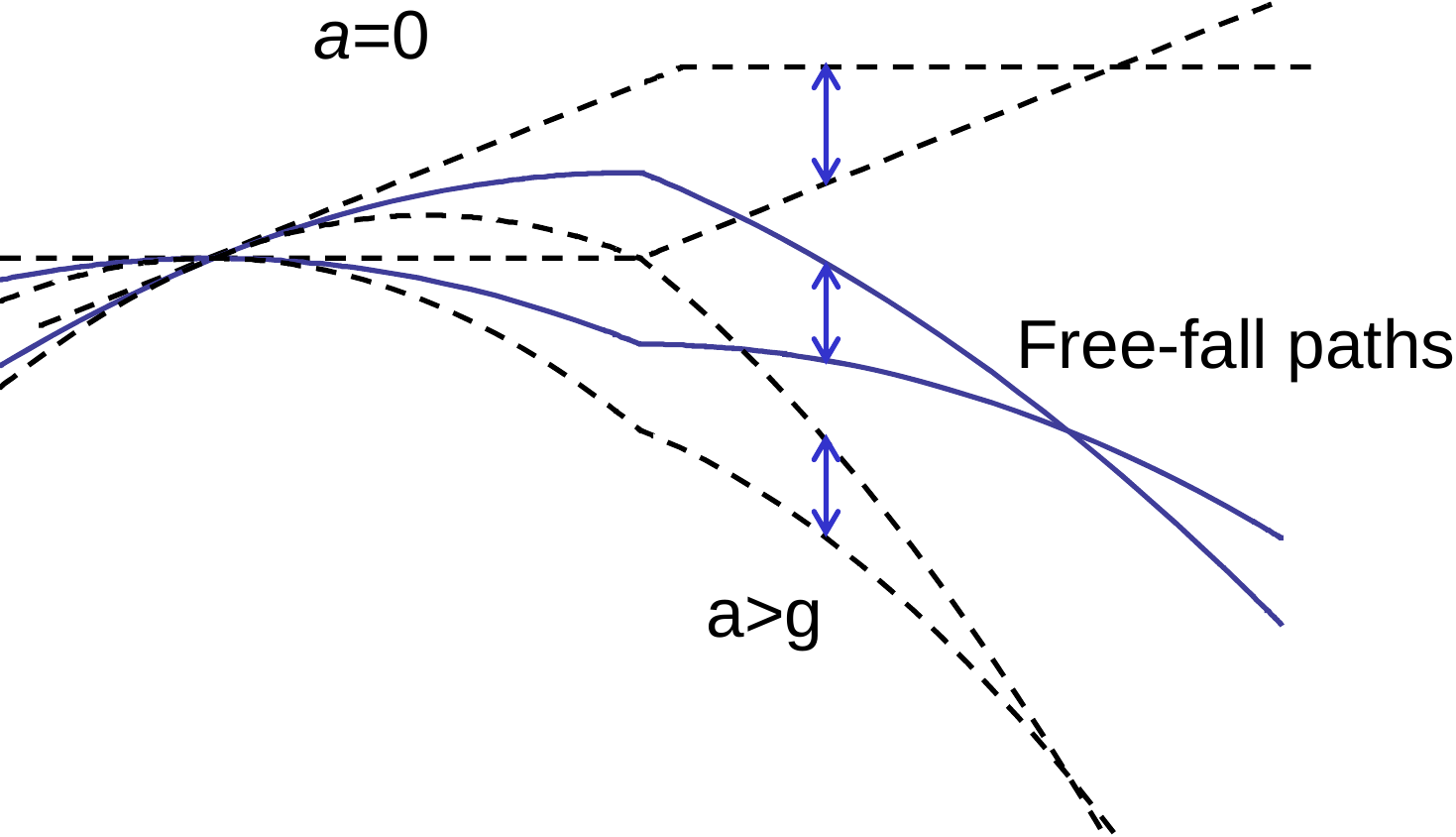,width=0.4\textwidth}\quad
\epsfig{file=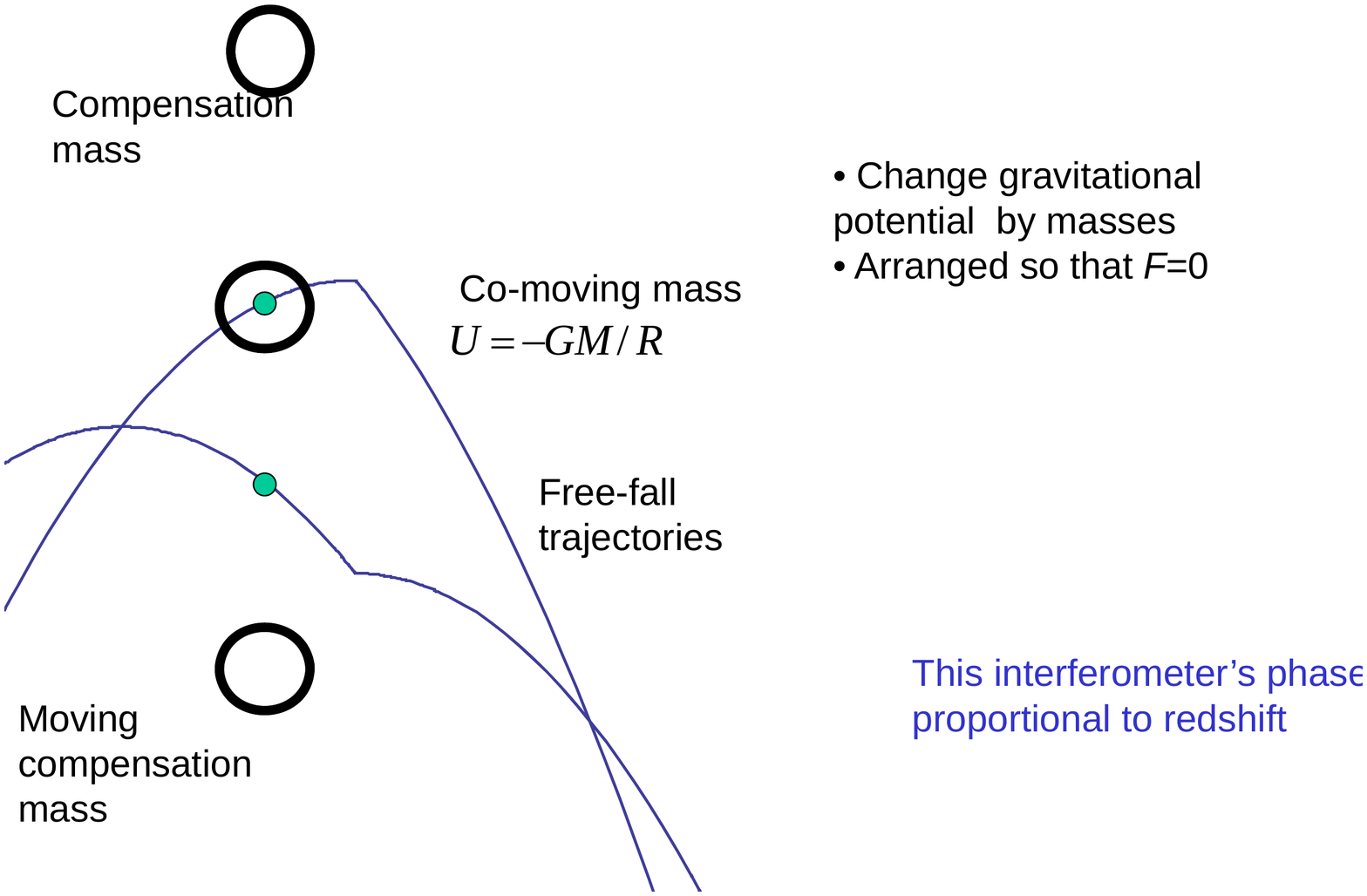,width=0.5\textwidth}
\caption{\label{MultiMZ} {\bf Left:} Mach-Zehnder interferometers in which the trajectories are modified by application of a common force while the potential difference between the paths and their separation is kept constant. This can be accomplished, e.g., with an optical lattice. The phase of the interferometer is not changed. {\bf Right:} Enclosing a trajectory with a hollow sphere of mass $M$ changes the potential at the wave packet's location by $-GM/R$ without applying a force. A lower compensation mass cancels the force due to the sphere on the lower wave packet, and an upper compensation mass cancels the force of the lower compensation mass on the upper wave packet. The phase of the interferometer is changed, but not the trajectories.}
\end{figure}

Consider, for example, the interferometer shown in Fig. \ref{MultiMZ}, left. It has the same trajectories as a conventional Mach-Zehnder, except that a common force is applied to the two wave packets so that the acceleration is not $g$ but can have any value. The force is applied in such a way that it doesn't affect the potential difference between the locations of the atom, which is possible using optical lattices. The interferometer will still measure the redshift and won't note the change of path. Conversely, Fig. \ref{MultiMZ}, right, shows how the potential can be changed without affecting the trajectories. The Mach-Zehnder atom interferometer will register this potential change even though the trajectories are completely unchanged. We will treat a similar  situation in  detail in Sec. \ref{GravABsect}.

These observations are inconsistent with an interpretation of the interferometer as a pure accelerometer but consistent with an interpretation as a redshift measurement. Both interpretations are simultaneously true in a simple gravitational potential. We conclude that the Mach-Zehnder atom interferometer always measures the integrated redshift along the two trajectories.

\section{Tests of relativity}
While the standard model of particle physics along with general relativity has been extremely successful, these theories are incompatible with each other, and there is strong observational evidence that they are incomplete. They are unable, e.g., to explain dark energy, or why the universe is dominated by matter when the theory exhibits perfect matter-antimatter (CPT-) symmetry, as any Lorentz-invariant, local field theory must. It is hoped that these theories can be unified and completed, perhaps by a version of string theory or loop quantum gravity. The natural energy scale for such theories is the Planck scale of $10^{19}\,$GeV, where corrections to general relativity and the standard model are expected to appear but where direct experimentation is impossible. One may, however, search for suppressed effects at lower energy scales in experiments of extreme precision. These effects will be minuscule and hard to discriminate against signals from conventional physics, except where the conventional physics signals are zero by an exact symmetry of the standard model. Examples for such symmetries are Lorentz and CPT symmetry. The numerous and extremely sensitive experimental searches for violations of them in flat space-time, however, have invariably failed to detect anomalies \cite{datatables}. By comparison, the Einstein Equivalence Principle (EEP) \cite{MTW} is a much less comprehensively tested symmetry and thus one of the most promising areas for finding low-energy signals of Planck-scale physics \cite{Damour2002}.

The EEP is the basis of gravitational theory \cite{Will2006,WillBook,Nobili} and holds that gravity affects all matter in exact proportion to its mass-energy: all objects experience the same acceleration of free fall $g,$ all clocks experience the same gravitational time dilation, and the laws of special relativity hold locally in inertial frames. Experimental tests of Lorentz invariance \cite{datatables}, local position invariance \cite{Vessot}, and the weak equivalence principle (WEP) \cite{Adelberger} have shown that nature adheres closely to this principle. If the EEP doesn't hold, general relativity cannot be valid. The EEP is or may be violated in many theories that attempt to join gravity with the standard model of particle physics - e.g., string theory, loop quantum gravity, higher dimensions, brane worlds - through new fields such as dilatons and moduli, or effective friction caused by quantum space-time foam.

\subsection{The standard model extension}
The significance of equivalence principle tests has been studied in the well-known parameterized post-Newtonian framework \cite{WillBook} and others \cite{Dimopoulos,Damour2002,DamourDonoghhue}. The gravitational standard model extension (SME) \cite{ColladayKostelecky,KosteleckyGravity,KosteleckyTassonPRL,KosteleckyTasson} offers important advantages: It is comprehensive, as it contains all known particles and interactions; it is consistent, as it preserves desirable features of the standard model such as conservation laws and the existence of a well-behaved flat space-time quantum field theory; it is predictive, as it can in principle describe the outcome of any experiment without any additional assumptions. It provides the most general way to describe Lorentz- and EEP-violations that preserves the above features and is in extensive use \cite{datatables}.

The SME is formulated from the standard model Lagrangian by adding all Lorentz- or CPT violating terms that can be formed from known fields and Lorentz tensors.  Different EEP tests will couple to different combinations of gravitational SME parameters. Using the standard model extension \cite{KosteleckyGravity,KosteleckyTasson} as a theoretical framework, we can answer, e.g., the following questions:
\begin{itemize}
\item Which parameters entering fundamental theories will a particular experiment measure? What influences the selection of the best species, like Rb/K or Rb/Rb? Can the Sun's gravitational field be used to perform additional measurements? How much will an experiment improve the overall constraints on equivalence principle violations? What are the implications for antimatter?
\item What is the significance of quantum tests of the equivalemce principle relative to tests using classical matter? Does gravity couple differently to particles of different spin? Or to particles exhibiting spin-orbit coupling?
\item How will use of species with different nuclear structure enhance the significance of particular tests?
\item What signals, if any, arise from the nonlinearity of general relativity? Does the validity of the EEP for particles in one rest frame guarantee its validity in frames in relative motion? Does its validity at one point imply its validity everywhere?
\end{itemize}


\subsubsection{The Fermionic sector}
The SME is constructed from the Lagrangians of the standard model and gravity by adding new interactions that violate Lorentz invariance and the Einstein Equivalence Principle. The non-gravitational Lagrangian density of a Dirac particle in the SME is
\bea
\mathcal L&=&\frac i2 \bar \psi \Gamma^\mu \overset{\leftrightarrow} D_\mu\psi-\bar\psi M\psi,\nonumber \\
M&=&m+a_\mu\gamma^\mu+b_\mu\gamma^5\gamma^\mu+\tfrac12H_{\mu\nu}\sigma^{\mu\nu},\nonumber \\
\Gamma_\nu&=&\gamma_\nu+c_{\mu\nu}\gamma^\mu+d_{\mu\nu}\gamma_5\gamma^\mu+e_\nu+if_{\nu}\gamma^5+\tfrac 12 g_{\lambda\mu\nu}\sigma^{\lambda\mu}.
\eea
We use a species specific notation $(a^w)_\mu, (b^w)_\mu,\ldots$, where $w$ can take the values n, p, and e denoting the neutron, the proton, and the electron, respectively. The Lorentz-violating interactions are encoded in eight Lorentz tensors $a-H$ known collectively as coefficients for Lorentz violation. Most of them lead to observable effects in flat space-time and have been constrained experimentally to levels well below those relevant here. The $a^\mu$ vector, however, can be removed from the flat space time equations of a single fermion via a redefinition of the energy scale and is unobservable. It becomes observable through effects in gravitational physics and is thus of particular interest.

The weak gravitational fields in the solar system can be described by a perturbation $h_{\mu\nu}$ to Minkowski spacetime. The perturbation is a function of the coefficients $a^\mu - H_{\mu\nu}$, via their contribution to the stress-energy tensor.  If, in addition, any of these coefficients has a non-metric coupling to gravity, those coefficients also become functions of $h_{\mu\nu}$.  In particular, $a_\mu=\bar a_\mu+\tilde a_\mu$ becomes the sum of its value in flat space-time $\bar a_\mu$ and a gravitationally-induced fluctuation $\tilde a_\mu$ (here, 'fluctuation' designates the change with gravitational potential, not random fluctuations) \cite{KosteleckyTasson}.  Although a nonzero $\bar a_\mu$ is unobservable on its own, the fluctuation $\tilde a_\mu$ induced by a non-metric coupling to gravity is observable.

For matter that is not spin-polarized, the $a$- and $c$-coefficients constitute a full description of EEP violation. For weak gravitational fields and slowly moving objects, it is sufficient to work with the temporal 0 and 00-components. This leaves six measurable coefficients $(a^p)_0, (a^n)_0, (a^e)_0, (c^p)_{00}, (c^n)_{00},$ and $(c^e)_{00}$. These violations of the EEP affect the free-fall trajectory for particles, as well as the phase shift $S/\hbar$ to the state of a quantum particle propagating along that (modified) trajectory, where $S$ is the action. The $c$-coefficients also change the binding energy of a composite particle, causing a position-dependence in the effective particle mass. These three effects combine to determine the leading order signal for atom interferometers \cite{redshiftPRL,Nuclear}. The effects in a particular experiment are set by the composition of the atoms in terms of protons, neutrons, and electrons, as well as by their inner structure, which determines how much the binding energy is affected by EEP-violation.
The effects of the $(a^w)_0$ are CPT-odd, or opposite for matter and antimatter, the effects of $(c^w)_{00}$ are CPT-even.  This means that experiments, despite using normal matter, will also be able to constrain anomalous physics of antimatter.

\subsubsection{The gravitational sector}
In a post-Newtonian approximation, the Lagrangian for the gravitational interaction between a central mass $M$ and a light point particle of mass $m$ in the SME is given by
\be\label{Lagrangian}
L=\frac12  m v^2+
G\frac{Mm}{2r}\left(2+3 \bar s^{00}+ \bar s^{jk} \hat r^j\hat r^k -3\bar s^{0j} v^j-\bar s^{0j}\hat r^j v^k \hat r^k\right).
\ee
For simplicity, we have taken $M$ to be at rest. We denote $\vec
r$ the separation between $M$ and $m$, pointing towards $m$. The
indices $j,k$ denote the spatial coordinates, $\vec v$ the
relative velocity, and $\hat r=\vec r/r$. The components of $\bar
s^{\mu\nu}=\bar s^{\nu\mu}$ specify Lorentz violation in gravity.
If they vanish, LLI is valid.

In principle, the components of $\bar s$ can be defined in any
inertial frame of reference. For experiments on Earth (as well as
on satellites), it is convenient to choose a Sun-centered
celestial equatorial reference frame \cite{KosteleckyMewesPRD}. 
The derivation of the time-dependent modulations of $g$ for an
observer on Earth involves taking into account the rotation and
orbit of the Earth; the Earth itself is modeled as a massive
sphere having a spherical moment of inertia of $I_\oplus\approx
M_\oplus R_\oplus^2/2$ \cite{WillBook} (not to be confused with
the conventional moment of inertia, which for Earth is about
$M_\oplus r_\oplus^2/3$). It suffices to consider the first order
in the Earth's orbital velocity $V_\oplus \simeq 10^{-4}c$. Bailey
and Kostelecky \cite{BaileyKostelecky} have studied this in
detail, and we refer the reader to this reference for the detailed
signal components in the purely gravitational sector.

\subsubsection{Electromagnetic sector}
\label{emsector}

An atom interferometer us also sensitive to Lorentz violation in the physics of electromagnetic fields, as it may cause variations of $k_{\rm eff}$. This physics is described by the Lagrangian density for the electromagnetic sector of the SME,
\be\label{SMEphotonLagrangian}
{\mathcal L}  =  - \frac 14 F^{\mu \nu} F_{\mu \nu} - \frac 14
(k_F)_{\kappa \lambda \mu \nu} F^{\kappa \lambda} F^{\mu \nu},
\ee
where $F^{\mu \nu}$ is the electromagnetic field tensor. The second term is proportional to a dimensionless tensor $(k_F)_{\kappa \lambda \mu \nu}$, which
vanishes, if Lorentz invariance holds on electrodynamics. The tensor has 19 independent components. The Maxwell equations in vacuum that are derived from the Eq. (\ref{SMEphotonLagrangian}) read
\be\label{inhomMaxwell}
\partial_\alpha F^\alpha_\mu + (k_F)_{\mu \alpha \beta \gamma} \partial^\alpha F^{\beta \gamma}
=
0,\quad \partial_\mu \tilde F^{\mu \nu} = 0,
\ee
where
\begin{equation}
\tilde F^{\mu \nu} = \frac12
\varepsilon^{\mu\nu\alpha\beta}F_{\alpha\beta}.
\end{equation}
They can be written in a 3+1 decomposition in analogy to the Maxwell equations in anisotropic media \cite{KosteleckyMewesPRD}. Lorentz violation in electrodynamics is thus analogous to electrodynamics in anisotropic media. It is convenient to define the linear combinations 
\begin{equation}\label{kappadef}
\begin{array}{ll} (\kappa_{DE})^{jk} = -2 (k_F)^{0j0k}, & (\kappa_{HB})^{jk} = \frac 12
\epsilon^{jpq} \epsilon^{krs} (k_F)^{pqrs}, \\ & \\
(\kappa_{DB})^{jk} = (k_F)^{0jpq} \epsilon^{kpq}, &
(\kappa_{HE})^{kj} = - (\kappa_{DB})^{jk} \nonumber
\end{array}
\end{equation}
and
\begin{eqnarray}\label{tildekappas}
(\tilde \kappa_{e+})^{jk} & = & \frac 12
(\kappa_{DE}+\kappa_{HB})^{jk}, \quad (\tilde \kappa_{o+})^{jk} =  \frac 12
(\kappa_{DB}+ \kappa_{HE})^{jk}, \quad \tilde \kappa_{tr}=\frac13(\kappa_{DE})^{ll}.\nonumber \\ (\tilde
\kappa_{e-})^{jk} & = & \frac 12 (\kappa_{DE}-\kappa_{HB})^{jk} -
\frac 13 \delta^{jk} (\kappa_{DE})^{ll},\quad (\tilde
\kappa_{o-})^{jk}  =  \frac 12 (\kappa_{DB}-\kappa_{HE})^{jk}.
\end{eqnarray}
The ten degrees of freedom of $\tilde\kappa_{o-}$ and $\tilde \kappa_{e+}$ encode birefringence; they are bounded to below $10^{-37}$ by observations of gamma-ray bursts \cite{KosteleckyMewesPRD,KosteleckyMewes}. The residual nine cause a dependence of the velocity of light on the direction of propagation. They are therefore relevant in interferometry experiments.

Finding the plane wave solutions yields the Lorentz-violating modification to the effective wavevector $k_{\rm eff}$ in the atom interferometer. Making the ansatz $F_{\mu\nu}(x)=F_{\mu\nu}(p)e^{-ik_\alpha x^\alpha}$ and inserting
into Eq. (\ref{inhomMaxwell}) one obtains the dispersion relation. Let
\begin{eqnarray}
\rho&=&-\frac12 \tilde k_\alpha{}^\alpha, \quad \sigma^2=\frac 12
(\tilde k_{\alpha\beta})^2-\rho^2,\nonumber \\ \tilde
k^{\alpha\beta}&=&(k_F)^{\alpha\mu\beta\nu}\hat p_\mu \hat p_\nu
\, , \quad \hat p^\mu=\frac{p^\mu}{|\vec p|}.
\end{eqnarray}
Then the dispersion relation is \cite{KosteleckyMewesPRD}
\begin{equation}
k^0_\pm=(1+\rho\pm\sigma)|\vec k|.
\end{equation}
The last term in this relation, which is proportional to $\sigma$, is purely polarization--dependent. Astrophysics experiments constrain such a birefringence to levels well below the levels relevant here \cite{KosteleckyMewes}. We can thus assume $\sigma=0$.

\subsection{Test of gravity's isotropy}
This subsection gives a summary of work that is described in detail in \cite{LVGrav,LVGravlong}.
Local Lorentz invariance (LLI) in the gravitational interaction
can be viewed as a prediction of the theory of general relativity,
rather than a pillar. And it is not a trivial consequence, given
that alternative theories of gravity have been put forward that do
not lead to LLI, yet agree with general relativity in their
predictions for the red-shift, perihelion shift, and time delay.
Experimental tests of the LLI in gravity are required to decide
between these theories \cite{Will71}.



\subsubsection{Hypothetical signal}

To obtain the explicit time--dependence of the signal, we transform the quantities from the sun--centered frame into the laboratory frame \cite{KosteleckyMewesPRD}. Adding the contributions of the electromagnetic and the gravitational sector yields the time--dependence of the interferometer phase as a Fourier series \cite{BaileyKostelecky}
\begin{equation}\label{Fourierseries} \frac{\delta
\varphi}{\varphi_0}= \sum_m C_m\cos(\om_m t+\phi_m)+D_m\sin(\om_m
t+\phi_m),
\end{equation}
consisting of signals at six frequencies $m\in\{\oo,2\oo,\oo\pm \Om,2\oo\pm \Om\}$, which are combinations of the frequencies of Earth's orbit $\Oo=2\pi/(1$\,y) and rotation $\om_\oplus\simeq 2\pi/(23.93$\,h). The amplitudes $C_m, D_m$ that are functions of the Lorentz violations, see Tab. \ref{VertInterf}. We define
\begin{equation}
i_4\sigma^{JK}=\is^{JK}-\kemin^{JK},\quad
i_4\sigma^{TJ}=\is^{TJ}+\frac12 \epsilon_{JKL}\kopl^{KL}.
\end{equation}

\begin{table}
\caption{\label{VertInterf} Signal components for vertical atom interferometers. $\chi$ is geographical colatitude. We denoted $i_4=1-3I_\oplus/(M_\oplus R_\oplus^2)\approx -1/2$.}
\begin{tabular}{ccc} \hline
Comp. & Amplitude & Phase \\ \hline  $C_{2\om}$ &
$\frac14\sin^2\chi[i_4(\bar s^{XX}-\bar
s^{YY})-(\kemin^{XX}-\kemin^{YY})]$ & $2\phi$ \\
$D_{2\om}$ & $\frac12 \sin^2\chi(\is^{XY}-\kemin^{XY})$ & $2\phi$ \\
$C_\om$ & $\frac12\sin2\chi(\is^{XZ}-\kemin^{XZ})$ & $\phi$ \\
$D_\om$ & $\frac12\sin2\chi(\is^{YZ}-\kemin^{YZ})$ & $\phi$ \\
$C_{2\om+\Om}$ & $-\frac14(\cos\eta-1)V_\oplus\sin^2\chi(\is^{TY}-\kopl^{XZ})$ & $2\phi$\\
$D_{2\om+\Om}$ &
$\frac14(\cos\eta-1)V_\oplus\sin^2\chi(\is^{TX}+\kopl^{YZ})$ &
$2\phi$
\\
$C_{2\om-\Om}$ & $-\frac14(\cos\eta+1)V_\oplus\sin^2\chi(\is^{TY}-\kopl^{XZ})$ & $2\phi$\\
$D_{2\om-\Om}$ &
$\frac14(\cos\eta+1)V_\oplus\sin^2\chi(\is^{TX}+\kopl^{YZ})$ &
$2\phi$
\\
$C_{\om+\Om}$ & $\frac14V_\oplus\sin\eta\sin^2\chi(\is^{TX}+\kopl^{YZ})$ & $\phi$\\
$D_{\om+\Om}$ &
$\frac14V_\oplus\sin^2\chi[(1-\cos\eta)(\is^{TZ}+\kopl^{XY})-\sin\eta(\is^{TY}-\kopl^{XZ})]$ & $\phi$\\
$C_{\om-\Om}$ & $\frac14V_\oplus\sin\eta\sin^2\chi(\is^{TX}+\kopl^{YZ})$ & $\phi$\\
$D_{\om-\Om}$ &
$\frac14V_\oplus\sin^2\chi[(1+\cos\eta)(\is^{TZ}+\kopl^{XY})+\sin\eta(\is^{TY}-\kopl^{XZ})]$ & $\phi$ \\ \hline
\end{tabular}
\end{table}

\subsubsection{Data analysis and results} \label{deconvolution}

\begin{figure}[t]
\centering \epsfig{file=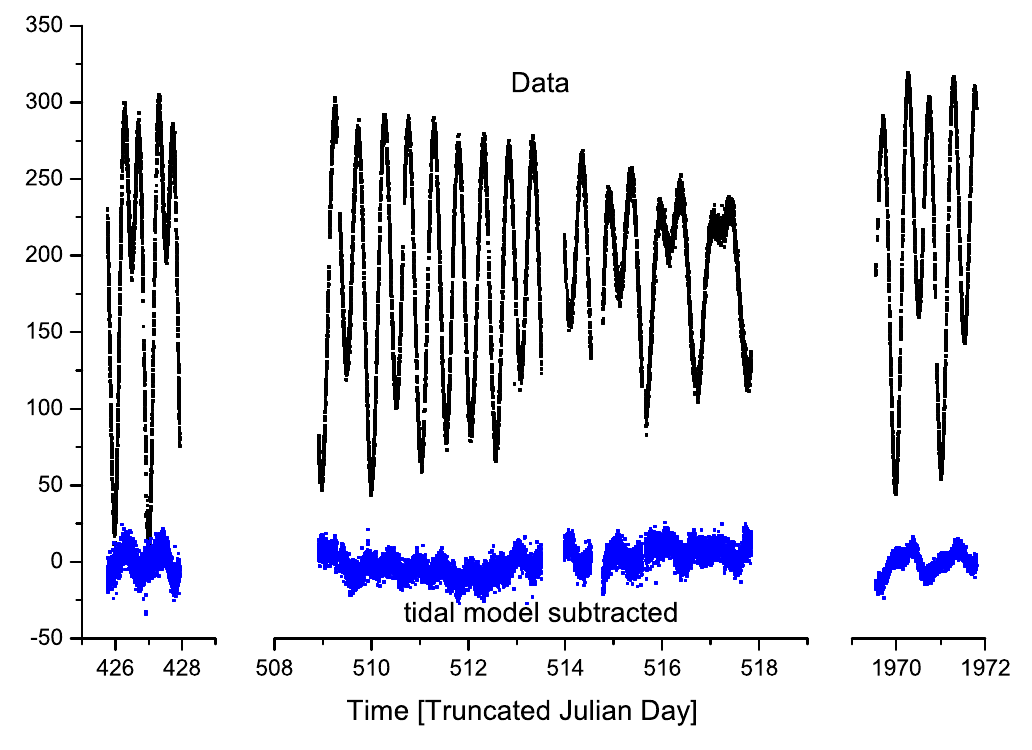, width=0.65\textwidth}
\caption{Data in $10^{-9}g$. Each point represents a 60-s scan of
one fringe (75-s after TJD1900). \label{alldata}}
\end{figure}

Fig. \ref{alldata} shows the data. It spans about 1500\,d, but is fragmented into three short segments. Major systematic effects in this experiment are tidal variations of the local gravitational acceleration. Subtraction of a Newtonian model \cite{Wenzel} and an additional model of the local tides \cite{Egbert} yields the residues shown at the bottom of Fig. \ref{alldata}.

Because of the highly fragmented data set, the Fourier components overlap. This overlap can be quantified by a covariance matrix. In order to obtain independent estimates for the parameters, we perform an overall fit assuming Gaussian statistics. The result is
\bea\label{finalbounds}
\sigma^{TX} &=& (-6.2\pm 5.1)\times 10^{-5}, \quad  \sigma^{TY} = (0.14\pm 5.4)\times 10^{-5}, \nonumber \\ \sigma^{TZ}&=& (2.8\pm 6.6) \times 10^{-5}, \quad \sigma^{XX}-\sigma^{YY} = (8.9\pm11)\times 10^{-9}, \nonumber \\ \quad \sigma^{XY} &=& (0.40\pm 3.9)\times 10^{-9}, \quad \sigma^{XZ} = (-5.3\pm 4.4)\times 10^{-9}, \nonumber \\
\sigma^{YZ} &=& (-0.66\pm 4.5)\times 10^{-9}.
\eea

Our experiment can be combined with the results of lunar laser ranging \cite{Battat}, if we assume that there is no Lorentz violation in electromagnetism. Tab. \ref{combinedbounds} lists the results thus obtained. They represent the most complete bounds on Lorentz violation in gravity, providing individual limits on the $\bar s$ as well as more components of $\bar s$ and higher resolution than either experiment. The only degrees of freedom of $\bar s^{JK}$ that are not bounded are $\bar s^{TT}$ and the trace, which do not lead to signals to first order in the Earth's orbital velocity.

\begin{table}
\centering\caption{\label{combinedbounds} Bounds resulting from
combining our data with the ones from lunar laser ranging as
reported by Battat {\em et al.} \cite{Battat}, assuming vanishing
Lorentz violation in electrodynamics.}
\begin{tabular}{cc}\hline
Coeff. & \\ \hline
$\bar s^{TX}$ & $(0.9\pm 6.2)\times 10^{-7}$ \\
$\bar s^{TY}$ & $(0.3\pm 1.3)\times 10^{-6}$ \\
$\bar s^{TZ}$ & $(-0.8\pm 3.8) \times 10^{-6}$ \\
$\bar s^{XX}-\bar s^{YY}$ & $(-2.3\pm1.6)\times 10^{-9}$  \\
$\bar s^{XX}+\bar s^{YY}-2\bar s^{ZZ}$ & $(3.5\pm 38)\times
10^{-9}$ \\
$\bar s^{XY}$ & $(-1.1\pm 1.5)\times 10^{-9}$ \\
$\bar s^{XZ}$ & $(-5.3\pm 1.4)\times 10^{-9}$ \\
$\bar s^{YZ}$ & $(1.3\pm 1.4)\times 10^{-9}$ \\
\hline
\end{tabular}
\end{table}

\subsection{Test of the Equivalence principle}


This section summarizes our initial anaylsis of equivalence principle tests in the SME \cite{redshiftPRL}. Without loss of generality, we may choose coordinates such that light propagates in the usual way through curved spacetime. 
The effects of EEP violation are then described by  
the $\alpha(\bar a^w_{\rm eff})_\mu$  and $(\bar c^w)_{\mu\nu}$ coefficients, which vanish if EEP is valid.\footnote{$\alpha$ is an arbitrary coupling constant that is attached to the $a-$coefficient by convention. In our context, since $a$ is never measurable separately. it is best to think of $\alpha(\bar a^w_{\rm eff})_\mu$ as one object.} The superscript $w$ takes the values $e,n,p$ indicating the electron, neutron, and proton, respectively.  The motion of a test particle of mass $m^{\rm T}$, up to $O(c^{-3})$, is that which extremizes the action~\cite{KosteleckyTasson}
\be
S = \int m^{\rm T}c\left( \sqrt{-\left(g_{\mu\nu}+2\bar c^{\rm T}_{\mu\nu}\right)dx^{\mu}dx^{\nu}}+\frac{1}{m^{\rm T}}\left(a^{\rm T}_{\rm eff}\right)_{\mu}dx^{\mu}\right),\label{eq:smeaction}
\ee
where $(a^{\rm T}_{\rm eff})_0= (1-2\phi \alpha)(\bar{a}^{\rm T}_{\rm eff})_0$, $(a^{\rm T}_{\rm eff})_j=(\bar a^{\rm T}_{\rm eff})_{j}$, and for composite particles with $N^{e}$ electrons, $N^{p}$ protons, and $N^{n}$ neutrons, 
\be\label{effectiveac}
(\bar c^{\rm T})_{\mu\nu}= \frac{1}{m^{\rm T}}\sum_wN^w m^w (\bar c^w)_{\mu\nu},\:\: (a_{\rm eff}^{\rm T})_\mu =\sum_w N^w (a_{\rm eff}^w)_\mu.
\ee
The metric $g_{\mu\nu}$ may also be modified by particle-independent gravity-sector corrections, as well as the $(\bar c^{\rm S})_{\mu\nu}$ and $(\bar a^{\rm S})_{\mu}$ terms 
in the action of the gravitational source body.  For experiments performed in the Earth's gravitational field, we may neglect such modifications as being common to all experiments.  
Here, we focus on an isotropic subset of the theory~\cite{KosteleckyTasson} and thereby upon the most poorly constrained flat-space observable $(\bar c^{w})_{00}$ terms and the $(\bar a_{\rm eff}^{w})_{0}$ terms, that are only detectable by gravitational experiments~\cite{ColladayKostelecky,KosteleckyTassonPRL}. 
The other $\bar c^{w}-$ and $\bar a^{w}-$ are respectively best constrained by non-gravitational experiments, or enter the signal as sidereal variations suppressed by $1/c$ and are neglected here.  
%

Expanding Eq. (\ref{eq:smeaction}) up to $O(c^{-2})$, dropping constant terms, and redefining $m^{\rm T}\rightarrow m^{\rm T}[1+\tfrac 53 (\bar c^{\rm T})_{00}]$ yields
\begin{equation}\label{lagrangesimplified}
S=\int m^{\rm T} c^{2}\left(\frac{\phi}{c^{2}}\left[1-\tfrac{2}{3}\left(c^{\rm T}\right)_{00}+\tfrac{2\alpha}{m^{\rm T}}\left(\bar{a}^{\rm T}_{\rm eff}\right)_{0}\right]
-\frac{v^{2}}{2c^{2}}\right)dt,
\end{equation}
where $v$ is the relative velocity of the Earth and the test particle.  Thus, at leading order, a combination of $\left(\bar c^{\rm T}\right)_{00}$ and $\alpha\left(\bar{a}^{\rm T}_{\rm eff}\right)_{0}$ coefficients rescale the particle's gravitational mass relative to its inertial mass.

\subsubsection{Gravity Probe A}
We begin with an analysis of gravity-probe A (GP-A).  This experiment compared a hydrogen maser on the ground to an identical one carried on a rocket along a ballistic trajectory \cite{Vessot}.  A first influence of EEP violation in this experiment arises through a change in the motion of an object used to map the gravitational potential $\phi$ as a function of position. The gravitational acceleration $g^{\rm T}$ of a test mass $m^{\rm T}$ is found by minimizing the action Eq.~(\ref{lagrangesimplified}),
\be\label{gt}
g^{\rm T} = g\left(1+\beta^{\rm T}\right), \quad
\beta^{\rm T}=\frac{2\alpha}{m^{\rm T}}(\bar a_{\rm eff}^{\rm T})_0-\frac 23 (\bar c^{\rm T})_{00},
\ee
where $(\bar a_{\rm eff}^{\rm T})_0$ and $(\bar c^{\rm T})_{00}$ are obtained from Eq.~(\ref{effectiveac}).  The test mass moves as if it were in the potential $\phi'=(1+\beta^{\rm T})\phi$.
We need not consider anomalies in the motion of the rocket, as these are removed by continuous monitoring of the rocket's trajectory.  EEP-violation also causes a position-dependent shift of hydrogen's $^2S_{1/2}$, $F=1\rightarrow F'=0$ hyperfine transition. 
The hyperfine splitting scales with the electron mass $m^e$ and the proton mass $m^p$ as $(m^e m^p)^2/(m^e+m^p)^3$.  In analogy with a previous treatment of the Bohr energy levels in hydrogen~\cite{KosteleckyTasson}, the hyperfine transition varies linearly with $\phi$ as
\begin{equation}
\xi_{\rm H}^{\rm hfs}=-\frac{2}{3}\frac{m^{p}\left(2\bar{c}^{e}_{00}-\bar{c}^{p}_{00}\right)+m^{e}\left(2\bar{c}^{p}_{00}-\bar{c}^{e}_{00}\right)}{m^{p}+m^{e}}.
\label{eq:hyperfinexi}
\end{equation}
Expressed in terms of the potential $\phi'$, 
the signal 
becomes
\begin{equation}\label{eq:gpa}
\frac{\delta f}{f_{0}}=\frac{\phi'{}_{s}-\phi'{}_{e}}{c^{2}}\left(1+\xi_{\rm H}^{\rm hfs}-\beta^{\rm SiO_{2}}\right)-\frac{v_{s}^{2}}{2c^{2}}.
\end{equation}


\subsubsection{Null Redshift Tests}
Null tests comparing clocks 1,2 with clock coefficients $\xi_{1,2}$ as they move together through a gravitational potential can yield bounds~\cite{KosteleckyTasson} on $\xi_1- \xi_2$.  One such experiment \cite{Ashby} resulted in $\xi_{\rm H}^{\rm hfs}-\xi_{\rm Cs}^{\rm hfs}=(0.1\pm1.4)\times 10^{-6}$; one using a strontium optical clock and a cesium microwave clock~\cite{Blatt} measured $|\xi_{\rm Cs}^{\rm hfs}- \xi_{\rm Sr}^{\rm opt}|< 3.5\times 10^{-6}$, and one~\cite{Fortier} using an optical clock based on $^{199}$Hg$^+$ vs. a microwave Cs clock measured $\xi_{\rm Hg^+}^{\rm opt}-\xi_{\rm Cs}^{\rm hfs}=(2.0\pm3.5)\times 10^{-6}$.  Our estimates of various optical clocks' sensitivities assume the clock transition energies scale as $(m^e m^{\rm atom})/(m^e+m^{\rm atom})$.  

\subsubsection{Nuclear Transitions}
The Pound-Rebka experiment~\cite{PoundRebka} measured the gravitational redshift of a $14.4$\,keV transition in stationary $^{57}$Fe nuclei.  
With $Z=26$, $^{57}$Fe has an unpaired valence neutron that makes a transition between different orbital angular momentum states.  Assuming the transition energy scales with the reduced mass of the neutron, the Pound-Rebka experiment constrains
\begin{equation}\label{eq:mossb}
\xi_{^{57}{\rm Fe}}^{\rm Mossb.}-\beta^{\rm grav}=-\frac{2}{3}\frac{m^{^{56}{\rm Fe}}c^{n}_{00}+m^{n}c^{^{56}{\rm Fe}}_{00}}{m^{^{57}{\rm Fe}}}-\beta^{\rm grav}.\end{equation}

\subsubsection{Matter-wave tests}
Determination of the EEP-violating phase in an AI proceeds by using the EEP-violating action Eq.~(\ref{lagrangesimplified}) to calculate the trajectories of the atom, and then integrating the phase accumutaed along that trajectory. 
To leading order, we obtain $\delta \varphi=(1+\beta^{\rm At})k gT^2$.  This reproduces the result obtained in~\cite{redshift}, with $\beta^{\rm At}$ given by Eq.~(\ref{gt}) specific to the atomic species.  
AIs are also sensitive to variations in the atoms' binding energy resulting from changes to the inertial mass of their constituent particles. We will consider this in detail later.  
Bloch oscillations \cite{Clade,Poli} are a special case of an AI where the atoms at rest and bound the same terms if they use the same species.

\subsubsection{Conclusion}


The constraints from the various experiments are sufficient to derive independent bounds on all parameter combinations relevant to neutral particles, see Tab. \ref{limits}. 
While some linear combinations of these parameters have been bounded in the past~\cite{datatables}, this is the first time that each has been bounded without assuming all others vanish. This closes any loopholes for renormalizable spin-independent EEP violations for neutral particles at $O(c^{-2})$ at the stated $1\sigma$ accuracies.  

\begin{table}
\centering
\caption{Limits ($\times10^6$), estimated by multivariate normal analysis  using results from the experiments discussed in the text, torsion balance tests~\cite{Gundlach2009}, and relative redshift measurements~\cite{KosteleckyTassonPRL,Blatt,Ashby,Fortier}, with $1\sigma$ uncertainties. The index $T$ replacing $0$ indicates these limits hold in the Sun-centered celestial equatorial frame \cite{datatables}.
\label{limits}}
\begin{tabular}{ccccc}\hline
$\alpha (\bar a^n_{\rm eff})_T$ & $\alpha (\bar a^{e+p}_{\rm eff})_T$  & $(\bar c^n)_{TT}$ & $(\bar c^p)_{TT}$ & $(\bar c^e)_{TT}$ \\
(GeV) & (GeV)\\
$4.3\pm3.7$ & $0.8\pm1.0$ & $7.6\pm6.7$ & $-3.3\pm3.5$ & $4.6\pm4.6$\\
\hline
\end{tabular}
\end{table}

Redshift and UFF tests differ in their style of execution, as the former compare proper times whereas the latter compare accelerations, but the EEP violations they constrain take the same form at $O(c^{-2})$, consistent with Schiff's conjecture.


\subsubsection{Influence of nuclear structure}\label{NucSec}

So far, the different types of matter used in EEP tests were characterized by just two degrees of freedom, their charge and mass number. This is justified as a first approximation, as the proton and neutron content of atomic nuclei makes up over 99\% of any normal isotopes' rest mass, and hence controls the bulk of its gravitational behavior. However, this means there are only two degrees of freedom, which makes it seemingly impossible to measure all two a-type and all three c-type coefficients. The reason why it is possible at all is the binding energy of nuclei, for which we had only a crude model. It is thus interesting to see how much better limits we can obtain by using a more sophisticated nuclear model. This is the subject of a paper that I worked out with my postdoc Michael Hohensee and Bob Wiringa of Argonne National Lab \cite{Nuclear}.


Using a nuclear shell model, we estimate the sensitivity of a variety of atomic nuclei to EEP violation for matter and antimatter. We also illustrate points of commonality between older representations of EEP violation based on neutron excess and baryon number, and that of the SME.  Existing experimental~\cite{adelberger,redshift,matterwaves,Vessot,PoundRebka,Kostelecky:2009a,Ashby, Blatt,Fortier,Hohensee:2013} limits on spin-independent EEP violation in matter and antimatter~\cite{redshiftPRL} yield limits on SME coefficients that are significantly tighter than previously thought. 
As before, we assume that anomalies affecting force-carrying virtual particles are negligible. We define our coordinates such that photons follow null geodesics, ensuring that electromagnetic fields do not violate EEP.



For a bound system of particles, the total Hamiltonian is a sum of single-particle Hamiltonians, plus an interaction energy $V_{\rm int}$ that is assumed to be free of EEP-violating terms. For a freely falling nucleus, e.g., the kinetic energy of the center of mass motion (with velocity $\bar{v}^{2}$) is small compared to the rest mass-energy. It is of similar order as the relevant change $\Delta U$ it explores in the gravitational potential.  Since its protons and neutrons are non-gravitationally bound, however, we cannot assume that the same is true for the kinetic energy of its constituent particles, which are in fact at the percent-level of the rest mass.  Thus, we include terms proportional to $v_{w,j}^{2}U/c^{2}$ in our Hamiltonian, where $v_{w,j}$ is the instantaneous velocity of the $j$th bound particle of species $w$.  
For any particular EEP test comparing the effects of gravity acting on systems $A$ and $B$, the observable anomaly is given by $\beta^{A}-\beta^{B}$, where $\beta^{A}$ and $\beta^{B}$ are the sensitivity coefficients of the two systems.  Since all high-precision tests of EEP are performed on charge-neutral systems, and since normal matter has a substantially similar ratio of proton to neutron content, the expression for $\beta^{A}-\beta^{B}$ can be usefully expressed in terms of an effective neutron excess $\widetilde{\Delta}_{j}$ and effective mass defect $\widetilde{m}'_{j}$
\begin{eqnarray}
\widetilde{\Delta}_{j}&\equiv&\frac{m^{n}}{m^{p}}\frac{m^{e}+m^{p}}{m^{n}}N_{j}^{n}-N_{j}^{p},\\
\widetilde{m}'_{j}&\equiv&m'_{j}-\frac{(m^{n}-m^{p})(m^{e}+m^{p})}{m^{n}}N_{j}^{p},
\end{eqnarray}
where $j\in\{A,B\}$. The EEP-violating observable can then be written in terms of linear combinations of the free particle ($\beta^{w}$) and anti-particle ($\beta^{\bar{w}}$) anomalies as
\bea
\beta^{A}-\beta^{B}=\frac{(m^{n})^{2}}{(m^{n})^{2}+(m^{e}+m^{p})^{2}}
\left[\left(\frac{\widetilde{\Delta}_{A}}{M_{A}}
-\frac{\widetilde{\Delta}_{B}}{M_{B}}\right)\left(\frac{\widetilde{\Delta}_{A}}{M_{A}}
-\frac{\widetilde{\Delta}_{B}}{M_{B}}\right)m^{p}\beta^{e+p-n}
\right. \nonumber \\ \left. -\left(\frac{\widetilde{m}'_{A}}{M_{A}}
-\frac{\widetilde{m}'_{B}}{M_{B}}\right)\beta^{e+p+n}\right] -\frac{1}{2}\sum_{w}\left(\frac{T^{w}_{A, {\rm int}}}{M^{A}c^{2}}-\frac{T^{w}_{B, {\rm int}}}{M^{B}c^{2}}\right)\left(\beta^{w}+\beta^{\bar{w}}\right)\label{eq:betadiffdefect},
\eea
where $T^{w}_{B, {\rm int}}$ are the bound kinetic energies of the particles, $M_{A}$ and $M_{B}$ are the masses of the two test bodies, and
\be
\beta^{e+p-n}\equiv \beta^{e+p}-\frac{m^{e}+m^{p}}{m^{n}}\beta^{n},\quad \beta^{e+p+n} \equiv \frac{m^{e}+m^{p}}{m^{n}}\beta^{e+p}+\beta^{n},\quad \beta^{e+p}\equiv\frac{m^{e}}{m^{p}}\beta^{e}+\beta^{p},
\ee
similar to definitions used in~\cite{Damour:1996}. We can define a similar set of terms $\beta^{\bar{e}+\bar{p}}$, $\beta^{\bar{e}+\bar{p}-\bar{n}}$, and $\beta^{\bar{e}+\bar{p}+\bar{n}}$ for antimatter. 
Thus the quantities $m^{p}\beta^{e+p-n}$ and $m^{n}\beta^{e+p+n}$ in the SME may be understood as parameterizing an anomalous gravitational coupling to a given particle's neutron-excess and total baryon number ``charges''~\cite{Damour:1996}.

\begin{figure}
\centering
\epsfig{file=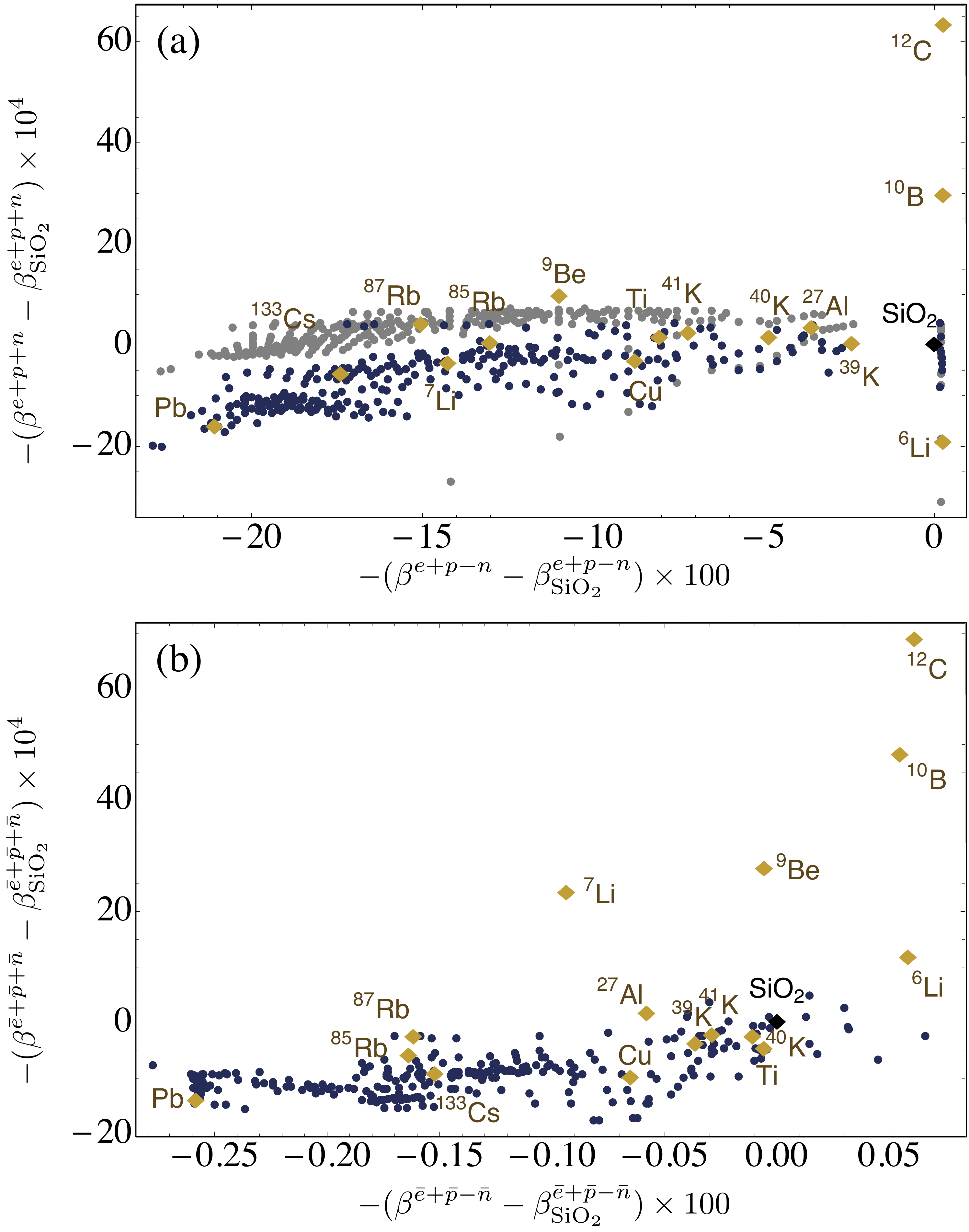,width=0.75\textwidth}
\caption{\label{fig:scatterplot} Scatterplot of the contribution of $\beta^{e+p\pm n}$ and $\beta^{\bar{e}+\bar{p}\pm\bar{n}}$ parameters to observable EEP violation in normal nuclides with lifetimes in excess of 1 Gyr, when compared to SiO$_{2}$.  Tests that compare two or more widely separated species are more sensitive than tests involving neighboring isotopes.  Plot (a) shows each species' relative sensitivity to matter-sector EEP-violation, and (b) depicts their sensitivities to antimatter-sector anomalies.  Gray points in (a) indicate the range of sensitivities obtained without accounting for nucleons' kinetic energies.  Sensitivities of $^{6}$Li, $^{7}$Li, $^{9}$Be, $^{10}$B, and $^{12}$C are taken from GFMC calculations, all others from a Woods-Saxon model.}
\end{figure}

To estimate the kinetic energy of protons and neutrons bound within a given nucleus, we model the nucleons as single particles bound within fixed, spherically symmetric rounded square well potentials.  These Woods-Saxon potentials~\cite{Woods:1954} are taken to be of the form developed by Schwierz \emph{et al.}~\cite{Schwierz:2007}.  Nuclide data is taken from Audi \emph{et al.}~\cite{Audi:2003}, and isotopic abundances (for deriving the EEP-violating signal in bulk materials) from Laeter \emph{et al.}~\cite{Laeter:2003}.  A complete summary of our calculated kinetic energies can be found in the Supplement to Ref \cite{Nuclear}. Using these estimates, we can determine the contribution of the matter-sector $\beta^{e+p\pm n}$ and antimatter-sector $\beta^{\bar{e}+\bar{p}\pm\bar{n}}$ parameters to any observed violation of EEP in the motion of two (normal matter) test masses.  These contributions are summarized in Fig.~\ref{fig:scatterplot}.  Species with particular relevance to existing or planned tests of EEP~\cite{Dimopoulos,DropTower,kasevich,STEQuest,SRPoem,lithium,GG} are explicitly labeled. Better estimates for nuclides with mass number below twelve are available from Green's function Monte-Carlo (GFMC) calculations \cite{Pieper:2001}.  They compare well (Fig.~\ref{fig:scatterplot}) with the corresponding predictions of our Woods-Saxon potential.

\subsection{Global limits}\label{GlobalLimits}


Using multivariate normal analysis of the results of an ensemble of EEP tests, including matter-wave~\cite{redshiftPRL,redshift,matterwaves}, clock comparison~\cite{PoundRebka,Vessot,Kostelecky:2009a,Ashby,Blatt,Fortier,Hohensee:2013}, and torsion pendulum experiments~\cite{adelberger}, we obtain limits on the five isotropic EEP-violating degrees of freedom that are observable in neutral systems, summarized in Tab.~\ref{tab:newlimits}.  The limits are stable against small variations in the estimated value of $T^{w}/Mc^{2}$ for the relevant nuclides, and are consistent with the limits obtained using substantially different nuclear models~\cite{WitekTBP}.
\begin{table}[t]
\caption{\label{tab:newlimits}Global limits ($\times 10^{6}$) on isotropic EEP-violation, obtained via multivariate normal analysis on the results of an ensemble of precision tests of EEP.  Limits are stated in the Sun-Centered, Celestial Equatorial Frame~\cite{datatables}, and are expressed in terms of the $\beta^{w}$ parameters as well as the individual $(\bar{c}^{w})_{TT}$ and $\alpha(\bar{a}^{w})_{T}$, with $(\bar{a}^{e+p})_{T}\equiv(\bar{a}^{e})_{T}+(\bar{a}^{p})_{T}$.  Also shown is the limit on the $1\sigma$ volume $\beta^{\Pi}$ of five-dimensional parameter space consistent with experiment. }
\begin{tabular}{lr|lr}\hline
$(\beta^{e-p}+\beta^{\bar{e}-\bar{p}})$ & $0.019\pm0.037$ & $(\bar{c}^{e})_{TT}$ & $-0.014\pm0.028$\\
$\beta^{e+p-n}$ & $-0.013\pm0.021$ & $(\bar{c}^{n})_{TT}$ & $1.1\pm1.4$\\
 $\beta^{e+p+n}$ &  $2.4\pm3.9$ & $(\bar{c}^{p})_{TT}$ & $0.24\pm0.30$\\
  $\beta^{\bar{e}+\bar{p}-\bar{n}}$ & $1.1\pm1.8$ & $\alpha(\bar{a}^{n})_{T}$ & $0.51\pm0.64$\\
   $\beta^{\bar{e}+\bar{p}+\bar{n}}$ & $-4.1\pm6.7$ & $\alpha(\bar{a}^{e+p})_{T}$ & $0.22\pm0.28$\\
\hline
\end{tabular}
\end{table}

Despite the fact that torsion pendulum tests~\cite{adelberger} set limits on specific combinations of $\beta$ parameters at the level of $10^{-12}$ (having constrained $\Delta g/g$ to the level of $10^{-14}$), the best bounds reported in Tab.~\ref{tab:newlimits} are at the level of $10^{-8}$.  Some combinations of the $\beta$'s are indeed constrained at the level of $10^{-9}$, $10^{-11}$ and $10^{-12}$, thanks to matter-wave interferometer and torsion pendulum results. But these limits are strongly correlated, leading to the lower accuracy of the global fit. 

The limits summarized in Tab. \ref{tab:newlimits} are highly significant. They rule out any observation of equivalence-principle violation in any theory that is compatible with the principles underlying the SME, unless the experimental sensitivity is high enough to evade them. No type of experiment (e.g., torsion balance, atom interferometer, or clock comparison) using any kind of matter may evade these bounds.

The precision of these bounds is limited by that of existing nuclear models, and uneven experimental coverage of EEP-violating parameter space.  New EEP tests with precision comparable to that of existing torsion pendulum experiments~\cite{DropTower,lithium,kasevich,STEQuest,SRPoem,GG} may substantially eliminate this model-dependent limitation.  Better nuclear modeling could also improve limits on EEP violation in the SME by up to eight orders of magnitude, the pursuit of which will be the subject of future work.

\section{Time, mass, and the fine structure constant}
Historically, time measurements have been based on oscillation frequencies in systems of particles, from the motion of celestial bodies to atomic transitions. Is that the simplest possible clock, i.e., is it impossible to measure time in absence of multi-particle systems? Relativity and quantum mechanics show that even a single particle of mass $m$ determines a Compton frequency $\om_C=mc^2/\hbar$. A clock referenced to the Compton frequency would enable high-precision mass measurements and a fundamental definition of the second. We demonstrate such a Compton clock using an optical frequency comb to self-reference a Ramsey-Bord\'e atom interferometer and synchronize an oscillator at a subharmonic of $\om_C$ \cite{CCC}. This directly demonstrates the connection between time and mass. It allows measurement of microscopic masses with $4\times 10^{-9}$ accuracy in the proposed revision to SI units. Together with the Avogadro project, it yields calibrated kilograms. Measuring $\om_C$ is equivalent to measuring $h/m$. From $\om_C$ or $h/m$, the fine structure constant can be calculated and thus be measured by atom interferometry \cite{Paris}. Since the topics of measuring time, mass, and the fine structure constant with atom interferometry are thus closely related, we describe them together in this chapter.

\subsection{Our atomic-fountain interferometer}
Atoms are assembled in a two-dimensional magneto-optical trap (2D-MOT),
loaded into a 3D-MOT and launched vertically upwards with a moving molasses. A sample having a measured 3-D temperature of 1.2$\,\mu$K is launched vertically to a height of 1-m in ultra-high vacuum every 2.1 seconds. Further preparation stages select a subset of atoms that have a narrow velocity distribution in the vertical direction corresponding to a temperature of 5.5\,nK and that are in the $F=3, m_F=0$ quantum state, which is magnetic-field insensitive to the leading order. We perform interferometry with $\sim 10^6$ atoms during the $\sim$1-s of free fall.

The free fall of the atoms causes a Doppler shift which we compensate for by ramping the laser frequency difference at a rate of $r=g\om_L/c\approx 11.5\,$MHz/s in the laser's rest frame. The ramp (provided by an Analog Devices AD9954 synthesizer) has a step size of $\sim 0.01\,\mu$s, i.e., is essentially smooth even on the time-scale of a single Bragg pulse. For fluorescence detection, the atoms are excited on the $F=4, m_F=4\rightarrow F'=5, m_{F'}=5$ cycling transition and their fluorescence is detected using a Hamamatsu R943-02 photomultiplier tube.


\subsubsection{Bragg diffraction}
In multiphoton Bragg diffraction, the atom coherently scatters $2n$ photons from a pair of antiparallel laser beams, without changing its internal state. The atom thereby acquires a kinetic energy of $4n^2\hbar \omega_r$, where $\omega_r=\hbar k^2/(2M)$ is the recoil frequency and $M$ the mass of the atom. Matching with the energy $n\hbar(\omega_1-\omega_2)$ lost by the laser field defines the resonance condition for the difference frequency $\omega_1-\omega_2$ of the beams.

\begin{figure}
\centering
\epsfig{file=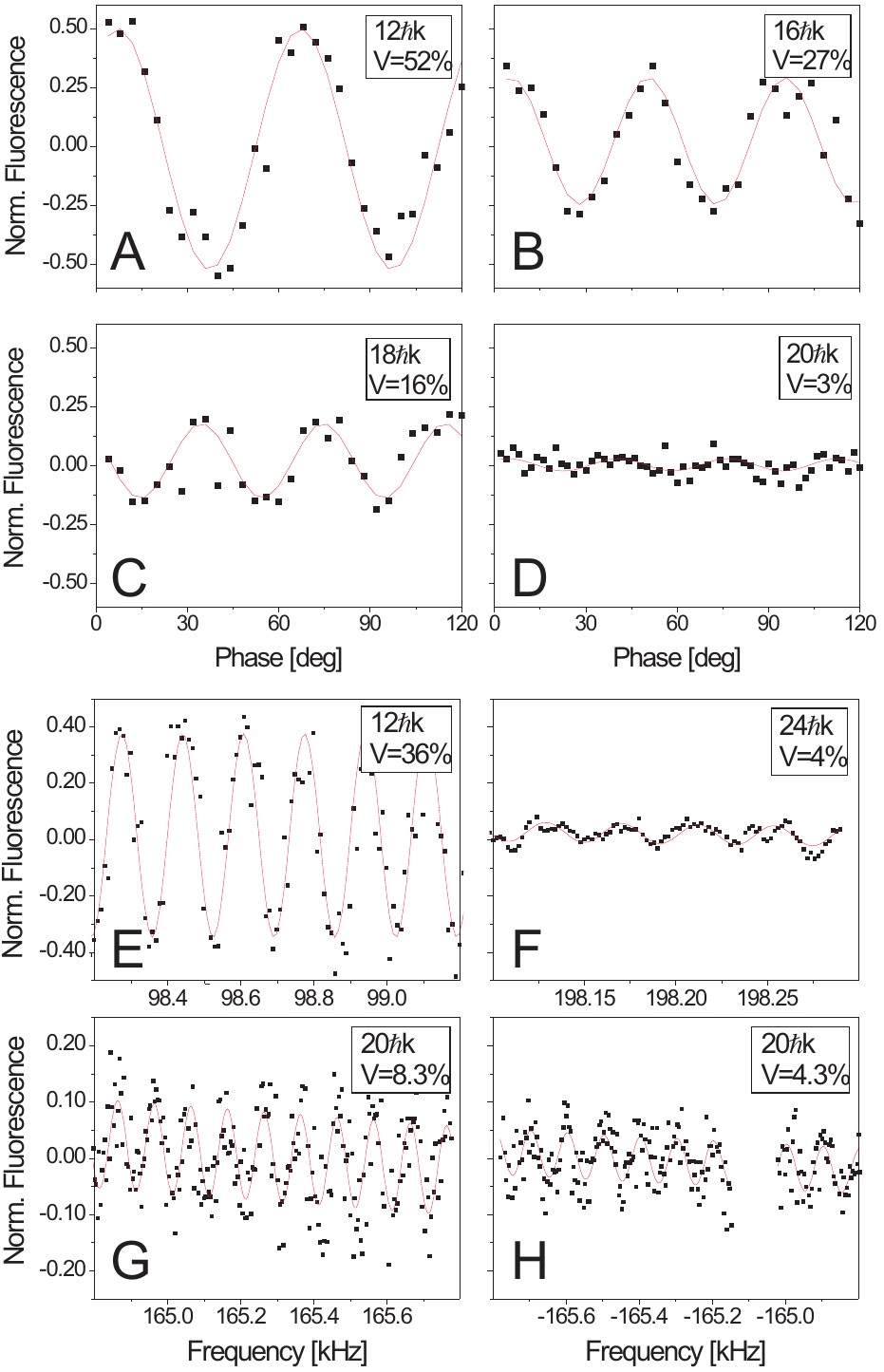, width=0.45\textwidth}
\caption{\label{MZGallery} A-D show MZ fringes with between 12 and
20$\hbar k$ momentum transfer; E and F are RB fringes with $12$
and $24\hbar k$. G and H show a conjugate $20\hbar k$ RB-pair.
Throughout, $T=1$\,ms, $T'=2\,$ms. Each data point is from a
single launch (that takes 2\,s), except for F, where 5-point
adjacent averaging was used. The lines represent a sinewave fit.}
\end{figure}

Bragg diffraction helps increase the signal, which scales quadratically with momentum transfer in recoil measurements, and is thus our method of choice. It also helps suppress the sensitivity to magnetic fields, as the atoms are in the same internal state in both interferometer arms. Fig. \ref{MZGallery} shows interference fringes measured with various degrees of high-order Bragg diffraction \cite{BraggPRL}.

\subsection{Simultaneous interferometers}
To make the Compton clock and recoil measurements independent of $g$ or the ramp rate $r$, we simultaneously operate a pair of conjugate interferometers, with the direction of the recoil reversed relative to each other. This also cancels accelerations from vibrations and is now a routine method, described in \cite{SCI}.



\subsubsection{Laser system}

High-powered laser beams are mandatory for driving high-order
multiphoton Bragg diffraction: The effective Rabi frequency \cite{Losses} $\Omega_{\rm eff} \approx \Omega^n/[(8\omega_r)^{n-1}(n-1)!^2]$ is a very strong function of the 2-photon Rabi frequency $\Omega$, and beams of large radius are required to accommodate the spread of the sample. We use a system of injection-locked Ti:sapphire lasers \cite{6Wlaser,Paris}. A first $\sim 1.2-$W Coherent 899 Ti:sapphire laser is frequency stabilized (``locked") to the 6S$_{1/2}$, $F=3\rightarrow 6P_{3/2}, F=4$ transition in a Cs vapor cell, with a blue detuning $\delta$ of 0-20\,GHz set by a microwave synthesizer. It injection locks a second one, which has no intracavity etalons or Brewster plate, and an output coupler with 10\% transmission (CVI part No. PR1-850-90-0537). 
Pumped with $20$\,W from a Coherent Innova 400 argon-ion laser, it provides a single-frequency output power of up to 7\,W. Acousto-optical modulators (AOMs) split the laser light into the top and bottom beams and shape them into Gaussian pulses, defined by arbitrary waveform generators (AWGs).

To reduce random wavefront aberrations, we minimize the number of optical surfaces. The beams reach the experiment via 5-m long, single-mode, polarization maintaining fibers and are collimated at a $1/e^2$ intensity radius of 8.6\,mm by a combination lens consisting of an achromatic doublet and an aplanatic meniscus. Polarization is cleaned by 2" polarizing
beam splitter (PBS) cubes and converted to $\sigma^+-\sigma^+$ by
zero-order $\lambda/4$ retardation plates having a specified
$\lambda/20$ flatness. 


The performance of Bragg beam splitters depends critically on the choice of the duration, envelope function, and intensity of the pulses \cite{Losses}. Our setup offers superior control of these. Short pulses, with their large Fourier width, reduce the sensitivity to the velocity spread of the atomic sample. However, below an FWHM of $n^{1/6}/[\omega_r(n-1)]$ for Gaussian pulses,
losses into other diffraction orders become significant. We use a pulse width (FWHM) of about $30-45\,\mu$s. At a detuning of 750\,MHz and a peak intensity of $0.5\,$W/cm$^2$ at the center of each beam, $30\hbar k$ momentum transfer was achieved at $>50\%$
efficiency.


\subsubsection{Coriolis compensation}
The Coriolis force does not just give rise to systematic effects. It also means that wave packets separate, causing non-closure of the interferometer. Compensation of the Earth's rotation with a rotating mirror alleviates this effect, increases contrast, and allows use of longer pulse separation times \cite{Coriolis}.

\subsection{The Compton clock: nonrelativistic treatment.}
The basic operation of the Compton clock can be described in a few lines: From the conventional nonrelativistic theory, Eq. (\ref{RBIphase}), a Ramsey-Bord\'e atom interferometer can be used to measure the recoil frequency $\om_r=\hbar k^2/(2m)$, where $k=\om_L/c$ is given by the laser frequency $\om_L$. If we can use feedback via a frequency comb to make $\om_L$ track a multiple of the recoil frequency, $\om_L=N\om_r$, we obtain $\om_r=\om_C/(2N^2)$. It is fascinating that this nonrelativistic result holds exactly in special relativity.

\subsection{Relativistic treatment}
A relativistic description of the Compton clock's operation is given in Ref. \cite{CCC}. It makes use of a rapidity parameter and hyperbolic functions. Here, we give a more elementary derivation.

\begin{figure}
\centering
\epsfig{file=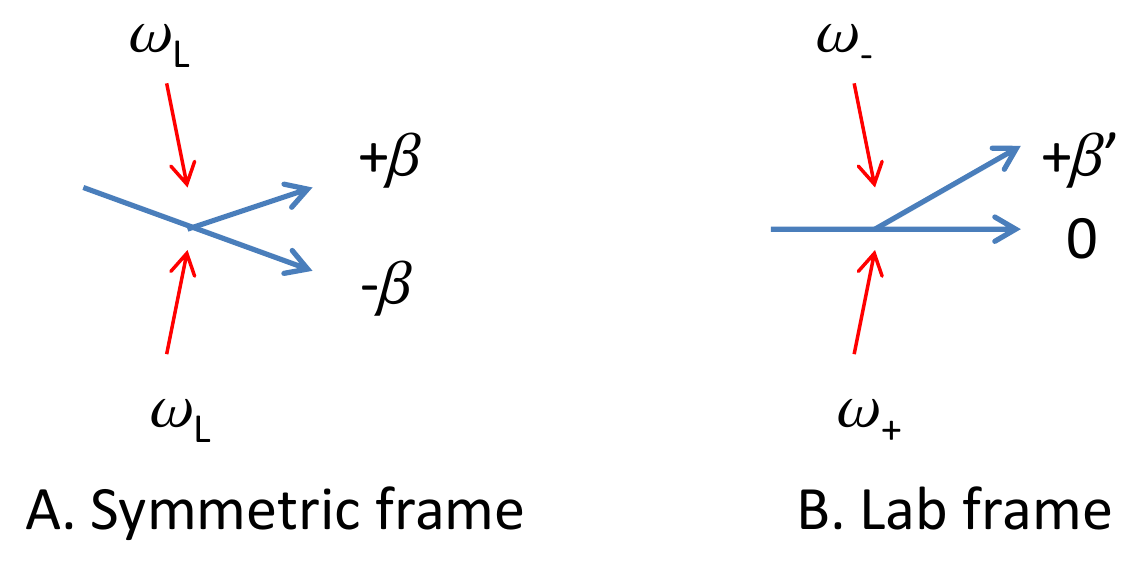,width=0.4\textwidth}\quad\quad
\epsfig{file=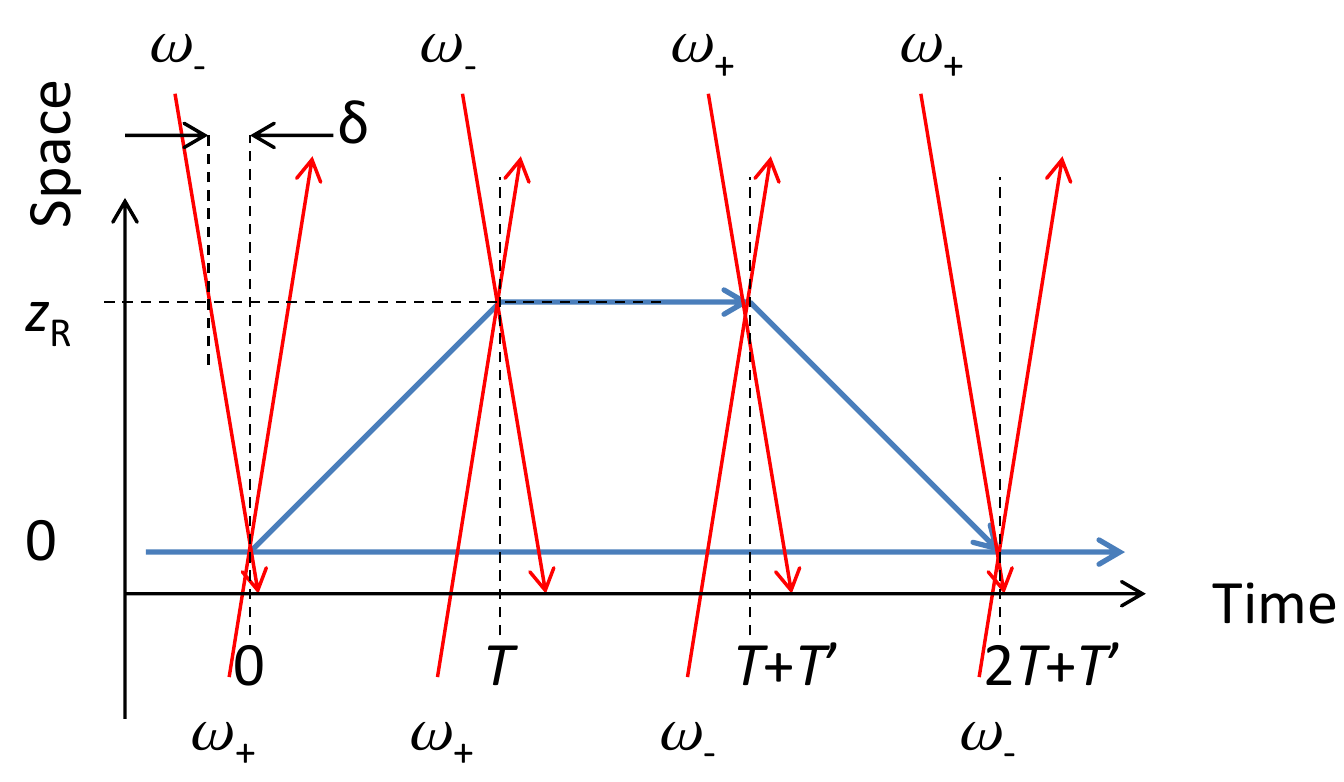,width=0.45\textwidth}
\caption{\label{BS} Beam splitter in the symmetric frame (A) and the laboratory frame (B). Right: full Ramsey-Bord\'e interferometer.}
\end{figure}

We first quantify the action of one beam splitter. It is easiest to start in a frame of reference in which the output momenta of the particle are both equal, Fig. \ref{BS} A. These momenta must be $\pm \hbar nk$, where $k$ is the wavenumber of each the laser beam and $n$ is the Bragg diffraction order, or half the number of photons transferred by each Bragg diffraction. By symmetry, in this frame the two lasers have equal frequencies $\om_L$ and wavenumbers $k_{\rm eff}=\om_L/c$. The atom's velocity in this frame satisfies $\beta \equiv \frac{v}{c}=n \frac{\om_L}{\om_C\gamma}$, where $\gamma=1/\sqrt{1-\beta^2}$, or
\be\label{omom}
\beta\gamma=n \frac{\om_L}{\om_C}.
\ee
We define the laboratory frame as the rest frame of the ingoing atom. It moves at a velocity of $\beta$ relative to the previous frame. The laser frequencies in the laboratory frame
\be\label{ompm}
\om_{\pm}=\om_L\sqrt{\frac{1\pm\beta}{1\mp\beta}}
\ee
are obtained using the Doppler formula (Fig. \ref{BS} B). The velocity of the moving output in this frame is obtained from the velocity addition formula
\be
\beta'=2\frac{\beta}{1+\beta^2},
\ee
and the $\gamma$-factor with this velocity is calculated to be
\be
\gamma'=(1-\beta'^2)^{-1/2}=\frac{1+\beta^2}{1-\beta^2}.
\ee

\subsubsection{Free evolution phase}
Fig. \ref{BS} (right) shows the entire interferometer. The time intervals $T, T'$ are the actual laboratory-frame durations of the atom's flights; the emission of the laser pulses has to be timed to achieve this, by taking into account the laser's propagation delay. The free evolution phase is
\be
\Delta\phi_{\rm F}=2\om_CT(\gamma'^{-1}-1)=-4\om_CT\frac{\beta^2}{1+\beta^2}
\ee

\subsubsection{Laser phase}
Whenever a photon is absorbed (emitted) by the atom, the photon's phase is added to (subtracted from) the matter wave phase. In the relativistic treatment, light travels on null geodesics; zero time elapses for the photons, and the photons do not accumulate phase while traveling. Calculation of the laser phase, however, has to take into account the propagation delay of the laser beams on their way from the laser to the interaction. The phase of the photon is the phase of the laser at the time it was emitted. We assume that the lasers are located directly at $z=0$ and $z=z_R$.\footnote{The reader is invited to show that the derived phase is independent of the location of the lasers.} We first note that the propagation delay of a laser beam between the upper and lower trajectory is
\be
\delta=\beta'T=\frac{2\beta}{1+\beta^2}T.
\ee
The oscillation frequencies of the laser are indicated in Fig. \ref{BS}. It is understood that the laser keeps oscillating at these constant frequencies between the initial and final pulse pair, respectively. The lasers are thus accumulating phase at $\om_\pm t$. Summing up the phases at the times the laser beams are emitted, with the appropriate sign (plus for absorption, minus for stimulated emission), yields
\be
\Delta \phi_{\rm I}=n[\om_-(T+\delta)+\om_+(T+T'-2T-T'+\delta)
 +\om_+(0-T+\delta)+\om_-(-T-T'+\delta+2T+T')]
\ee
which we simplify to
\be
\Delta \phi_{\rm I}=2n \om_L\left(\sqrt{\frac{1-\beta}{1+\beta}}(\delta+T)
+\sqrt{\frac{1+\beta}{1-\beta}}(\delta-T)\right).
\ee
Now note that we can replace $\om_L=(1/n)\om_C \beta\gamma$. This yields, after some algebra,
\be\label{Cclaserphase}
\Delta \phi_{\rm I}=4\om_CT\frac{\beta^2}{1+\beta^2}
\ee
It is evident that, for the appropriately chosen laser frequencies given by Eq. (\ref{ompm}), the laser phase cancels the free evolution phase,
\be
\Delta\phi=\Delta\phi_{\rm F}+\Delta\phi_{\rm I}=0.
\ee
In the experiment, one adjusts the laser frequency changes $\om_\pm$ such that the interferometer phase vanishes. This cancellation happens when
\be\label{omm}
\om_m\equiv \om_+-\om_-=\om_L\frac{2\beta}{\sqrt{1-\beta^2}}
\ee
Thus, $\omega_m$ provides a measurement of the free evolution phase $-4\om_CT\beta^2/(1-\beta^2)$. The frequency comb is used to make sure that
\be
\om_L=N\om_m.
\ee
From Eq. (\ref{omm}), we then have
\be
\frac 1N=\frac{2\beta}{\sqrt{1-\beta^2}}.
\ee
We solve for $\beta=1/\sqrt{1+4N^2}$ (choosing the positive solution) and obtain
\be
\beta\gamma=\frac{\beta}{\sqrt{1-\beta^2}}=\frac{1}{\sqrt{1+4N^2}}\frac{1}{\sqrt{1-\frac{1}{1+4N^2}}}
=\frac{1}{2N}
\ee
Because of Eq. (\ref{omom}), $\beta\gamma=n\om_L/\om_C$, we find $\om_m=\om_C/(2nN^2)$, which is equivalent to the elegant relation
\be\label{Comptoneq}
\om_C:\om_L:\om_m=2nN^2:2nN:1,
\ee
valid to all relativistic orders.

\subsection{Experiment}
\begin{figure}
\centering
\epsfig{file=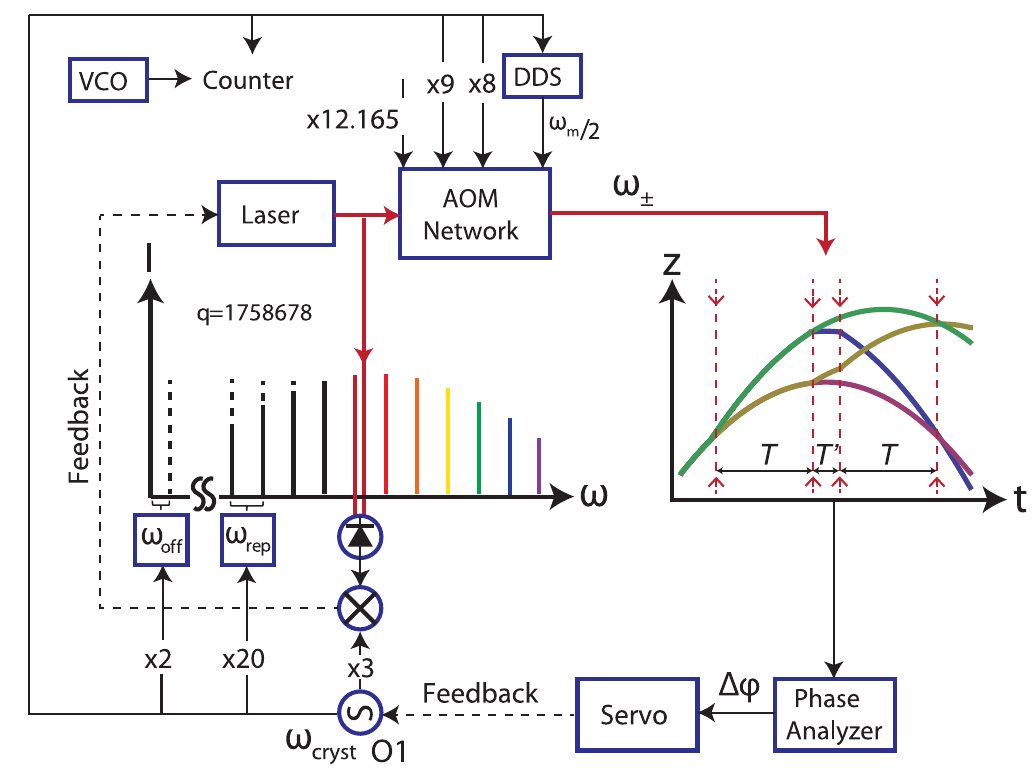,width=0.75\textwidth}
\caption{\label{CCCsetup} Schematic of the Compton clock.}
\end{figure}

\begin{figure}
\centering
\epsfig{file=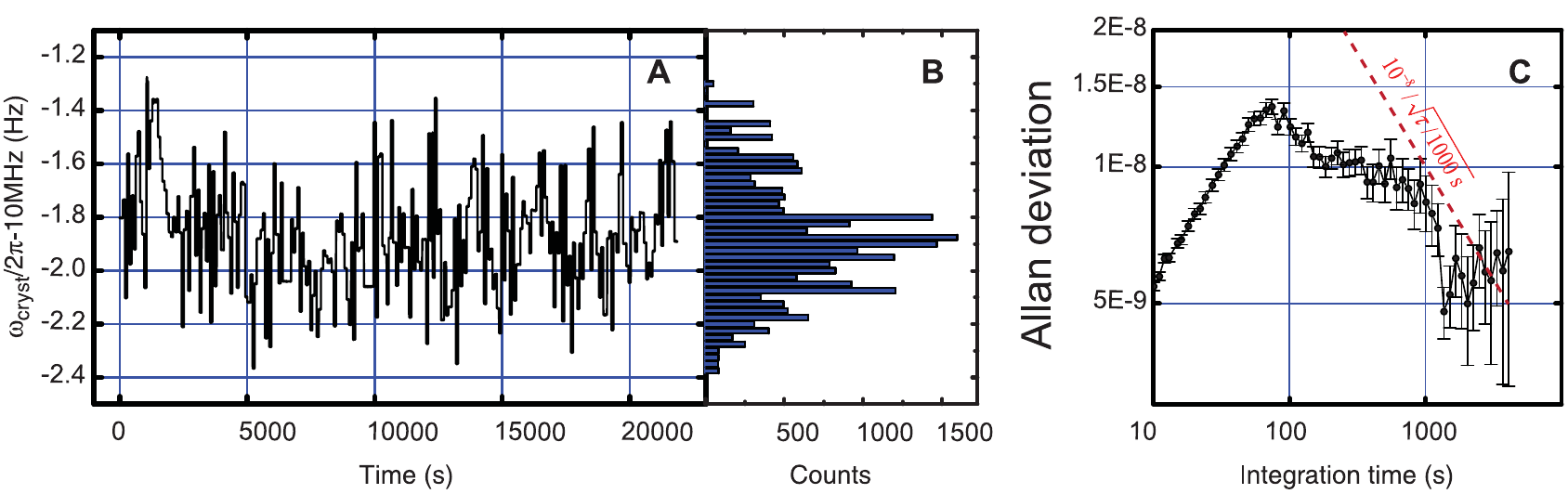,width=\textwidth}
\caption{\label{CCCdata} Compton clock performance. (A) Frequency minus 10 MHz versus time plotted over 6 hours. (B) Histogram of data (bin size = 0.025 Hz). (C) Root Allan variance (RAV) of the data in (A). It is below $10^{-8}/(\tau/1000 {\rm s})^{1/2}$ for integration times $\tau$ between 100\,s and 1\,hour. The slope between $\tau = 10-100\,$s is an artifact of the 80\,s update cycle of the experiment.}
\end{figure}

Fig. \ref{CCCsetup} shows the setup of the clock, which we have already described in \cite{CCC}. Oscillator O1 is the frequency reference for all signal generators and the optical frequency comb. The laser used to address the atom interferometer is phase-locked to the comb. Shown
in the diagram are the trajectories of the simultaneous conjugate interferometers. The phase measurement from the atom interferometer provides an error signal to stabilize O1. We have compared the Compton clock to a Rubidium frequency standard for about 6\,hours, see Fig. \ref{CCCdata}. The agreement of the measured frequency with the one expected from the cesium mass confirms our understanding of the clock within the experimental error. The leading order systematic effects are discussed in \cite{CCC} and summarized in Tab. \ref{CCsyst}.

\begin{table}
\caption{\label{CCsyst} Systematic effects of the Compton clock}
\begin{tabular}{ccc}\hline
Influence & Offset (ppb) & Error bar (ppb) \\ \hline
Gravity gradient & 15 & 1 \\
Beam splitter phase shift & 340.4 & 3.1 \\
Gouy phase & 1.9 & 0.1 \\
Counterpropagation angle & -1.5 & 1.1 \\
Magnetic fields & 0 & 0.2 \\ \hline
\end{tabular}
\end{table}

\subsubsection{Is there a ``clock ticking at the Compton frequency"?}
Rather than philosophizing over the meaning of the term ``ticking," let's make a simple observation: In a perfect conventional atomic clock, the frequency of the atomic transition is the only dimensional quantity that determines the frequency of the output of the clock. Besides that, there may only be known numerical ratios given, e.g., by frequency dividers. In a perfect Compton clock, the Compton frequency is the only dimensional quantity that determines the output frequency, besides numerical ratios.

For example, if a cesium atomic clock is to deliver a reference frequency of $\nu_{\rm ref}=10$\,MHz, the frequency of the hyperfine transition of 9,192,631,770\,Hz is divided by a divisor of $\eta=9,192,631,770/10,000,000=919.263177$. This number is given by the settings of various phase-locked loops and frequency dividers. In a practical example, a stable crystal oscillator at $\nu_{\rm ref}$ might be multiplied by a factor of $\eta_1=18$ to 180\,MHz using electronics. The $\eta_2=51^{\rm th}$ harmonic of that frequency is used as a reference for phase-locking a Dielectric Resonator Oscillator (DRO), with an intermediate frequency of $12.631770$\,MHz, obtained from $\nu_{\rm ref}$ by multiplication with a factor of $\eta_3=12,631,770/10,000,000$ using a direct digital synthesizer (DDS).  frequency of this DRO is stabilized to the atomic transition via Ramsey spectroscopy. We thus obtain $\nu_{\rm hfs}=\nu_{\rm ref}(\eta_1\eta_2+\eta_3)$. This number is known from the construction of the apparatus. If two cesium atomic clocks are compared, they will deliver the same frequency provided that $\eta$ is set to the same value.

If a Compton clock is to deliver a frequency of $\nu_{\rm ref}=10$\,MHz, the Compton frequency of a Cesium atom of 2,993,486,252$\times 10^{16}$\,Hz\footnote{For simplicity, we don't write error bars in this paragraph} is to be divided by $2.993486252\times 10^{18}$. In our clock, a stable crystal oscillator at $\nu_{\rm ref}$ is multiplied by a factor of $N_c=35,173,594.165$ to give $\nu_L\approx 351$\,THz. This factor is given by $20\times 1758678+2+3+29.165$, where the summands listed in order of appearance represent: the harmonic generated by the frequency comb, the comb offset of 20\,MHz due to carrier-envelope phase, the beat frequency of 30\,MHz in the laser lock, and the combined shifts of three acousto-optical modulators. All frequencies are directly proportional to $\nu_{\rm ref}$, as they are generated by multiplying $\nu_{\rm ref}$ . Finally, $\nu_m=2\nu_{\rm ref} N_{\rm DDS}$, where $N_{\rm DDS}=2,326,621,801,616/2^{48}$ is given by a DDS. Colsing the feedback loop, we obtain $\nu_{\rm ref}=\nu_0/(4n \eta_c^2/N_{\rm DDS})$ using Eq. (\ref{Comptoneq}). If two cesium Compton clocks are compared, they will deliver the same frequency provided that $\eta$ is set to the same value.

A common misconception is that the Compton clock is somehow referenced to the  internal structure of the cesium atom through its transition frequencies. However, with the factors $N_c$ as stated above, the lasers in the clock are actually 5-15\,GHz blue detuned from the $F=3\rightarrow F'=4$ line in Cs. So the interpretation is off by ten thousand ppb, or thousands of $\sigma$. The internal structure of the atom is used only to enhance its polarizability. The clock could actually run using elementary particles such as electrons, which have no internal structure. See section \ref{chargedpart}.

A related misconception is that the Compton clock is unable to deliver an output signal at the Compton frequency itself, even in principle. However, choosing $n=2$ and $N=1/2$, we obtain $\om_m=\om_C$. Tab. \ref{cclocks} lists a few notable combinations of $n$ and $N$ for Compton clocks. For the numerical examples, we have assumed the clock uses an electron (see Sec. \ref{chargedpart} for more on interferometry with electrons). Table \ref{CCsim} makes a comparison between a conventional clock and a Ramsey-Bord\'e Compton clock.

\begin{table}
\caption{\label{cclocks} A few notable examples for electron Compton clocks. In the first example, $\om_m=\om_C$; in the second, $\om_L=\om_C$. The third example is notable because its lasers are near 255\,eV; coherent radiation at such energies has already been generated via high harmonic generation. Entries marked $\ast$ are too long to be included. The irrational frequency ratios can be approximated to any desired accuracy using direct digital synthesis.}
\begin{tabular}{ccccccc}\hline
    & $\beta$ & $\beta'$ & $\om_m/\om_C$ & $\om_L/\om_C$ & $\om_+/\om_C$ & $\om_-/\om_C$ \\ \hline
    & $\frac{1}{\sqrt{1+4N^2}}$ & $\frac{\sqrt{1+4N^2}}{1+2N^2}$ & $1/(2nN^2)$ & $1/(2nN)$ & $\ast$ & $\ast$ \\
$n=2, N=1/2$
 & 0.707 & 0.943 & 1 & 0.5 & 1.207 & 0.207 \\
$n=1, N=1/2$
 & 0.707 & 0.943 & 2 & 1 & 2.414 & 0.414 \\
$n=1, N=10^3$ & 0.0005 & 0.001 & $5\times 10^{-7}$ & 0.0005 & 0.00050025 & 0.00049975 \\ \hline
\end{tabular}
\end{table}

\begin{table}
\caption{\label{CCsim} Comparison of a conventional atomic clock based on Ramsey spectrocopy to the Compton clock}
\begin{tabular}{p{6.5cm}|p{6.5cm}}\hline
Ramsey atomic clock & Ramsey-Borde Compton clock \\ \hline
An oscillator is locked by zeroing the central fringe of an interference pattern. Its frequency is thereby aligned to the frequency corresponding to the level splitting. & An oscillator is locked by zeroing the central fringe of an interference pattern. Through self-referencing, the recoil frequency becomes a subharmonic of the Compton frequency via Eq. (\ref{Comptoneq}). \\ \hline
The phase $\om T$ accumulated by the microwave oscillator between the two interactions cancels the phase accumulated between the quantum states, making the clock independent of the exact value of $T$. & The phase Eq. (\ref{Cclaserphase}) accumulated between the counterpropagating laser frequencies cancels the phase accumulated between the quantum states, making the clock independent of the exact value of $T$. \\ \hline The total phase is due to the free evolution of the quantum system and the microwave interaction term. & The total phase is due to the free evolution and the light-atom interaction term. \\ \hline The transition is between two internal states of the atom, induced by photons. & The transition is between momentum states, of the atom (maintaining same internal states), induced by  multi-photon Bragg transitions. \\ \hline
The output frequency is equal to the transition frequency & The output frequency is equal to Eq. (\ref{Comptoneq}) and can in principle be equal to the Compton frequency, e.g., for $n=2, N=1/2$. \\ \hline
The output frequency is determined exclusively by the energy-level splitting of the atom and a known frequency divisor. & The output frequency is determined exclusively by the Compton frequency of the atom and a known frequency divisor. \\ \hline Two atomic clocks using the same transition will deliver the same frequency, up to the frequency divisor. Knowledge of e.g., $\hbar, c, \alpha,\ldots$ or anything but the transition frequency is not needed to predict the output frequency. & Two Compton clocks using the same particle will deliver the same frequency, up to the frequency divisor. Knowledge of e.g., $\hbar, c, \alpha,\ldots$ or anything but the Compton frequency is not needed to predict the output frequency. \\ \hline All of the above subject to experimental error. & All of the above subject to experimental error. \\ \hline
\end{tabular}
\end{table}

\subsection{The fine structure constant}\label{finestructureconstant}

The fine structure constant $\alpha$ describes the strength of the electromagnetic force on fundamental particles and is ubiquitous in physics. It determines the structure and hierarchy of matter, from nuclear matter over atoms and simple molecules to biological macromolecules and bulk matter.

Precise knowledge of $\alpha$ will impact many fields of science. Today's best value - with a precision of 0.25 parts per billion - is derived from a measurement of the gyromagnetic anomaly g-2 of the electron and its prediction in terms by the theory of quantum electrodynamics, arguably the most precise prediction made in all of science. Unfortunately, $\alpha$ is not known from independent experiments to the same precision. Thus, what could be the most precise test of QED is hampered by our lack of knowledge of $\alpha$.

\begin{figure}
\centering
\epsfig{file=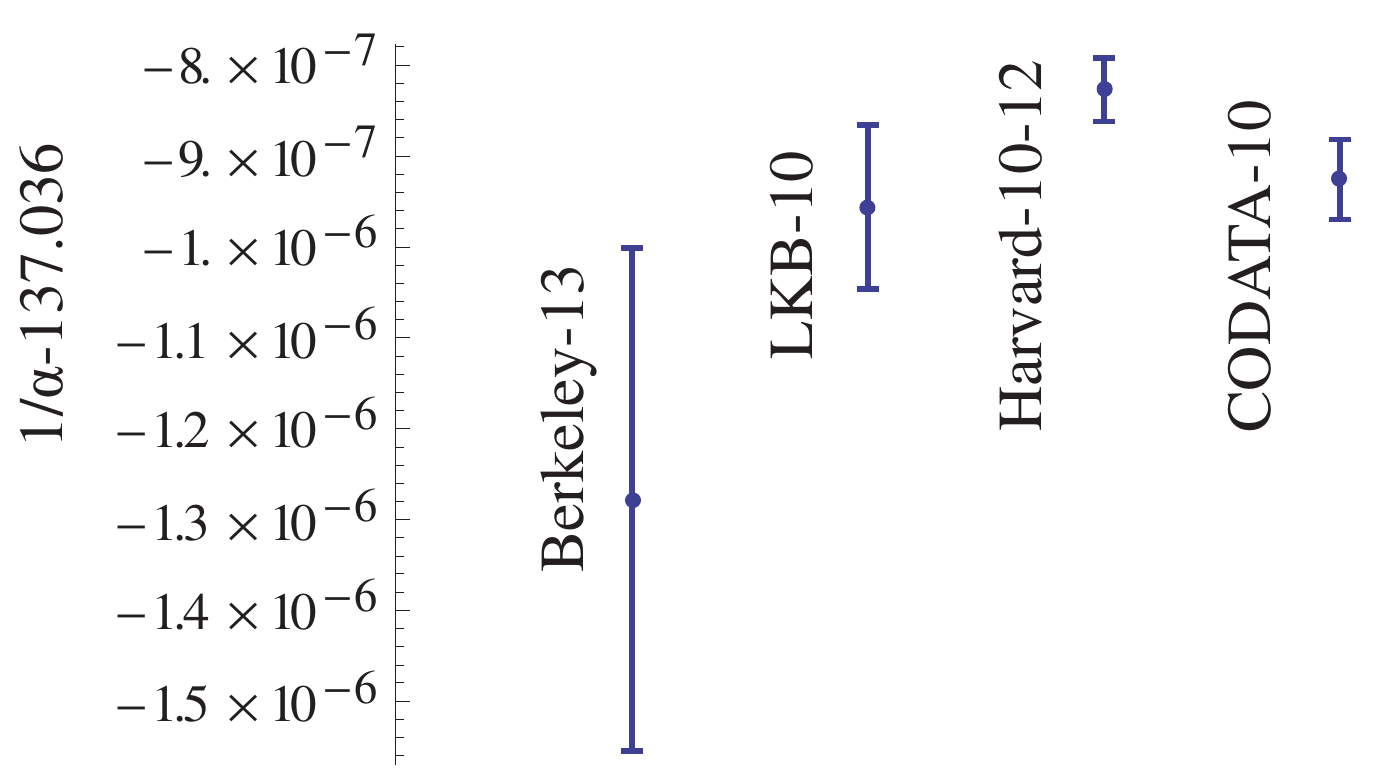,width=0.4\textwidth}
\caption{\label{Alphameas} Comparison of $\alpha^{-1}$ as measured by us, at Laboratoire Kaster-Brossel \cite{Bouchendira2011,Bouchendira2013} and the Harvard-Kinoshita collaboration \cite{Hanneke2008,Hanneke2011,Aoyama2012}}
\end{figure}

The best current measurement of $\alpha$ based on atom interferometry \cite{Bouchendira2011,Bouchendira2013} reaches 0.66\,ppb. From the Compton clock measurement, the values
\be
\om_C(133{\rm Cs})=(2.993486252±12)\times 10^{16}\,{\rm Hz} \nonumber 
\ee
can be calculated, from which the fine structure constant can be derived according to
\be
\alpha^2=\frac{2R_\infty c}{\om_C} \frac{A_r(133{\rm Cs})}{A_r(e)}.
\ee
We may use the CODATA values \cite{CODATA} for the Rydberg constant $R_\infty$ and the relative atomic mass $A_r(e)$ of the electron.
For the relative cesium mass, we use the unweighted arithmetic mean of two recent measurements \cite{Bradley,Mount} as $A_r(133{\rm Cs})=132.905451947(24)$. We obtain
\be
\alpha=7.297\,352\,589(15)\times 10^{-3}\, \quad [2.0\,{\rm ppb}].
\ee
The measurement is mostly limited mostly by the beam splitter phase shift (1.6\,ppb in $\alpha$) and statistics $\sim 1.3\,$ppb. Fig. \ref{Alphameas} shows a comparison of the two best atom interferometry measurements of $\alpha$ alongside the result derived from Gabrielse's measurement of $g-2$ \cite{Hanneke2008,Hanneke2011} in \cite{Aoyama2012} and the latest CODATA adjusted value.

\subsection{Further improvements}
\begin{figure}
\centering
\epsfig{file=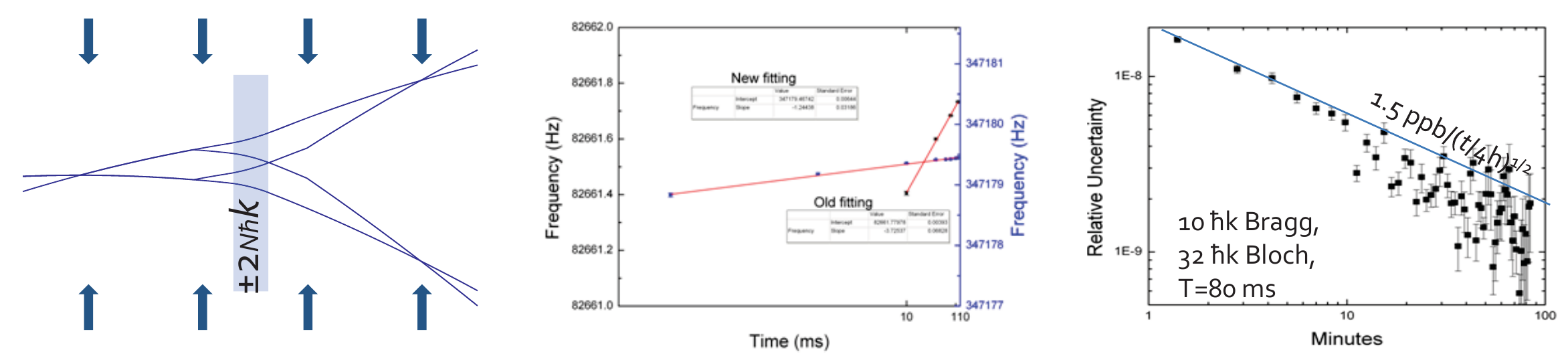,width=\textwidth}
\caption{\label{BlochInterf} {\bf Left:} Interferometer using Bloch oscillations to accelerate the conjugate interferometers away from one another. {\bf Middle:} Beam splitter phase shift. {\bf Right:} Allan variance.}
\end{figure}

In order to improve this, we have implemented Bloch oscillations to accelerate the two interferometer further apart from one another by transferring $\pm 2 N'\hbar k$ momentum to the upper and lower interferometer, respectively, see Fig \ref{BlochInterf}. This method is similar to the one used by \cite{Bouchendira2011,Bouchendira2013}, but we use high-order Bragg diffraction as a beam splitter.

We have thus increased our measured frequency $\om_m$ about 4-fold, while the absolute value of the beam splitter phase shift has been decreased two-fold. Thus, the relative error caused by the beam splitters in now reduced by a factor of about 8.

The larger signal, along with a better signal to noise ratio that we achieved by using Raman sideband cooling to increase the atom number, has reduced the statistical error. With $n=5$, $N'=16$ and $T=80\,$ms, we achieve 0.48\,ppb in $\alpha$ in six hours. The root Allan deviation shows the $1/\sqrt{\tau}$ behavior that is typical of white noise. The data can be analyzed using different ellipse fitting methods, which further reduces the statistical error to 0.33\,ppb in $\alpha$ in 6 hours of integration. Assuming white noise, this is four times better than  during the Compton clock, and makes our interferometer the lowest-noise instrument for measuring the recoil frequency and the fine structure constant. We are thus confident that a sub-part per billion measurement of $\alpha$ is within reach, both from a signal to noise standpoint and considering systematic effects.

\subsection{Atom interferometry and the SI: Mass standards}

The clock (and in fact any Ramsey-Bord\'e atom interferometer) can be used for the opposite purpose, measuring mass by measuring the Compton frequency. In 2011, the General Conference on Weights and Measures (CGPM-2011) considered a revision to the SI units that would assign an exact value to the Planck constant \cite{CGPM}. The kilogram would then be referenced to the second through the defined values of the Planck constant and the speed of light. Microscopic masses could be related to the fine structure constant or $h/M$; macroscopic masses could be measured using the Watt balance \cite{Wattbalance1,Wattbalance2}.

Atom interferometry would provide an absolute measurement of the cesium atom's mass. Other microscopic masses can be related to the cesium mass by mass spectroscopy. The link to macroscopic masses can be made by Avogadro spheres: silicon crystals of accurately measured volume V and lattice constant $a$ \cite{Avogadro}. Present data yields the spheres' mass with an overall accuracy of 30\,ppb so that they would constitute the most accurately calibrated macroscopic masses under the proposed CGPM-2011 redefinition - a testament to the precision achieved in constructing Avogadro spheres.

While any method for measuring microscopic mass can be employed, the Compton clock offers a transparent connection between the second and a microscopic mass based on simple physical principles and without requiring auxiliary measurements. The method outlined here offers a different set of systematic effects as compared to Watt balances, thus serving as an important test of the overall consistency of the laws of physics and experimental methods.

\section{Atom interferometer in a cavity}

\subsection{Gravitational Aharonov-Bohm effect}\label{GravABsect}



The wave function of a particle is measurably phase shifted by $\phi_A=-\frac e\hbar \int \vec A \cdot d\vec l$ or $\phi_V=\frac e\hbar \int V dt$ in the presence of a vector potential $\vec A$ or an electrostatic potential $V$, even in the absence of any classical force. This is the essence of the Aharonov-Bohm (AB) effect \cite{EhrenbergSiday1949,AharonovBohm1959}.

Very few experiments have been able to detect a gravitational influence on quantum systems. The first were neutron interferometers \cite{Colella,Rauch}. Later, atom interferometers \cite{KasevichChu,PritchardReview,Petersmetrologia} and Bloch oscillation experiments \cite{Peik,Poli,Clade} were developed. Perhaps the latest addition to this list is the observation of neutron quantum states in Earth's gravitational field \cite{Nesvizhevsky,Vankov}. In all these experiments, a gravitational force (proportional to the local gravitational acceleration $g$) is acting on the quantum system.

A realization of a gravitational AB effect will demonstrate the influence of a gravitational potential on a quantum system even when the potential does not cause any classical force. The effect shares the features of its electromagnetic cousin in being non-dispersive, non-local, and topological. The experiment will demonstrate that knowledge of the classical gravitational field $g$ acting locally on a particle is not sufficient to predict the particle's quantum-mechanical behavior. Paraphrasing R. P. Feynman \cite{FeynmanLect}, for a long time it was believed that the electromagnetic scalar and vector potentials $V$ and $\vec A$  were not ``real," since they could be replaced by the $\vec E$ and $\vec B$ fields in the description of any observable phenomenon then known. However, the AB effects in quantum mechanics have established the fact that $V$ and $\vec A$ are ``real" in that sense, and the fields $\vec E$ and $\vec B$ are slowly disappearing from the modern expression of physical laws.

While the electromagnetic Aharononv-Bohm effect is typically presented as a closed subject in textbooks, we are completely lacking an analogous experimental demonstration for gravity. One reason for this gap is the relative weakness of gravity, making it hard to tailor the special gravitational potential needed for a demonstration of gravity's AB effect in the laboratory, and create a signal of measurable size.

\subsubsection{The Aharonov-Bohm (AB) effect}

\begin{figure}\centering
\epsfig{file=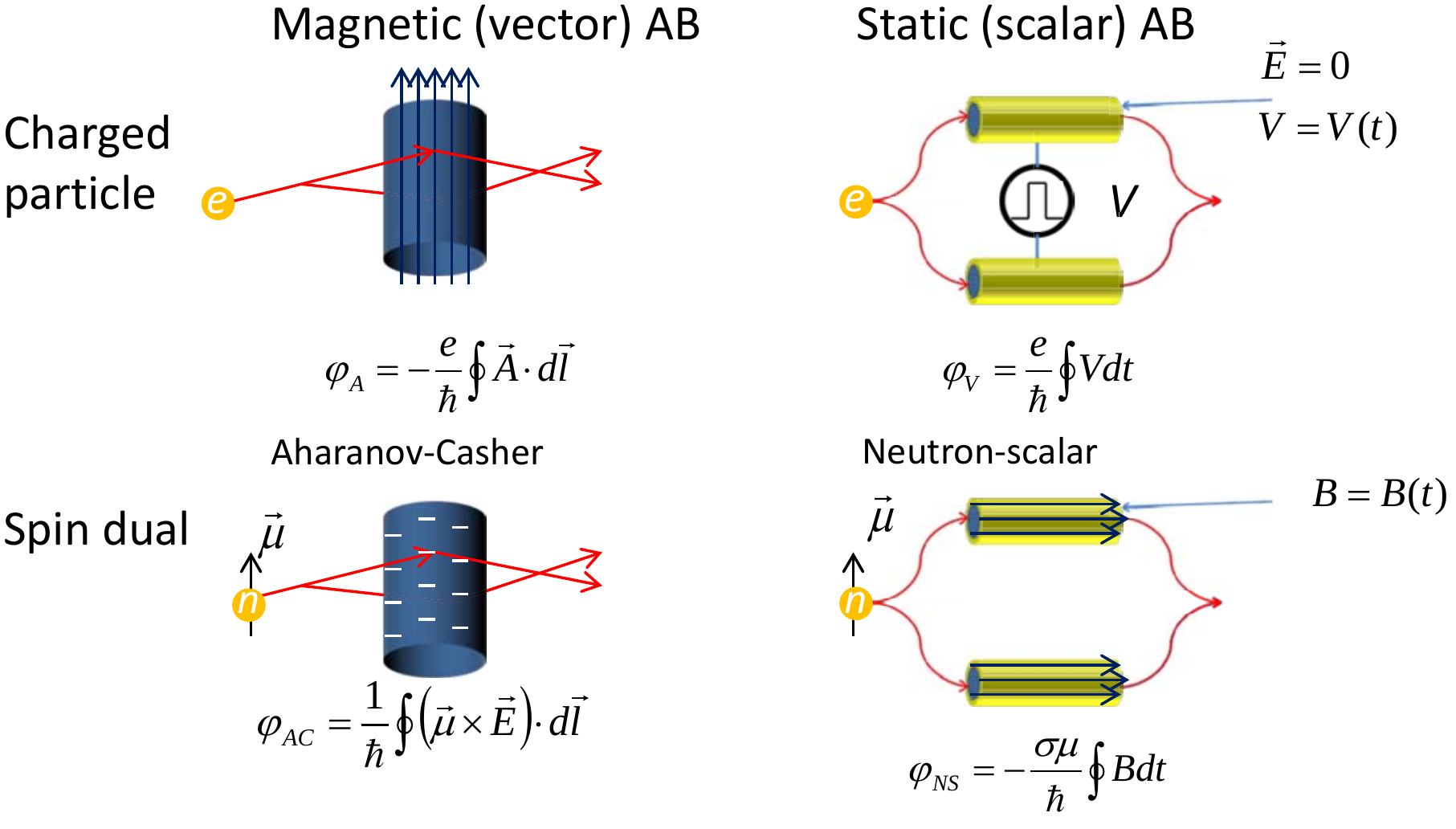,width=4in}
\caption{\footnotesize Magnetic (vector-) and static (scalar)-AB effects and their spin duals}\label{VariousAB}
\end{figure}

Because of its importance fundamental physics (\cite{Feynman}) and its non-obvious nature, the effect keeps generating a vast literature \cite{Olariu,PeshkinRMP,Batelaan}. In the magnetic (vector-) AB effect, a charged particle may take either of two paths around a region in which a magnetic field exists, but may not enter this region (Fig. \ref{VariousAB}). When the two paths interfere, the probability of detecting the particle in either output of the interferometer is given as $\cos^2(\phi_A/2)$, even thought the particle never encounters any magnetic field or Lorentz force. The effect has been closely scrutinized and confirmed using a great variety of experiments \cite{ChambersMollenstedt,MollenstedtBayh,Tonomura,Osakabe,Caprez,Zeilinger85,Badurek}. The electrostatic (scalar) AB effect occurs for a charge that passes a pair of Faraday shields. Once the particle has entered the shields, a voltage $V$ is switched on, and switched off before the particle exits. The particle picks up the phase $\phi_V$. This ``type I)" static AB experiment, in which the particle never experiences a force arising from the voltage $V$, has not yet been realized. However, the phase $\phi_V$ was measured in a ``type II" experiment, wherein the particle does encounter electrostatic fields which are arranged such as to not displace the particle \cite{MatteucciPozzi}. Phase shifts in absence of classical forces can also occur for neutral particles with a magnetic moment $\vec \mu$. In these effects, called spin duals to the AB effect, the role of the vector and scalar potentials is played by effective potentials $\vec\mu\times\vec E$ and $\vec B\cdot \mu$, respectively. The spin-dual experiments of the magnetic and static \cite{Allmann} AB effects have been realized with neutrons.

The AB effects are \cite{Sjokvist} {\em nondispersive,} i.e., not caused by distortion or movement of the wave packet \cite{Badurek}. They are also {\em nonlocal} and {\em topological} \cite{Peshkin}: No number of local measurements at any location in which the particle is allowed to exist is sufficient to predict the effect. For example, no measurements of $\vec E$ or $\vec B$ in those regions will be able to predict $\phi_A, \phi_V$. Rather, the region in which the particle is allowed to exist must be multiply connected to obtain a nonzero effect.  For example, if the interferometer enclosing the magnetic field was reduced to a single path (and therefore a simply-connected region in which the particle is allowed), no AB effect could be measured.

\subsubsection{Gravitational Aharonov-Bohm effect}
\begin{figure}
\centering
\epsfig{file=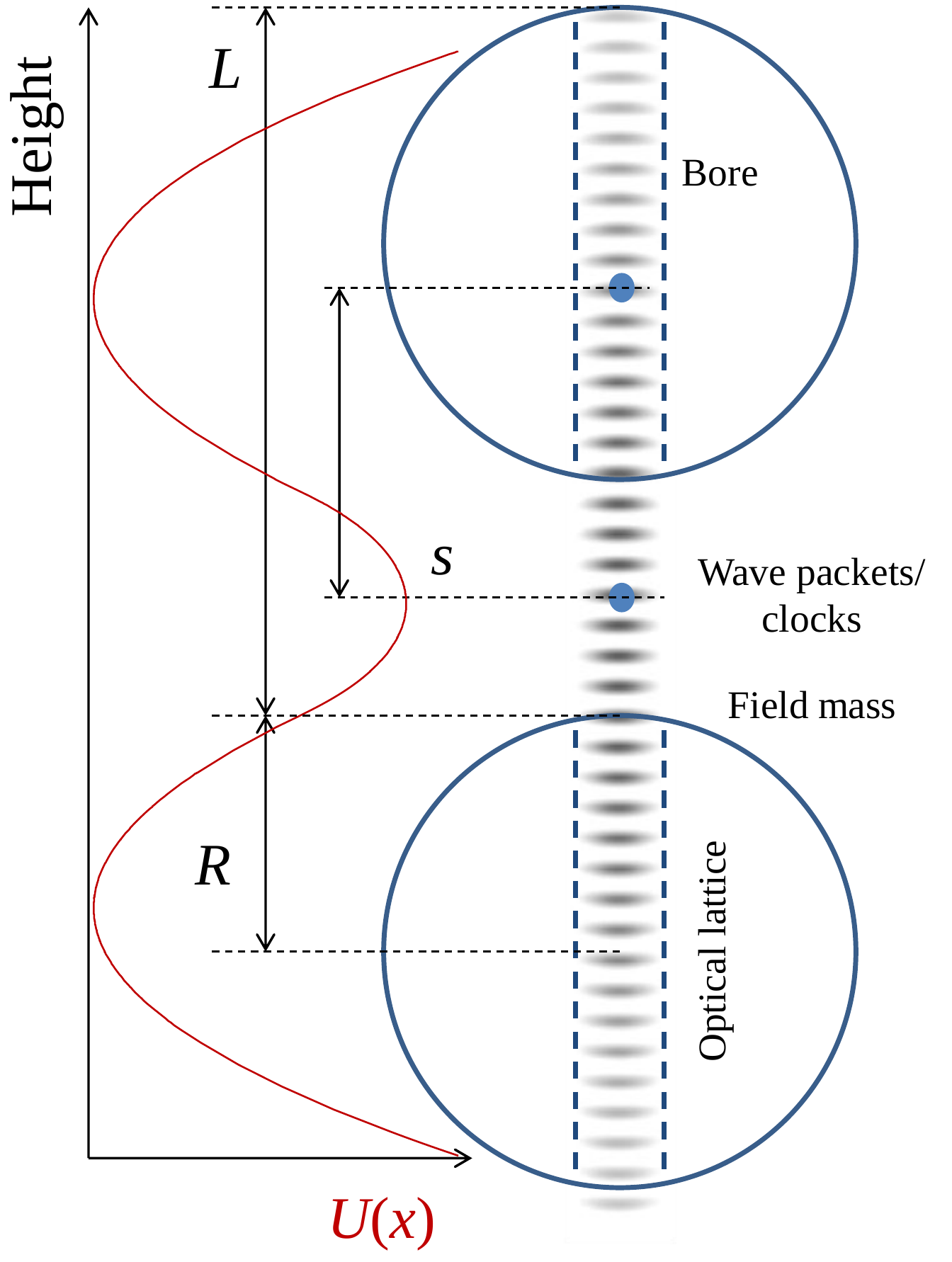,width=1.5in}
\epsfig{file=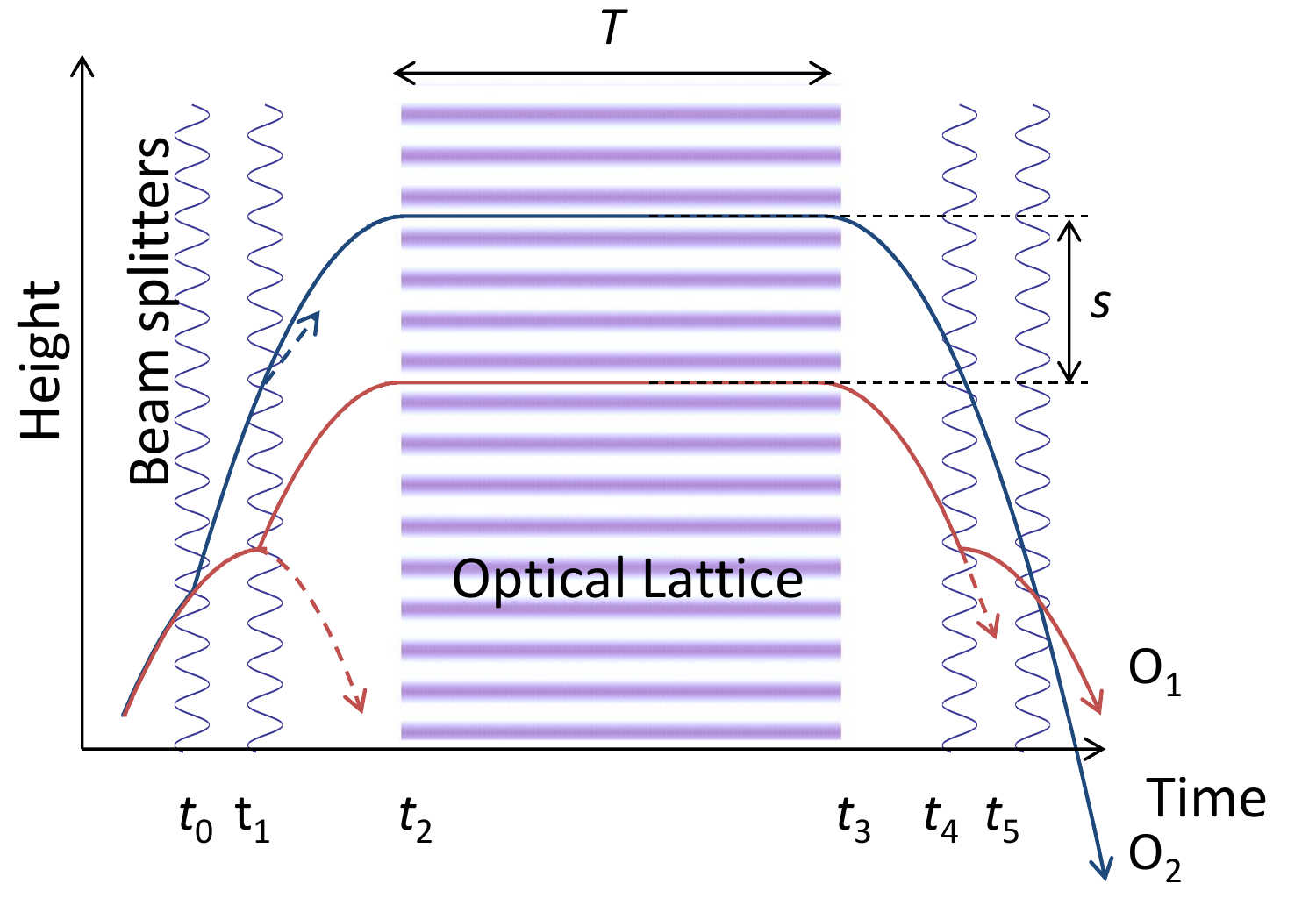,width=2.9in}
\caption{\footnotesize {\bf Left:} Setup. Spherical source masses create a gravitational potential $U(x)$. Partial atomic wave packets are brought to the saddle points of $U(x)$, where they accumulate phase shift given by the potential difference. {\bf Right:} Atom's trajectories versus time. Atoms from a MOT are launched upwards. $\pi/2$ pulses at times $t_0, t_1$ create a pair of wave packets that are brought to rest at $t_2$ by gravity. The optical lattice is switched on to hold the atoms, and the field-generating masses are brought in. The masses are removed and the atoms released at $t_3$. The waves interfere at the time $t_5$ of the final $\pi/2$ pulse. The dashed lines are examples for trajectories that do not interfere.}\label{setup}
\end{figure}

Gravitational analogs to the AB effect, broadly defined as phase shifts due to a gravitational potential $U$ in the absence of a gravitational acceleration or force \cite{Sagnac,stringy}, have also been of great interest, but to date no experimental realization of a gravitational AB effect \cite{Ho1997,Zeilinger1983} has been suggested that would produce a signal of measurable size. Here, we suggest a feasible experiment, see Fig. \ref{setup} \cite{GravAB}. It uses matter waves to probe the proper time in a multiply connected region of space-time comprised by two arms of an interferometer, Fig. \ref{setup}). The force caused by artificial gravitational field-generating masses vanishes in the space-time region in which the matter wave is allowed to exist. Using cold atoms held in an optical lattice, even the minuscule gravitational potential difference $\Delta U/c^2 \sim 1.6 \times 10^{-27}$ (Figure \ref{setup}) will produce a measurable phase difference \cite{redshift,redshiftPRL} $\phi_G=\om_C \int(\Delta U/c^2) dt$, owing to the long ($\sim 1\,$s) coherence times possible in such a system, and the large value of the atom's Compton frequency, $\om_C=mc^2/\hbar$.

This gravitostatic AB effect shares the distinguishing features of its electrostatic cousin: It is nondispersive, as the field-generating masses do not cause a force, motion, or distortion of the wave packet during the interaction time $T$ (residual phase shifts generated while the masses or the atoms are in motion are suppressed by varying the time $T$, keeping everything else constant). It is nonlocal, as is obvious from the sources' force-free configuration: no num,ber of local measurements (e.g., gravimeters) confined to the neighborhood of the atoms could register the field-generating masses' presence or predict the gravitostatic AB effect. It is also topological; the multiply connected region is the {\em space-time area} enclosed by the interferometer in Fig. \ref{setup}. This follows immediately from the fact that the interferometer phase is proportional to the line integral of a gauge-dependent integrand~\cite{Peshkin} (here, the local gravitational potential). The atoms' wave functions are confined by an optical lattice to the multiply-connected region comprised by the two arms of the interferometer (Figure \ref{setup}), wherein all gravitational forces due to the source masses in this region vanish, and so the interferometer will measure their induced topological phase.

\subsubsection{Signal size} The potential difference $\Delta U$ for a given $s$ has a relatively flat maximum for a sphere radius of $s=1.14R$. In this case, the distance between the spheres' centers is $L=2.62R$, and $\Delta U=1.17G\rho s^2$. The AB phase shift is then conveniently expressed as
\be\label{deltaphiG}
\delta \phi_G=0.33\left(\frac{s}{\rm cm}\right)^2\left(\frac{\rho}{20\,  {\rm g/cm}^3}\right)\left(\frac{m}{m_{\rm Cs}}\right)\left(\frac{T}{{\rm s}}\right),
\ee
where $m_{\rm Cs}$ the mass of Cs atoms. Were this force-free gravitational redshift, or gravitostatic AB effect, to be measured by atomic clocks, it would require km-sized source masses. Alternatively, clocks could be located at different Lagrange points of the Earth-Moon system.  Laboratory-scale tests, however, can make use of matter-wave clocks.

\subsubsection{Relation to other proposed gravitational AB effects} The gravitostatic AB effect considered here requires that there be no classical forces acting on the atoms, which is equivalent to vanishing Christoffel symbols in the atom's rest frame. Other definitions~\cite{Dowker1967,strings,combinations} go further and require a vanishing Riemann tensor. Since the Riemann tensor does not vanish in our experiment, rapidly moving particles may still feel a force, though the force acting on the atoms at rest is zero.


\subsection{Newton's gravitational constant $G$}

\begin{figure}
\centering
\epsfig{file=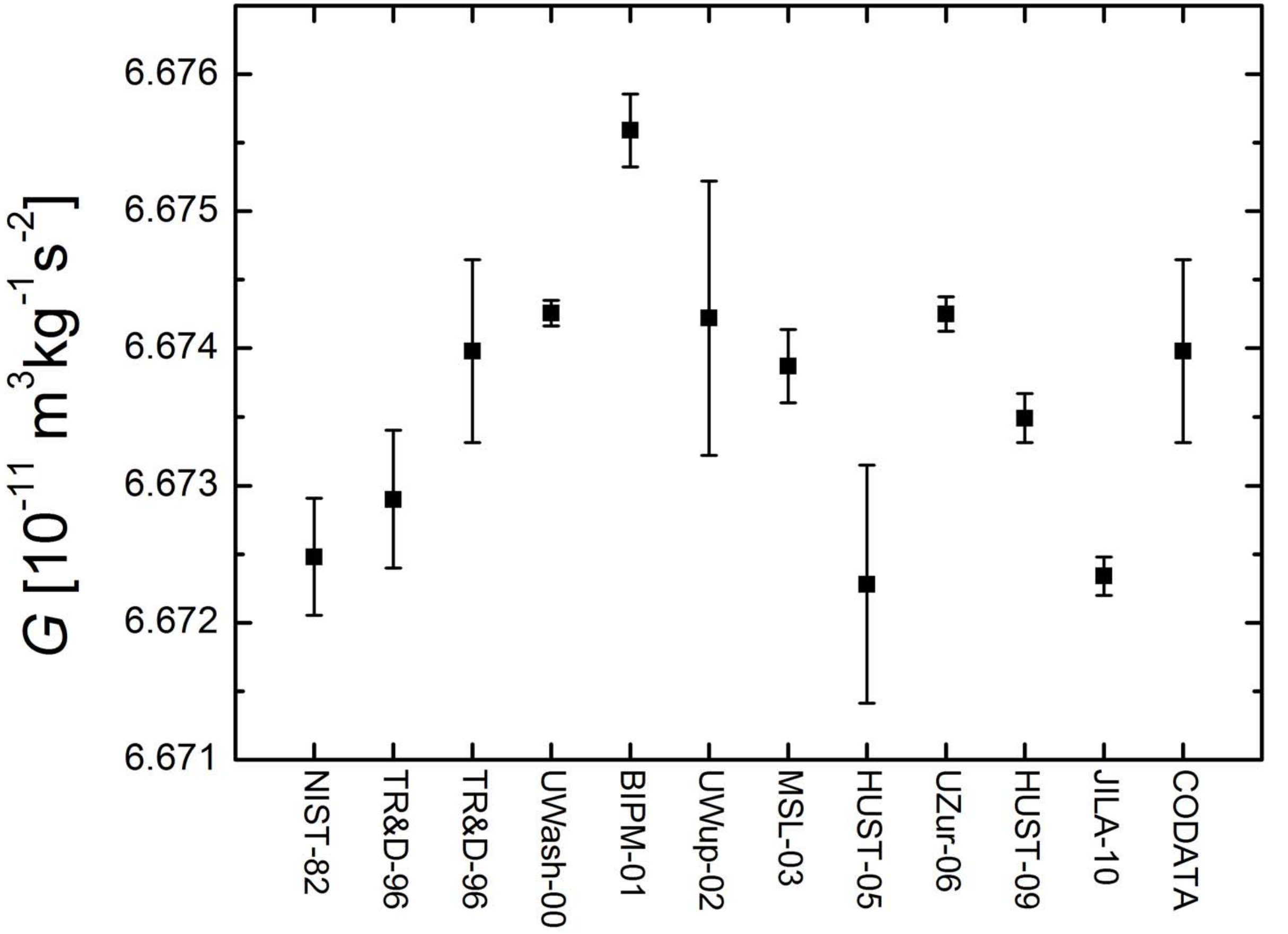,width=2.9in}
\caption{\footnotesize Measurements of $G$ entering the CODATA-2010 adjustment \cite{CODATA}.}\label{Gmeas}
\end{figure}
According to the latest adjustment by the Committee on Data for Science and Technology (CODATA), the gravitational constant $G=6.673 84(80)\times 10^{-11}$m$^3$kg$^{-1}$s$^{-2}$ is known with a relative error of 120\,ppm. Since $G$ has no known relationship to any other fundamental constant, it can only be measured directly. Fig. \ref{Gmeas} shows the CODATA value along with the 11 input data. They have all been derived from mechanical experiments with macroscopic masses. 
The only measurements of $G$ that are not based on mechanical measurements are based on atom interferometers. In principle, they promise a largely independent set of systematic effects. This makes them interesting, even if they are not yet accurate enough to feature in the CODATA adjustment.


The group of Mark Kasevich (Stanford) measured $G=6.693(21)\times 10^{-11}$m$^3$kg$^{-1}$s$^{-2}$ \cite{Fixler}. They combined two Mach-Zehnder cesium atom interferometers separated vertically by $\sim 1.3$\,m, addressed by the same laser beams. This gradiometer setup cancels the signal due to Earth's gravity. A lead test mass is moved up and down by $\sim 28\,$cm, causing a differential acceleration modulation of $\sim 30\times 10^{-9}g$ between the interferometers, modulating the measured phase shift of the atomic matter waves by $\sim 100$\,mrad peak to peak. 
The leading systematic effects include the atom's initial position (1.88 parts per throusand, ppt) and velocity (1.85 ppt), magnetic fields (1 ppt), rotations (0.98 ppt),
source position (0.82 ppt), source mass density (0.36 ppt) and dimensions (0.34 ppt), gravimeter separation (0.19 ppt), and source mass homogeneity (0.16 ppt).

Guglielmo Tino (Florence, Italy) measured $G=6.667(11)\times 10^{-11}$m$^3$kg$^{-1}$s$^{-2}$ \cite{Lamporesi} using a similar setup with $^{87}$Rb atoms and a combination of {\em two} annular  tungsten test masses (516\,kg total mass). The two masses move in a push-pull configuration so that their center of mass remains nearly stationary, suppressing distortions of the setup by the substantial forces needed to support the masses. The masses induce a differential phase shift modulation of 600\,mrad (peak to peak). Most of the 1.6\,ppt error is statistical. The leading systematic errors in parts in $10^5$ are position errors (36), the atom's initial velocity (23), source masses mass (9) and homogeneity (2.1), mass of the support platforms (8), gravity gradient (1), magnetic fields (3), and atom launch direction (6).

The gravitostatic AB effect could become the first measurement of $G$ not based on force, but on the gravitational potential difference between saddle points. 
This has important advantages: Near the saddle points, the potential is constant up to quadratic terms, suppressing errors arising from the uncertainty of the relative position of the source masses and the atoms. 
The small (several hundred grams compared to hundreds of kg) source masses can be made of single-crystalline, high-purity (99.99-99.999\%) heavy metals (tungsten, tantalum). 
This will reduce errors due to inhomogeneity and impurity of the masses. The spheres can be precisely manufactured and their shape characterized with optical means with high precision. They can be moved on lightweight supports whose mass will have negligible influence on the gravitational field, see below. While small source masses usually mean small signals, the test particles in the AB measurement are located very close to the field masses. We can thus use  $\sim 100$\,g test masses that to achieve the same phase shift that Ref. \cite{Lamporesi} achieved with 516\,kg.


\subsection{Experimental setup}\label{atomsource}
\begin{figure}
\centering
\epsfig{file=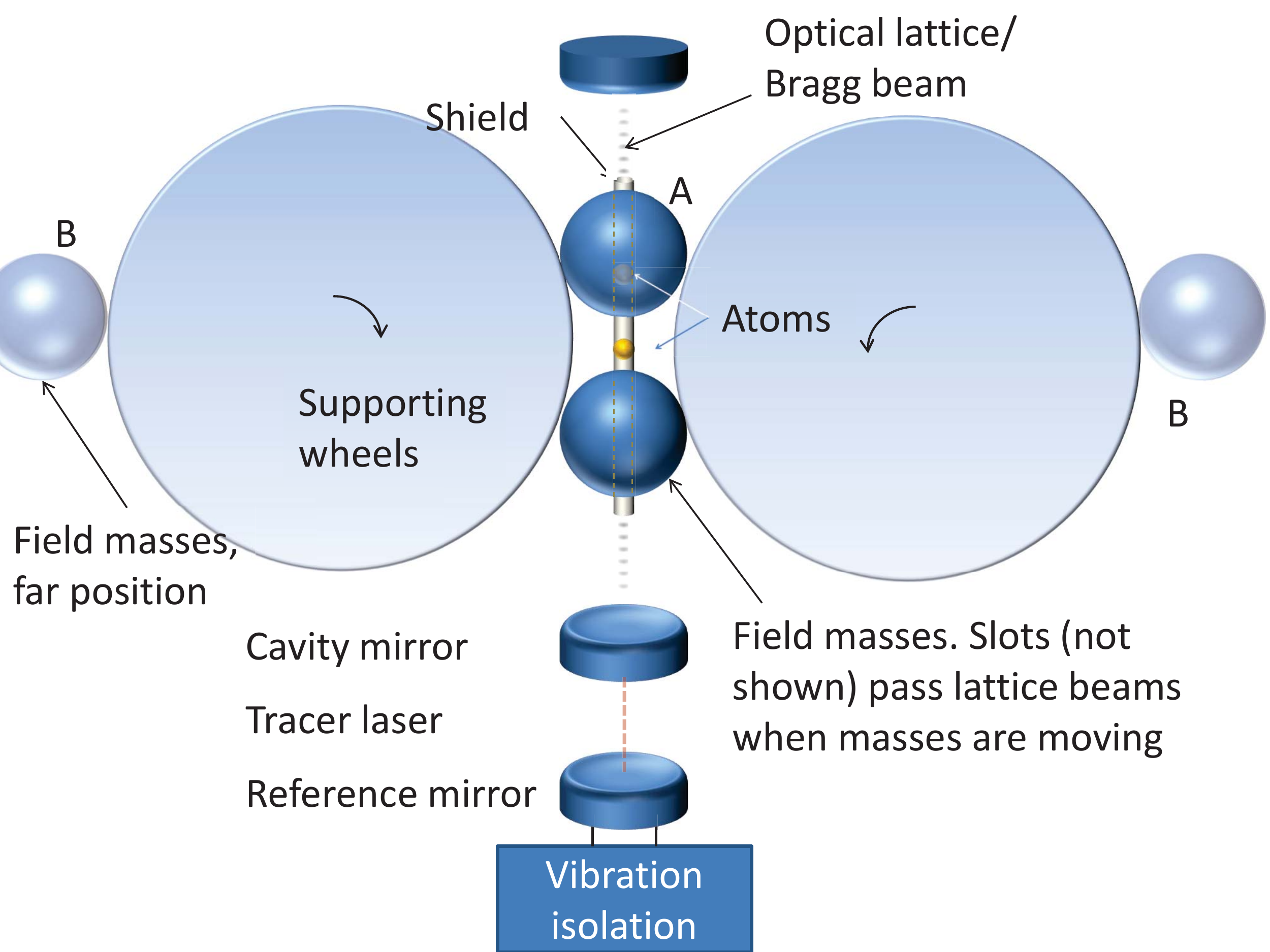,width=2.9in}
\caption{\label{wheels} Field-generating masses (slotted for passing the shield and the laser beam) move on the ``mass wheels." At position A, the masses create the potential difference $\Delta U$ between the atom's two positions; the masses at B, this difference is zero by symmetry. The wheels themselves are rotationally symmetric and do not change the potential difference. The Bragg diffration/optical lattice beam is resonantly enhanced and mode filtered in an optical cavity. The position of the entire cavity is actively referenced to a reference mirror on an active vibration isolation.}
\end{figure}

Figure \ref{wheels} shows two identical tungsten spheres whose combined gravitational potential has a saddle point between the spheres ($x_A=0$) and two lower potential saddle points at $x=\pm x_B$, close to the individual spheres' centers. In the close position, the spheres generate the potential shown in Fig. \ref{setup}. In the far position, the spheres are positioned symmetrically, such that they cause zero potential difference between the atoms' positions. The spheres are mounted to wheels which move them between the near and far positions. For an ideal circular geometry of these wheels, the wheel's rotation will not affect the gravitational potential, in contrast to the heavy support structures of previous atomic experiments. The wheels will be made of optical-grade glass, a low-density material whose homogeneity and dimensions can be characterized optically to high precision.

The laser beams that interact with the atomic sample will be resonant in an optical cavity. The cavity enhances the intensity of the laser beams by a factor of $\mathcal F/\pi$, where the Finesse $\mathcal F\approx 1000$. With just 10\,mW of laser power impinging the cavity, we will reach over 1\,kW/cm$^2$ intensity. This allows the use of large momentum transfer (LMT) beam splitters \cite{Losses,BraggPRL,BBB} which make the atom interact with a large number of photons. The cavity also avoids laser wavefront distortions, which we believe is the major decoherence mechanism in our large-scale atomic fountain. We thus expect to be able to split the partial waves by a large distance $s$. For the purpose of measuring the AB effect, we shall assume $s=10\,$mm.

At $10^{11}$ atoms/cm$^3$, about $10^7$ atoms are loaded into the spherical volume of 0.3\,mm diameter that can be trapped in the optical lattice. The density is reduced by $1/1000$ by selection of the $m_F=0$ state, of a subrecoil velocity group, and by expansion of the cloud  between $t_0$ and $t_2$. This leads to a shot noise of 10\,mrad per run, that averages to about 0.2\,mrad, or 0.7 parts per thousand of the expected signal, in one hour.  For the future project to measure $G$, we will increase the available atom number by Raman sideband cooling in an optical lattice, and a larger lattice beam.

The interferometer geometry needed for the experiment has already been demonstrated \cite{Charriere}. The atoms are launched as shown in Fig. \ref{setup}. At the time $t_0$, they interact with a pair of counterpropagating laser beams having wavenumbers $k_1, k_2$.

In our case, multiphoton Bragg diffraction is used to split the wave packets \cite{Losses,BraggPRL}. The possible Bragg diffraction order $n$ scales roughly with the 1/4 power of the laser intensity, if the detuning is adjusted so as to keep spontaneous emission constant \cite{Losses}. Our previous experiments \cite{SCI,BBB,Coriolis,CCC} reach $2n = 24$ photons with $0.5 \,$W/cm$^2$. Scaling to the 1\,kW/cm$^2$ intensity possible in the cavity in the proposed experiment, we expect a maximum momentum transfer of $2n\approx 24\times (1000/0.5)^{1/4}\approx 160$, while 32 is sufficient for this project.

Gravity brings the wave packets to rest at $t_2$. At this point, an optical lattice is switched on, which traps the atoms, keeping them at their respective positions. The field-generating masses are now brought in by rotating the wheels, and the matter wave packets accumulate relative phase due to gravity for a time $T$. The field-generating masses are moved out, the lattice is switched off, and the laser pulse sequence repeated to interfere the partial wave packets. The populations in the two outputs of the interferometer are detected by exciting the $F=4\rightarrow F'=5$ cycling transition of the atoms and observing their fluorescence with a camera or a photomultiplier tube.

Since we cannot turn off Earth's gravity, a true type I AB test (characterized by the complete elimination of {\em any} force acting on the wave packet~\cite{Batelaan}) would only be realizable in microgravity.  Nevertheless, such an experiment can be approximated in the laboratory using an apparatus to move the source masses into place after the wavepackets have reached their respective holding positions $x_{A}$ and $x_{B}$, with the masses' trajectories selected such that they produce no significant forces at any time. The effect of Earth's gravity can then be suppressed by comparing measurements made with and without the source masses.

When the states are interfered at $t_3$, the phase difference $\Delta \phi=\phi_A-\phi_B$ can be measured by detecting the population in the outputs of the interferometer, which is given by $\cos^2 \Delta \phi/2$. The phase difference $\Delta \phi$ consists of the AB phase produced by the source masses, $\delta \phi_G=m \Delta U T/\hbar$, analogous to the electrostatic AB effect \cite{Batelaan}.

Experimental techniques similar to the ones used for the proposed experiment have already been demonstrated separately by us: Interferometers that use optical lattices not only to hold, but even to accelerate, the atoms, have already been demonstrated experimentally by us \cite{BBB} and another group \cite{Charriere}. The needed splitting between the wave packets of $\sim 1\,$cm has already been reached by us \cite{Coriolis}.

\subsection{Systematic effects}

\begin{table}
\centering
\caption{\label{parameters} Dimensions as they enter our estimates for systematic effects. Present: setup proposed here. The required wave-packet separation has been demonstrated by us in \cite{Coriolis}, while still obtaining interference fringes. Future: tentative design for a precision measurement of $G$. This design will be validated by the operation of the ``present" setup.}
\begin{tabular}{lcc|c}\hline
Parameter & & Present & Future \\ \hline
Atomic species & & Cs & Cs/Sr\\
Wave packet separation & $s$ & 11.4\,mm & 25\,mm\\
Sphere radius & $R$ & 10\,mm & 22\,mm \\
Interaction time & $T$ & 1\,s & 10\,s \\
Time & $t_0=-t_5$ & -0.6\,s & -10.1\,s \\
Time & $t_1=-t_4$ & -0.7\,s & -10.2\,s \\
Momentum transfer & $\hbar k$ & 32 & 64 \\
Atomic density & $n$ & $10^8$/cm$^3$ & $10^8$/cm$^3$\\
Density balance & $\Delta n/n$ & $10^{-3}$ & $10^{-3}$\\
Atom posit., long. & $\sigma_x$ & 0.05\,mm & 0.05\,mm \\
Atom posit., trans. & $\sigma_r$ & 0.03\,mm & 0.03\,mm \\
Magnetic bias & & (10$\pm 0.1$)mG & (10$\pm 0.1)$\,mG \\
Lattice depth & $V_0$ & 10\,kHz & 5\,kHz \\
Lattice beam waist & $w_0$ & 0.3\,mm & 1\,mm \\
Inner radius of shield & & 1.2\,mm & 5\,mm \\
Finesse & $\mathcal F$ & 300 &  10,000\\
Repetition rate & $r$ & 20/minute & 4/minute \\
\hline
\end{tabular}
\end{table}

The proposed experiments are well within the range of technical possibilities. Many systematic effects have been described by us in \cite{GravAB}. 
Tab. \ref{parameters} lists the dimensions and parameters of the setup used for the proposed experiment, alongside those of a future setup for a precision measurement of $G$. The systematic effects for the present experiment are summarized in Tab. \ref{tab:systematics}. The ``present" scenario can give a demonstration of the gravitational AB effect to 7.5$\sigma$ significance (13\% error bar). This is based on a conservative estimate of the influence of magnetic fields and will likely be better. The ``future" one may achieve a $10^{-4}$ measurement of $G$.

\subsubsection{Zeeman effect}\label{Zeeman}
Choosing $m_F=0$ quantum states eliminates the linear Zeeman effect. The quadratic Zeeman effect for cesium is given by $\gamma^{(2)}=2\pi\times 430\,$Hz/G$^2$. Most backgrounds cancel when comparing the phase with and without the source masses in place. The source masses will not be ferromagnetic, as their residual iron content may be suppressed to the level of parts per million. Using a field of $B_0=10$\,mG to set the quantization axis, a $\delta B=1$\,mG variation due to the field masses causes a phase shift $\delta \phi=\gamma^{(2)} T 2 B_0\delta B$ of 0.05\,rad for the ``present" scenario. For the ``future" precision experiment, we assume that this can be reduced $\sim 10^3$ fold, using very pure materials for the field masses, magnetic shielding by a thin tube of mu metal(Fig. \ref{wheels}) and/or using both the $F=3$ and $F=4$ states, which feature opposite quadratic Zeeman effect.

\begin{table}
\caption{\label{tab:systematics} \footnotesize Leading systematic errors in parts per thousand (ppt) or ppm of the signal $\phi_G$, $407\,$mrad for the present setup and $19.5\,$rad in the future setup. The table gives the cause, the section in which it is described, and the magnitude in the present and future scenarios. Additional systematic effects from Ref. \cite{GravAB} are included; the ones marked $\ast$ cancel when comparing the experiment with and without source mass.}
\begin{tabular}{lrrr}\hline
Effect & Sec. & Present & Future \\
&  & ppt & ppm \\ \hline
Source mass magnetism & \ref{Zeeman} & $\pm 125$ & $\pm 25$ \\
AC Stark (Gaussian beam) & \ref{acstark}& $\pm 31$ & $\pm 59$ \\
AC Stark (fringes) & \ref{acstark}& $\pm 14$ & $\pm 1$ \\
Vibrations & \ref{vibe} & $\pm 28\sqrt{\rm hour}$ & $\pm 6.5\sqrt{\rm day}$ \\
Mean field shift & \ref{meanfield} & $\pm 2.3$ & $\pm 48$ \\
Field mass position & \ref{masspos} & $-0.09\pm 0.10$ & $-20\pm 21$ \\
Shot noise & \ref{atomsource} & $\pm 1.0\sqrt{\rm hour}$ & $\pm 0.4\sqrt{\rm day}$ \\
Rotational vibrations & \ref{rotationnoise} & $\pm 3.6\sqrt{\rm hour}$ & $\pm 23\sqrt{\rm day}$ \\
Source masses & \ref{homogen} & & 12 \\
Dispersive (Earth's gravity) & \cite{GravAB} & $\ast$ & $\ast$ \\
Quadratic potential & \cite{GravAB} & $2\times 10^{-3}$ & 2 \\
Dispersive (field mass) & \cite{GravAB} & $2\times 10^{-5}$ & 0.02 \\\hline
{\bf Total uncertainty} & & {\bf 132} (1 hour) & {\bf 87} (1 day) \\
 \hline
\end{tabular}
\end{table}

\subsubsection{Lattice potential/AC Stark effect}\label{acstark}
The lattice potential $V_0\cos^2 kx$ causes a phase shift that is mostly common to both interferometer arms and, moreover, cancels when comparing the phase with and without the field masses. The residual differential shift is estimated from the intensity difference of the laser beam at the atom's location for a Gaussian beam. In addition, while the laser beam inside the cavity is very nearly Gaussian, it diffracts at boundaries, e.g. the shield (Fig. \ref{wheels}). If the shield has a radius of $r_s$, a fraction of $e^{-2 r_s/w_0^2}$ of the beam power is diffracted. In the pessimistic scenario that half of the scattered power enters the shield, its interference with the cavity mode will cause quasi-random intensity variations with an amplitude of $\sqrt{2}e^{-r_s^2/w_0^2}$. For the values in Tab. \ref{tab:systematics} we conservatively assume that they do not average out between atoms.



\subsubsection{Vibrations}\label{vibe}

\begin{figure}
\centering
\epsfig{file=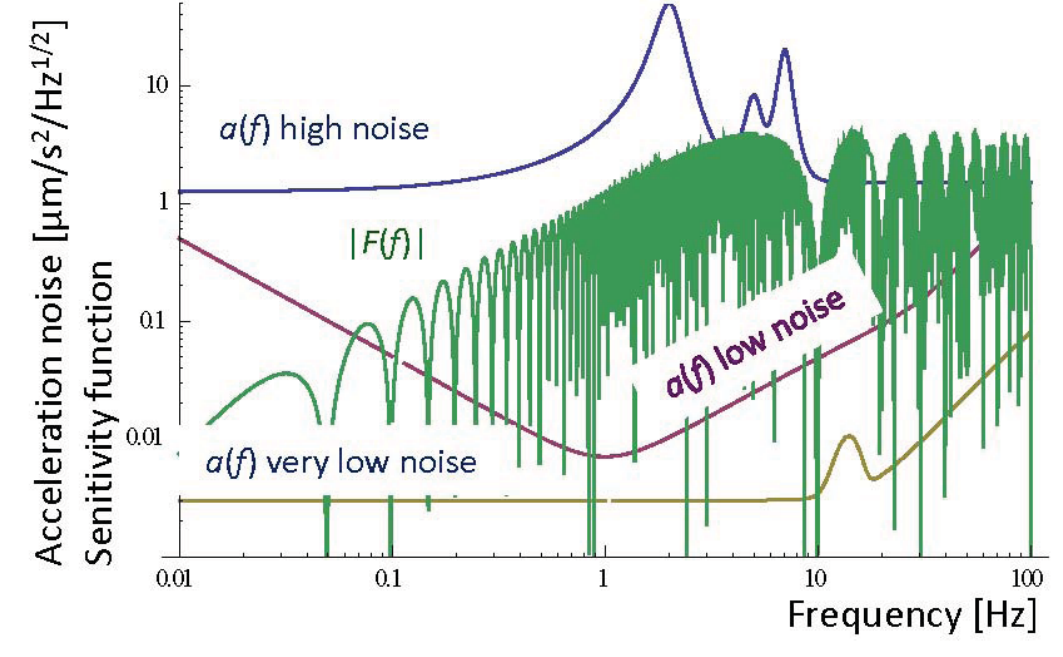,width=2.4in}
\caption{\footnotesize Acceleration noise, modeled after the measurements in Ref. \cite{PetersRSI} plotted versus frequency. The position noise $A(f)$ is obtained by dividing acceleration noise by $\omega^2$. The high noise model is representative of a setup without vibration isolation, the low noise model respresents the performance of the vibration isolator of Ref. \cite{PetersRSI}. The very low noise model represents the improved vibration isolator reported in ref. \cite{Petersmetrologia}. Also shown is the absolute value $|F(f)|$ of the sensitivity function of our interferometer in the ``future" scenario of Tab. \ref{parameters}.}\label{NoiseModel}
\end{figure}

Phase noise is caused predominantly by vibrations of optical elements, in particular the cavity mirrors. The influence of vibrations is modeled by a sensitivity function $F(\omega)$ \cite{Cheinet}. In the short-pulse regime, when the laser pulses are much shorter than $1/\om$, the sensitivity function is calculated by summing up the phase changes transferred from the moving laser to the atom at each laser pulse. For a simple Mach-Zehnder interferometer, e.g., $F=4\sin^2(\om T/2)$ \cite{Cheinet}. (The finite pulse duration $\tau$ of the laser causes an additional reduction of $F$ for frequencies exceeding $1/\tau$.) 
For our interferometer, we choose the origin of the time coordinate in the middle of Fig. \ref{setup} (right), such that $t_0=-t_5, t_1=-t_4, t_2=-t_3$. 
The optical lattice section does not, to leading order, transfer phase noise to the atomic wave function because any action of the lattice is common-mode to both interferometer arms. (Vibrations much stronger than those relevant here would drive transitions to higher lattice bands, causing loss of contrast.) We find $F(\om)=-[x(t_0)+x(t_1)+x(t_4)-x(t_5)]$ or
\be
F(\om)=4\sin\left(\om\frac{t_0+t_1}{2}\right)\sin\left(\om\frac{t_0-t_1}{2}\right).
\ee
The second sine function means that frequencies below $1/|t_0-t_1|$ are suppressed, which simplifies the vibration isolation considerably.

The ``high noise" and ``low noise" graphs in Fig. \ref{NoiseModel} show an analytic representation of acceleration noise $a(\omega)$ measured on a typical optical table (next to an operating air conditioning unit) with and without active vibration isolation \cite{PetersRSI}, alongside the sensitivity function $F$. We obtain the total noise affecting the experiment by noting that the position noise $A(\omega)=a(\omega)/\omega^2$ and integrating $k_{\rm eff}[\int_0^\infty (a(\om)F(\om)/\om^2)^2d\om]^{1/2}$.


In the present scenario, the high-noise model predicts an effective rms phase noise of $10^5$\,rad, making it impossible to see interference. Vibration isolation can reduce this noise: The low noise model (Fig. \ref{NoiseModel}) predicts an effective rms phase noise of 0.39\,rad so that fringes can be observed. The experiment can be repeated 1200 times per hour, allowing to average the vibration-induced noise down to 28 parts per thousand of the expected signal. For the future scenario we expect to be able to reduce the noise to the level of Ref. \cite{Petersmetrologia}, as represented by the ``very low noise model." The predicted effective phase noise is then 0.1\,rad. 


\subsubsection{Mean field shift}\label{meanfield}
Atom-atom interactions produce a phase shift $4\pi\hbar a nT/m$, where $a\sim 3000a_0$ is the scattering length of Cs atoms ($a_0$ is the Bohr radius) and $n$ is the atomic density. 
Systematic effects are caused by a difference in atomic densities $\Delta n$ correlated with the position of the field-generating masses. We know of no mechanism that would cause such correlations and thus assume they are $0.1\%$ or less, obtaining the estimate given in Tab \ref{tab:systematics}. 

\subsubsection{Test masses and atom positions}\label{masspos}
We consider the gravitational potential due to test masses in the geometry of Fig. \ref{setup}, i.e, $L=2.62R$. We assume that deviations of the atom positions from the maximum and minimum saddle points $(\delta x)_{\rm max}, (\delta x)_{\rm min}$ longitudinally and $(\delta r)_{\rm max}, (\delta r)_{\rm min}$ transversally are taken  from Gaussian distributions centered at zero with variances $\sigma_z$, $\sigma_r$, respectively. Second-order expansion of the potential 
leads to an expectation value $\langle \delta \phi\rangle$ of the phase shift of $(0.04\sigma_r^2-3.89\sigma_x^2)/R^2$ and a variance of $\sqrt{5.76\sigma_r^4+16.11\sigma_x^4}/R^2$ and leads to the estimates in Tab. \ref{tab:systematics}.

\subsubsection{Rotations}\label{rotationnoise}
The one-dimensional vibration-isolation system cannot eliminate rotational vibrations. Ideally, the planned interferometer does not enclose a spatial area (rather than a space-time area), as there is no horizontal velocity component of the atoms. The interferometer is thus insensitive to rotations $\vec \Omega$ to leading order. The residual influence scales with the induced displacement $\vec \delta_r$ of the optical lattices relative to the (inertial) matter wave packets, which causes a phase shift $\vec k_{\rm eff}\cdot  \vec \delta_r$. During the optical lattice holding time $T$, the atoms do not move relative to the lattice, so the relevant time scale is given by $t_1-t_0$. We thus arrive at an order-of-magnitude phase shift of $(\vec k_{\rm eff}\times \vec v_0)\cdot \Omega (t_1-t_0)$. For an estimate of $v_0$, the velocity $v_0$ must be less than $w_0/T$ if the atoms are to be held in the lattice of waist $w_0$.  
Rotational noise is typically much lower than Earth's rotation rate $\Omega_\oplus$ \cite{Coriolis}. Assuming vibrational noise amounts to less than $\Omega_\oplus/10$ rms, we obtain the estimates in Tab. \ref{tab:systematics}.

The constant rotation rate of the Earth is irrelevant, as it is suppressed when taking the difference between experiments with the source masses and without.

\subsubsection{Source masses}\label{homogen}
In \cite{Lamporesi}, the most precise measurement of $G$ with atom interferometry to date, source masses consisted of 24 tungsten cylinders and contributed an error of 90\,ppm through their mass and 21\,ppm through their homogeneity. Additional errors (80\,ppm) were due to the mass of the support platforms. The small (hundreds of grams versus hundreds of kg), spherical source masses can be made out of high-purity, crystalline material and thus extremely homogenous and their shape well controlled. While the measurement of $G$ will not be performed as part of the propsed work, we expect that these errors can be improved tenfold.





%

\subsection{Matter waves and the measurement of proper time}

The experiment will unambiguously demonstrate that matter-waves are not merely classical point masses that provide quantum measurements of the gravitational acceleration: a matter-wave is subject to the same gravitational redshift and time-dilation effects that apply to a conventional clock, even if it is constrained to a space time region of vanishing gravitational force. The proposed experiment will be the first demonstration of a force-free gravitational redshift, and the first experimental demonstration of a gravitostatic AB effect.  The effect is non-dispersive and topological, and thus impossible to ascribe to any local influences on the wave packet. This rules out interpretations that ascribe the phase entirely to the phase of the diffraction gratings at the positions of the wave packets \cite{WolfCQG,JoeSam,Giulini}.



\section{Antimatter interferometry}

The Antihydrogen Laser Physics Apparatus (ALPHA) has reported groundbreaking results in antihydrogen physics over the past few years: the world's first trapped antihydrogen \cite{Andresen2010}, confinement of antihydrogen for 1,000 seconds or more \cite{AndresenNatPhys}, and hyperfine spectroscopy of antihydrogen \cite{AmoleNature}. It has already conducted a pioneering measurement which constrained the gravitational acceleration of antihydrogen to a range of $-0.7\ldots +1.1\,$km/s$^2$ \cite{Antig}. The methods used, however, are not suitable for precision measurements; at best, they may tell us whether antihydrogen will rise or fall in Earth's gravitational field. The apparatus described here \cite{AntiHint} should be capable of precision at initially the 1\% level of precision and, with upgrades, the $10^{-6}$ level. We assume that anti-atoms will be laser-cooled to a temperature of 20\,mK \cite{Donnan2013} and that a new, vertical trap will be built at CERN.

The interferometer is designed to overcome the challenges posed by extremely scarce anti-atoms (about 300 per month compared to millions per second in a conventional atom interferometer), large thermal velocity of atoms, large uncertainty of the atoms' initial position, and the Lyman-alpha wavelength of 121 nm of hydrogen atoms, for which resonant lasers of sufficient power are unavailable. These challenges are addressed by combining innovations from atomic physics, antihydrogen trapping physics, and laser technology: use of high-energy, far red-detuned lasers overcomes the need for a high-powered Lyman-alpha laser (an available low-power Lyman alpha laser will be used for laser cooling at ALPHA); and magnetic confinement and recycling of atoms during the interferometer stage help overcome the challenges presented by dilute samples of rare atoms.

\subsection{The equivalence principle for antimatter}

Experiments show that normal, neutral matter closely adheres to the EEP, see \cite{WillBook,Will2006}. Neutral antimatter, however, has only recently been trapped by ALPHA and the Antihydrogen Trap ATRAP \cite{Atrap} at CERN, while tests of the EEP for charged particles \cite{Witteborn1967,Witteborn1968} have been inconclusive. As a result, the EEP has been directly confirmed neither for antimatter, except for ALPHA's pioneering coarse test nor for charged particles. A direct test of the EEP for charged particles as well as for antimatter is thus a very promising avenue to detect low-energy signatures of physics beyond the standard model.

In absence of any direct experimental measurement, indirect limits on equivalence principle violations for antimatter have been proposed. However, none of these arguments are universally accepted \cite{Goldmann1986,Nieto1994,Chardin1997,Fischler2008}. They rely on assumptions about the gravitational interactions of virtual antimatter, on postulates such as CPT invariance, or on other theoretical premises. None is entirely free of loopholes, and therefore none is a substitute for a direct measurement \cite{Nieto1991}. We will breifly review these here:

\subsubsection{Energy conservation}	A simple gedankenexperiment is based on energy conservation \cite{Morrison1958}: One annihilates a particle with an antiparticle and sends the photons up in a gravitational potential. There, the photons are used to make a new particle-antiparticle pair, which is then dropped to the original level. The gravitational redshift to the photons is known from experiment, as is the acceleration of free fall of the normal-matter particle. If we furthermore assume that the entire process conserves energy, and that there are no forces besides gravity, we can conclude that the gravitational potential energy of the antiparticle is the same as the one of the particle. However, this argument fails if the difference in the acceleration of antimatter and matter is caused by a "fifth force," as is generally the case in theories beyond the standard model that predict such a difference. The potential energy stored in the fifth force field leads to global conservation of energy \cite{Nieto1991}.

\subsubsection{Supernova 1987A}  Neutrinos and antineutrinos from supernova 1987A were simultaneously observed on earth after travelling for $10^5$ light years through the gravitational potential of the Galaxy. This implies that any difference in the gravitational interaction with neutrinos and antineutrinos should be below the percent level \cite{Longo1988,Krauss1988,Pakvasa1989}. However, the ability of neutrino detectors to distinguish neutrinos and antineutrinos is limited. Moreover, the mass of the observed neutrinos consists almost entirely of kinetic energy. While these tests are good confirmations of the equivalence principle for kinetic energy, they are quite insensitive to matter/antimatter anomalies \cite{Unnikrishnan2012}.

\subsubsection{Virtual antiparticles} 	L. I. Schiff \cite{Schiff1958} noted that atoms of normal matter contain a certain fraction of antimatter due to vacuum fluctuations. The relative contribution of such virtual antimatter to the mass of atoms can be estimated, and varies slightly from species to species (it tends to be larger for heavy atoms). Thus, the observed fact that all normal-matter atoms fall at the same rate to high precision rules out large equivalence principle violations for antimatter. However, the renormalization techniques used by Schiff can be criticized and the results be seen as inconclusive \cite{Nieto1991}.

\subsubsection{The Kaon system / electrons and positrons in Penning traps} The neutral Kaon $K_2^0$, which does not decay into pions, is a coherent superposition of $K_0$ and its antiparticle, which do decay into pions. M. L. Good \cite{Good1961} noted that the time evolutions of the wave functions of $K_0$ and its antiparticle experience redshifts in the gravitational potential. If the redshifts were different, the superposition would dephase, allowing a decay of the neutral K20 into pions. Since this has not been observed, the gravitational masses of K0 and its antiparticle should be equal. A similar argument has been made about the equivalence of the redshift of the cyclotron frequencies of positrons and antipositrons in Penning traps \cite{Hughes1991}. However, the arguments involve the absolute gravitational potential, an unphysical quantity. The sensitivity varies between $10^{-8}$ and $10^{-15}$ depending on whether the potential due to the earth, the sun, the galaxy or the entire universe are used. One may avoid the use of absolute potentials by resorting to the yearly variation of the sun's gravitational potential on earth that results from the earth's elliptic orbit. Because this variation is weak ($\Delta U/c^2\sim 10^{-11}$), no interesting limits are obtained.

\subsubsection{Bound kinetic energy} Finally, there are our bounds based on kinetic energy, see section \ref{NucSec}. These limits become invalid if there are anomalies associated with the particles mediating the binding forces, or if the deviations from general relativity and the standard model are outside the scope of the SME.

\subsection{Previous experiments with charged particles}

Gravitational experiments with charged particles are extremely difficult. An electric field of only 10-11 V/m will swamp any gravitational force on electrons. Nevertheless, Fairbank and Witteborn attempted to measure the acceleration of free fall for electrons \cite{Witteborn1967}. A pulsed source of particles was located at the bottom of a metallic electrostatic shield. The particles travel vertically upwards to the top, where they are detected. A particle needs to have a minimum initial velocity in order to reach the detector. These minimum-velocity particles are the last to arrive at the detector after the pulse has been emitted, and their time of flight is characteristic of their acceleration of free fall.

The electrostatic shield causes a fundamental problem. It contains a free electron gas which will move in response to any field, until it generates an electric field whose action on the electrons cancels the original field. This applies to the motion of the electron gas in response to gravity: the electron gas falls under its own weight until it generates an electric field whose force cancels the original gravitational force. The free fall we would like to observe will be subject to this electrostatic force and, thus, not fall. This makes it impossible to observe the electron's g. Indeed, the observed acceleration was consistent with zero, but the result has been criticized as an artifact, as it seems unlikely that the requisite freedom from stray electric fields was indeed achieved \cite{Nieto1991}. An attempt to measure the acceleration of free fall of a positron, on the other hand, should result in an acceleration of 2g. Such an experiment was proposed \cite{Witteborn1968}, but not realized.

As a result, there has been no experiment measuring the acceleration of free fall of a charged particle, be it matter or antimatter. This is a serious gap in the verification of the equivalence principle. For example, because neutral atoms always contain the same number of protons and electrons, it is impossible to separately determine the acceleration of free fall of these particles from existing data. This leads to a gap in the determination of parameters of the Standard Model Extension describing EEP violation \cite{datatables}. It should be noted that no indirect ways to mend this gap have been proposed to date. Thus, it is as urgent to test the EEP for charged elementary particles as it is to test it with antimatter.

\subsection{Setup}
\begin{figure}[t]
\centering
\epsfig{file=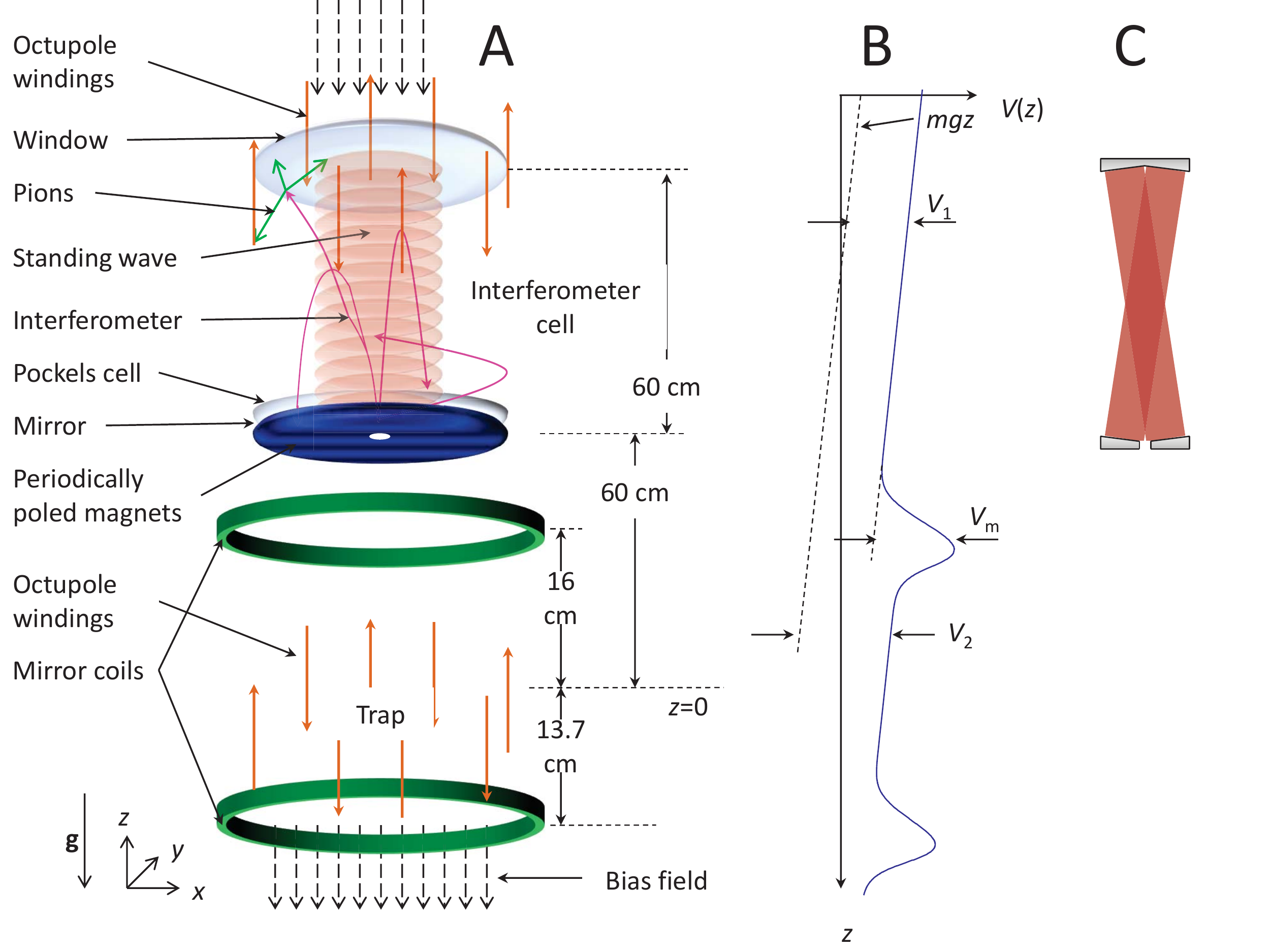,width=0.75\textwidth}
\caption{\label{interf} (A) Schematic. Atoms are extracted from the vertical magnetic trap (bottom) into the interferometer cell (top) by adiabatically lowering the trapping potentials, creating an antihydrogen fountain. The octupole is wound onto these walls of the vacuum chamber, which have an inner radius of 2.22\,cm. (B): Potential, not to scale. (C) Schematic of an off-axis multipass cell.
}
\end{figure}

The setup (Fig. \ref{interf}) is described in greater detail in \cite{AntiHint}. It  consists of two joined magnetic traps, the lower ``trap" region wherein antihydrogen atoms are produced and laser cooled, and the upper ``interferometer cell" wherein interferometry is performed. These traps are similar to the one currently used by ALPHA, but oriented vertically. Atoms are laser-cooled to 20\,mK in the trap \cite{Donnan2013} and then adiabatically released into the interferometry cell. Interferometry is performed using a powerful off-resonant laser, retroreflected using a mirror that divides the interferometer cell and the trap. Atoms that have received momentum transfer from the laser are energetic enough to leave the trap and  detected by their annihilation products at the vacuum chamber walls.

Atoms are adiabatically released from the trap region into the interferometer cell. To achieve a nearly constant average vertical velocity, the trap solenoid is turned off completely while the upper mirror coil is ramped linearly. This results in particles entering the interferometer cell with the velocity distributions as narrow as 0.4 m/s rms vertically and 5 m/s horizontally. These figures can be improved further by optimizing the magnetic field configurations and ramp time constants.

The atoms enter the interferometer cell through an aperture. They can be prevented from colliding with the walls by periodically poled refrigerator magnets, see Fig. \ref{interf} A. Use of a large central aperture with a slightly tilted laser beam, an off-axis multipass cell, see Fig. \ref{interf} C, or bringing in the  laser beams from top and bottom avoids this bottleneck. The atom's upwards velocity when they enter the interferometer cell is chosen such that gravity will turn them around (86\,cm above the trap center for the atom's average velocity) before they reach the top of the interferometer cell, unless they receive an upwards momentum kick from the interaction with photons from the laser.

The interferometer cell is basically another magnetic trap. The overall potential seen by an atom depends on the radius coordinate $r$ as $\sqrt{V_6(r/\rho)^6+V_1^2}$, where $V_6$ and $\rho$ are constants. The atom interferometer is formed by the atoms' interaction with counterpropagating pulses from a laser whose wavelength is far off-resonant with any atomic transition. For a far-detuned infrared laser, the two-photon Rabi frequency $\Om^{(2)}=\alpha I/(2\eps_0\hbar c)$ is given by the atom's dc polarizability $\alpha$, the laser intensity $I$ and the vacuum permittivity $\eps_0$. For hydrogen, $\alpha=(9/2) 4\pi \eps_0 a_0^3$ exactly, so that $\Om^{(2)}=9\pi a_0^3 I/(\hbar c)$, where $a_0$ is the Bohr radius. Since the dc polarizability is nonzero for any atom, the interferometer can work with any species.

An ideal interferometer would have a contrast of one. In practice, this ideal contrast is not realized, e.g., when laser pulses miss the atom. In our apparatus, however, such atoms keep orbiting in the trap and thus have a chance of $P_b$ to encounter the laser beam again and take part in an interferometer. Fig. \ref{interf} shows the simulated fringes. The simulation takes into account the geometry of the trap, the laser beam, and all magnetic fields; the 3-dimensional motion of the atoms, and the quantum mechanics of the beam splitters. It starts with tracing the paths of a laser-cooled sample of antihydrogen at 20\,mK in the trap for 0.1\,s and then simulating the adiabatic release from the trap. The atom-light interaction is modeled by numerically integrating the Schr\"odinger equation using the $|a,2n \hbar k\rangle$ ($n=-5,\ldots 5$) states as basis states, fully accounting for the Doppler shift of the laser frequencies as seen by the moving atoms.  

The observed contrast decay is due to magnetic field gradients caused by the mirror coils. It can be avoided by reducing such gradients, e.g., using multiplet mirror coils. A laser of shorter wavelength, e.g., 532\,nm, will increase the initial contrast to $\sim 50\%$, as the larger recoil velocity has a more favorable ratio to the vertical velocity spread. Short wavelengths also lead to a larger measured signal, allowing better resolution.

\begin{figure}
\centering
\epsfig{file=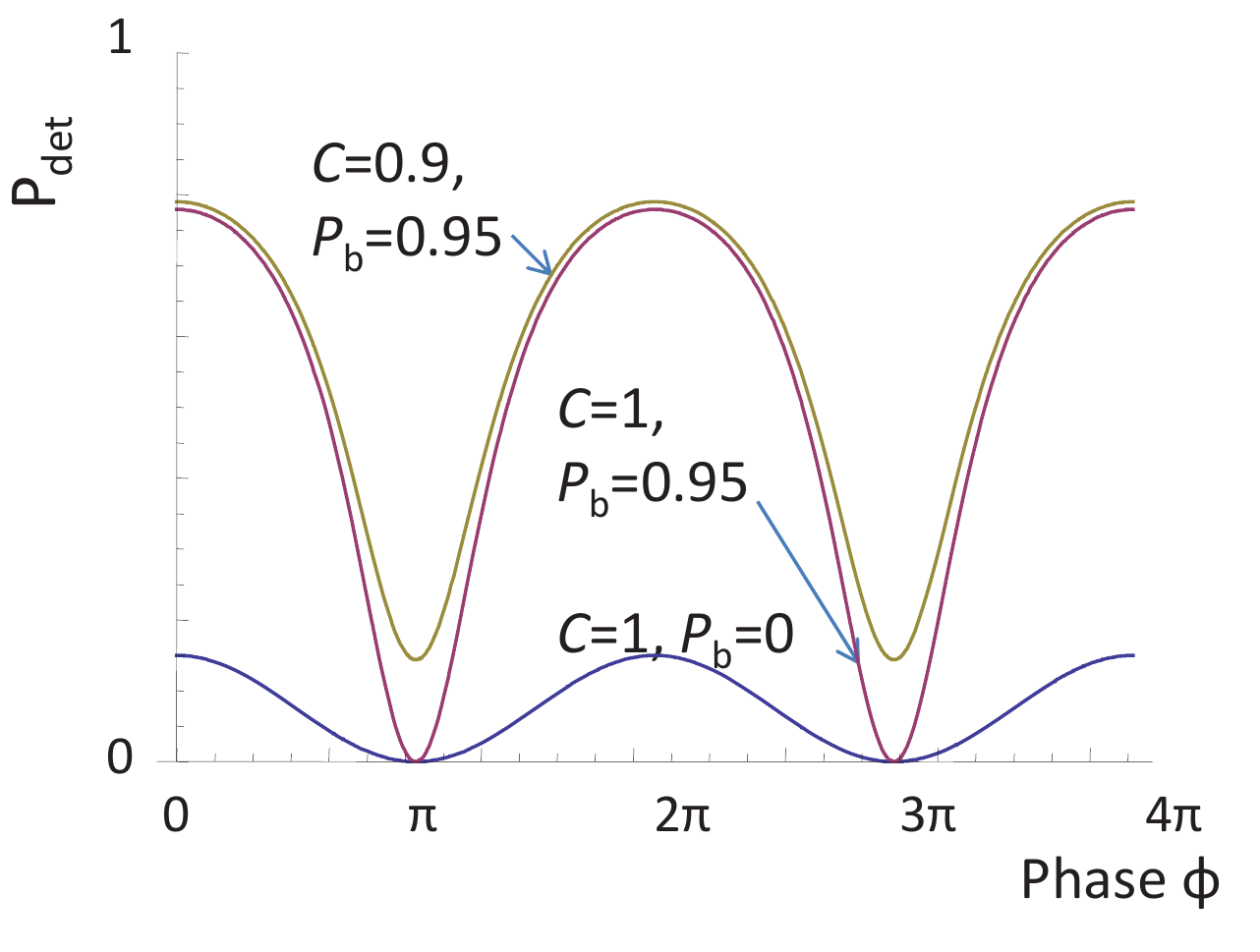,width=0.4\textwidth}
\epsfig{file=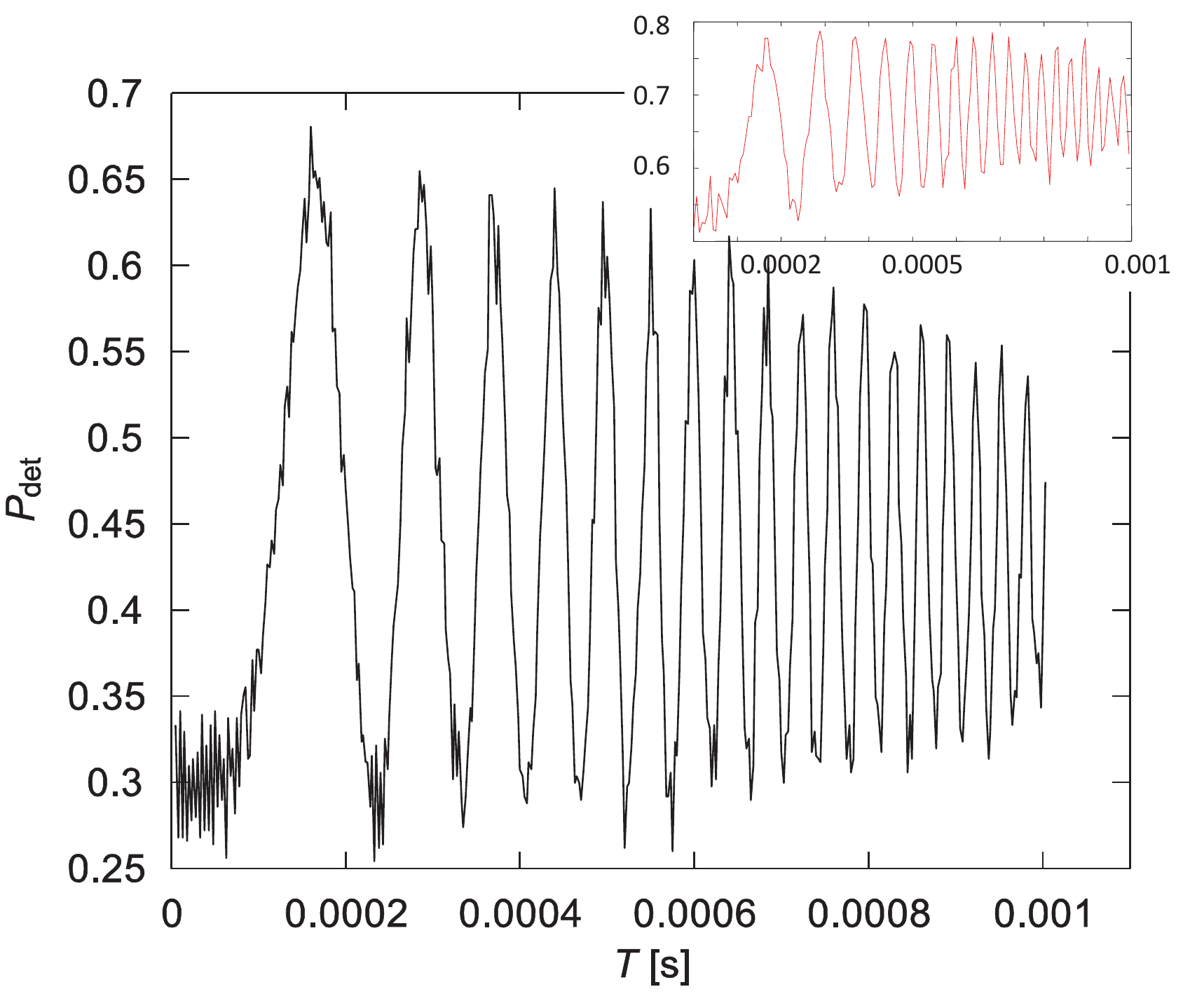,width=0.4\textwidth}
\caption{\label{Recycling} Left: Fringes of a simple interferometer. Atom recycling leads to higher visibility and sharpens the features. Right: Simulation of the full atom interferometer. The number of atoms detected at the top of the interferometer cell versus pulse separation time $T$ shows the expected $\sin^2(kgT^2)$ signature. Inset: simulation taking into account a 5-mm diameter aperture in the mirror, with 256\,s adiabatic release time. The laser beam has 1064-nm wavelength and 1\,cm radius with a flat-top intensity profile. The pulses have a Gaussian time envelope with a $\sigma=250\,$ns time constant and a $\pi-$pulse energy of 7.4\,J. }
\end{figure}

\begin{figure}
\centering
\epsfig{file=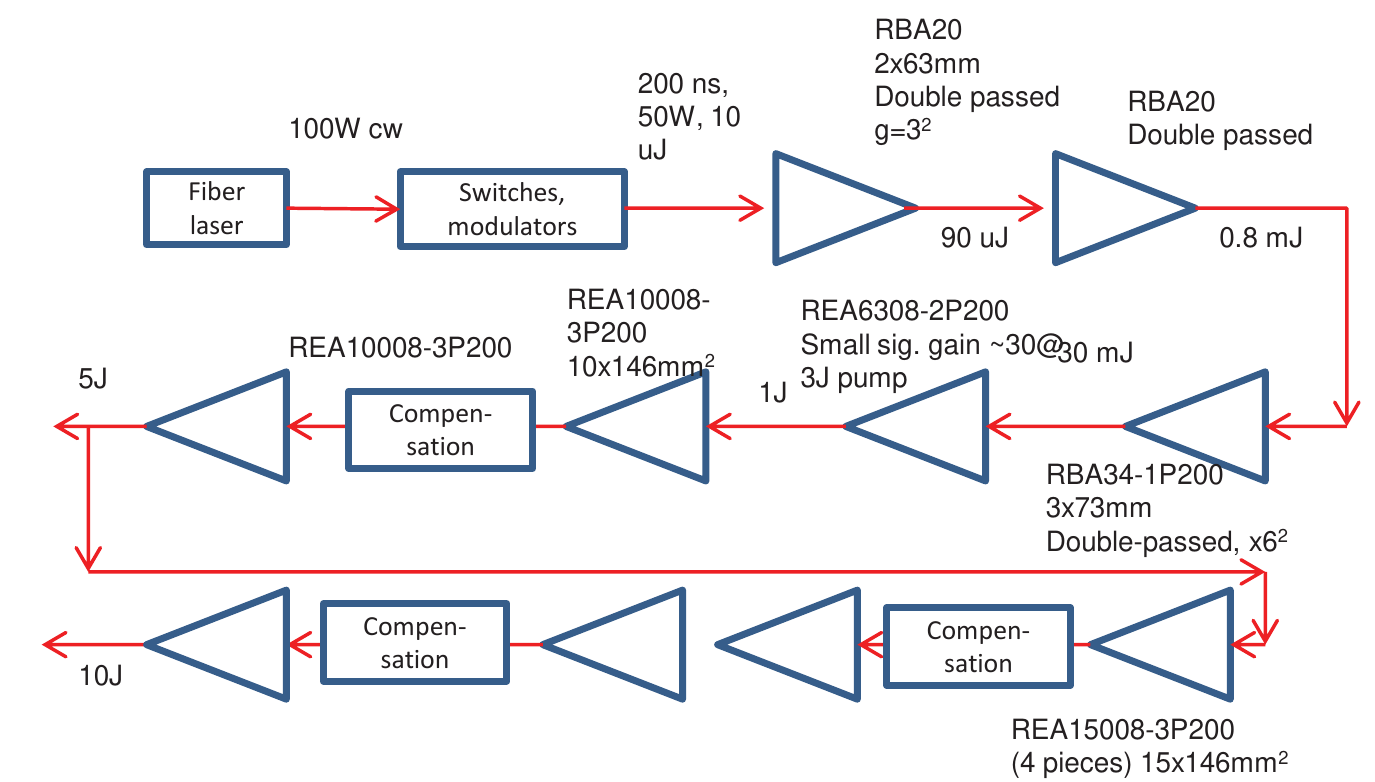,width=0.4\textwidth}
\caption{\label{AntiHintLaser} Laser system for experiments with (anti-)hydrogen or charged particles.}
\end{figure}

The laser system (Fig. \ref{AntiHintLaser}) starts with a 100-W fiber laser at 1064 nm, that is first modulated to shape pulses and then amplified in several states of diode-pumped Nd:YAG amplifier modules. The low-power stages are double passed for sufficient gain.

\subsection{Interferometry with charged particles}\label{chargedpart}

Using the same laser, and sharing the cryostat with the hydrogen interferometer, the experiment can work with electrons, protons as well as their antiparticles.
\begin{figure}
\centering
\epsfig{file=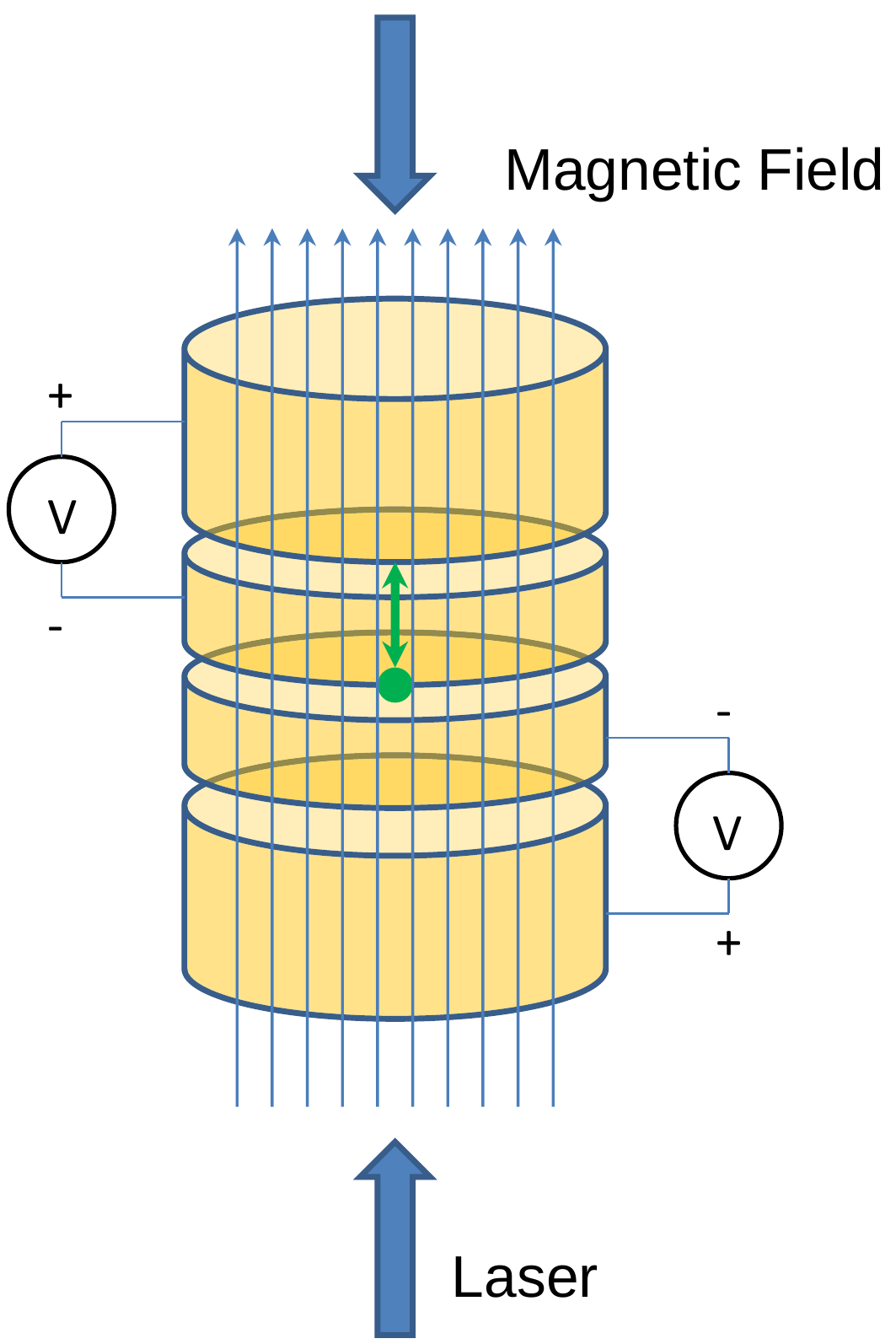,width=0.2\textwidth}\quad
\epsfig{file=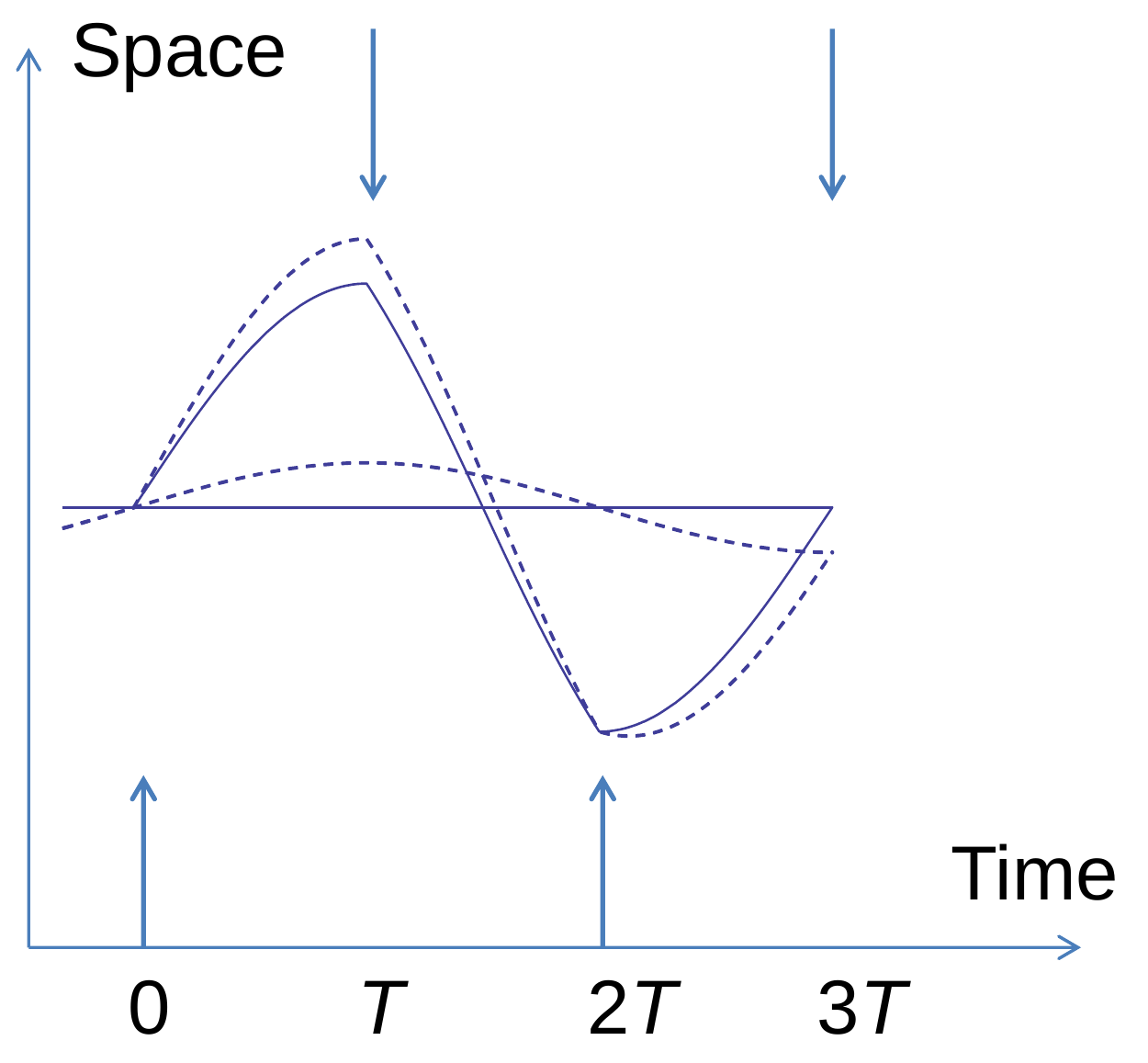,width=0.3\textwidth}\quad
\epsfig{file=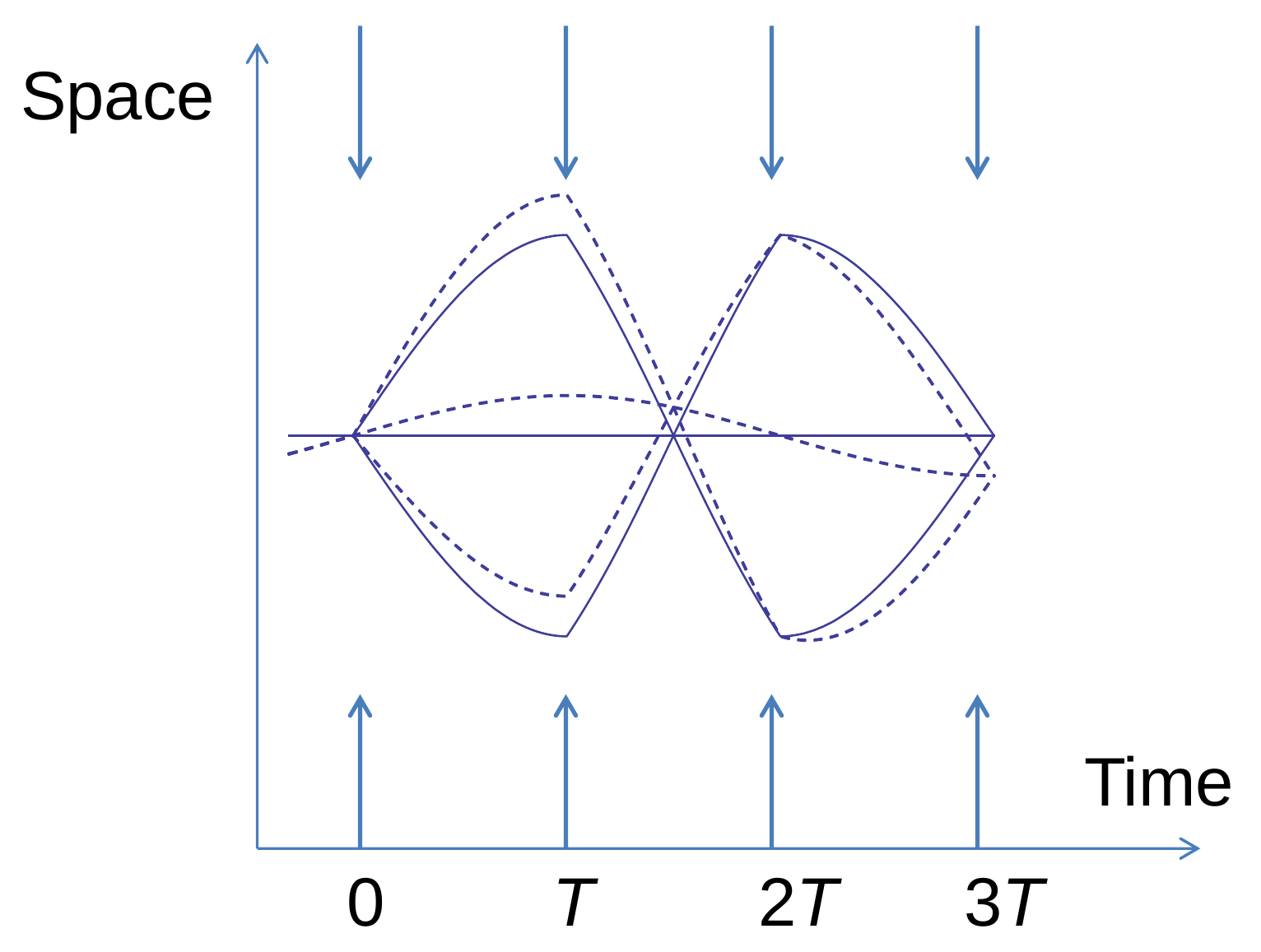,width=0.35\textwidth}
\caption{\label{Penning} {\bf Left:} Penning trap suitable for interferometry with charged particles. {\bf Middle} Trajectories. Dashed lines represent interferometer with initially moving electron, arrows represent the direction of the laser pulses' momentum transfer. The four pulses are $\pi/2$ pulses transferring $2\hbar k$ of momentum. For a harmonic axial potential, the interferometer closes regardless of initial motion of the electron. {\bf Right:} Double diffraction trajectories. Dashed lines represent interferometer with initially moving electron.}
\end{figure}

The particle is held is a Penning-Malmberg trap (Fig. \ref{Penning}, left). Multiple compensation electrodes allow trap anharmonicities to be compensated for, so that the $z^4$ and $z^6$ terms may individually be set to zero. The axial motion is harmonic with an axial frequency $\om_z$.\footnote{We here assume that the axial motion is completely decoupled from the radial motion.}  The particle is Bragg diffracted by a laser through the Kapitza-Dirac effect \cite{Freimund2002}. Multiple diffractions form an interferometer (Fig. \ref{Penning}). If the pulse separation time satisfies $T=\pi/(2\om_z)$, a closed interferometer is obtained independent of the initial motion of the particle, and the phase difference between the upper and lower interferometer $\Phi=(8\pi)\om_rT$  is independent of initial motion.

The two-photon Rabi frequency for a charged particle is given by
\be
\Om_{R,e} =\frac{q^2I}{\hbar \epsilon_0 c \om_L^2 m}= \frac{256}{81}\frac{\lambda^2}{\lambda_0^2}\Om_{R,H}
\ee
where $q$ is the charge, $I$ the laser intensity, $\epsilon_0$ the vacuum permittivity, $m$ the particle mass, $\Om_{R,H}$ the Rabi frequency for a hydrogen atom, and $\lambda_0$ the Lyman-$\alpha$ wavelength. Thus, for a 266-nm quadrupled YAG laser, the electron Rabi frequency 15 times as large as the one for hydrogen, for a 1064-nm laser even 250 times.

The major systematic effect is from patch charges. Very clean gold-coated surfaces at 4\,K can have patch potentials of tens of microvolts. They can be further  suppressed by free electrons on a thin helium film \cite{Etz1984}. Such potentials can be considered a perturbation of the trap potential. At leading order, they change the axial frequency, and can thus be measured and taken out. Higher terms cause anharmonicities, but the first two of them can be cancelled by adjusting the trap. Thus, high-precision interferometry should be possible despite patch effects.

\subsubsection{Testing the equivalence principle for charged particles}
One particular application is an equivalence principle test with electrons and positrons. An electron Compton clock would deliver a measurable frequency (several GHz for electrons) determined solely by the mass of the electron \cite{CCC}. Monitoring its frequency over the course of a year will subject it to a gravitational redshift due to the change in the Sun's gravitational potential on Earth. If the equivalence principle holds for electrons, this redshift should be the same as the one to a conventional atomic clock. Comparing them will thus yield a test of the equivalence principle for a charged particle. Because the test does not rely on monitoring the electron's free fall, it is not affected by the shielding issue of the Fairbank-Witteborn experiment (see above).  Rather, it is based on the action of the gravitational potential on the electron rest-mass clock, which cannot be shielded.

\subsubsection{Phase calculation for a single interferometer}
We first consider the interferometer of Fig. \ref{Penning}, middle, with a harmonic potential, initial position and velocity zero, and perfect timing. Thus,
\be
z(t)=A\left\{\begin{array}{lr} \sin\om_z t & 0<t<T \\ \sqrt{2}\cos(\om_z t-\pi/4) & T<t<2T \\ \cos\om_z t & 2T<t<3T \end{array} \right. ,\quad T=\frac{\pi}{2\om_z}.
\ee
where $A=\hbar k/(m\om_z)$. The free evolution phase is
\be
\Delta \phi_{\rm F}=\frac{\hbar k^2T}{m\pi}=\frac 4\pi \om_r T.
\ee
The laser phase is $\sum (\pm) k z=(-k)(+A)+(+k)(-A)=-2Ak=-2\frac{\hbar k^2}{m\om_z}=-\frac 8\pi \om_rT$. The total phase is
\be
\Delta\phi=-\frac4\pi \om_rT
\ee

\subsubsection{Nonzero initial position and velocity}
If we assume an initial motion of the electron according to $z(t)=A_0\sin(\om_z t+\phi_0)$ at $t<0$, the trajectories are
\bea\label{traj1}
z_1(t)&=&\left\{\begin{array}{lr} A_0\sin(\om_z t+\phi_0)+A\sin\om_z t, & 0<t<T \\
A_0\sin(\om_z t+\phi_0)+A[\sin\om_z t -\sin\om_z (t-T)], & T<t<2T\\
A_0\sin(\om_z t+\phi_0)+A[\sin\om_z t & \\ -\sin\om_z (t-T)+\sin\om_z(t-2T)]. & T<t<2T\end{array}\right. , \nonumber \\
z_2(t)&=&A_0\sin(\om_z t+\phi_0),
\eea
where $A=\hbar k/(m \om_z)$ and $A_0=\hbar k_0/(m \om_z)$. For $T=\pi/(2\om_z)$, the interferometer closes for all $A_0, \phi_0$. We calculate the free evolution phase as
\begin{eqnarray}
\Delta \phi_{\rm F}=\frac{\om_r}{\om_z}\left[3\cos\om_z T-4\cos2\om_zT-2\cos4\om_zT \ +\cos5\om_zT +\frac{k_0}{k}\left(2\cos(\om_zT+\phi_0)\right.\right. \nonumber \\ \left.\left. -2\cos(2\om_zT+\phi_0)
+4\cos(3\om_zT+\phi_0) -2\cos(4\om_zT+\phi_0) +2\cos(5\om_zT+\phi_0)\right)\right]\sin\om_z T. \nonumber
\end{eqnarray}
Proceeding as above, we calculate the total phase. In order to study the effect of slight timing errors or uncertainty in $omega_z$, we substitute $\om_z\rightarrow \om_z+\delta_z$, where $\delta_z$ is small.
\bea
\Delta \phi&=& \frac 4\pi \om_r T\left(1-\frac{k_0}{k}(\cos\phi_0-\sin\phi_0)\right) -\frac{4}{\pi^2}\om_r \delta_z T^2  \left\{2(\pi-1)\right.\nonumber \\ && \left. +\frac{k_0}{k}[2(\pi+1)\cos\phi_0+(\pi-2)\sin\phi_0]\right\}
\eea
In principle one could scan $T$ and zero the $\delta-$dependent term.


\subsubsection{Double diffraction}

Since the above methods are sensitive to the atom's initial oscillation, we'll try a more complicated scheme with double diffraction, i.e., two interferometers in which the recoil directions are reversed relative to each other. The trajectories of the first interferometer are given by Eq. (\ref{traj1}), the third trajectory is
\be\label{traj3}
z_3(t)=\left\{\begin{array}{lr} A_0\sin(\om_z t+\phi_0)-A\sin\om_z t, & 0<t<T \\
A_0\sin(\om_z t+\phi_0)-A[\sin\om_z t &\\ -\sin\om_z (t-T)], & T<t<2T\\
A_0\sin(\om_z t+\phi_0)-A[\sin\om_z t & \\ -\sin\om_z (t-T)+\sin\om_z(t-2T)]. & T<t<2T\end{array}\right.
\ee
Each interferometer has a phase $\Delta\phi^A=\Delta\phi_{\rm F}^A+\Delta \phi_{\rm I}^A, \Delta\phi^B=\Delta\phi_{\rm F}^B+\Delta \phi_{\rm I}^B$. Their sum,
\bea
\Phi=\Delta\phi^A+\Delta\phi^B \\
=2\frac{\om_r}{\om_z}\sin\tfrac 12 \om_zT \left(3\cos\tfrac 12 \om_zT-\cos\tfrac 32 \om_zT+2\cos\tfrac 72 \om_zT -\cos\tfrac 92 \om_zT+\cos\tfrac{11}{2} \om_zT\right) \nonumber
\eea
is completely independent of initial electron motion. We let $\om_z=\pi/(2T)+\delta_z$ and expand
\be
\Phi=\frac 8\pi \om_rT+\frac{16}{\pi^2}(\pi-1)T^2\delta_z+\frac{4}{\pi^3}\om_rT^3\delta_z^2(8-8\pi+7\pi^2)+...
\ee

\subsubsection{Trap anharmonicity}
Let's say the trap potential is
\be
V=\frac12 m \om_z^2 d^2 \left(\frac{z^2}{d^2}+D_3 \frac{z^3}{d^3}+D_4\frac{z^4}{d^4}+\ldots\right),
\ee
where $d$ is the typical trap size and $D_{3,4,...}$ are coefficients. We assume the anharmonicity is low so it can be treated perturbatively, integrating the anharmonic parts over the unperturbed trajectories. The result is a shift
\bea
\Delta\Phi&=&8\frac{\om_r^2T^2}{k^2 \pi^3r^2}[4D_3k_0\pi d(\cos\phi_0-\sin\phi_0) \\&&+D_4\om_r T(9\pi-16+24\frac{k_0^2}{k^2}(\pi-1)+6\frac{k_0^2}{k^2}\pi\sin2\phi_0)]\nonumber
\eea
which is zero if there is no initial motion, $k_0=0$. For $\om_r=2\pi\times 1.2\,$GHz, $T=2.8\,$ms, $k=2\times 2\pi/1064$\,nm, and $r=2\,$cm, we obtain
\be
\frac{\Delta\Phi}{\Phi_0}=-1.7\times 10^{-9}D_3 \frac{k_0}{{\rm m}^{-1}} (\sin\phi_0-\cos\phi_0) +3.2\times 10^{-4} D_4
\ee
where we have neglected terms proportional to $k_0^2$. The coefficient $D_4$ can be of order $10^{-4}$ and $D_3$ is expected to be much lower for symmetry reasons, so we expect the error due to anharmonicity to be better than 32\,ppb. That can be improved with a larger trap.

\subsubsection{Non-closure of the interferometer}

We treat the trajectory change perturbatively. Neglecting damping, the classical trajectory of the electron satisfies
\be
\ddot z_i(t)+\om_z^2(t) z=F_i(t)/m
\ee
Using a Green's function $G(t,t')=1/(m\om_z)\sin\om_z(t-t')$, the resulting displacement $\delta_i(t)$ of the trajectory $z_i$ at time $t$ is given by
\be
\delta_i(t)=\frac{1}{m\om_z}\int_{-\infty}^tF_i(t')\sin\om_z(t-t')dt.
\ee
The force is due to the $z_i^4$ term in the potential, $F_i(t)=2m\om_z^2 D_4 z_i^3(t)/d^2$ with $z_i(t)$ given by Eqs. (\ref{traj1},\ref{traj3}), respectively, for the three trajectories. In particular, we are interested in $\delta_i\equiv \delta_i(3T)$, i.e., the displacement of the wave packets when they interfere at $t=3T$. We find a gap between $z_1$ and $z_2$ of
\be
\delta_1(3T)-\delta_2(3T)=\frac{32\sqrt{2} D_4 v_r^2 v_0 T^3}{d^2 \pi^3}\sin(\phi_0-\pi/4)+O(k_0^2),
\ee
where we assume $\om_z=\pi/(2T)$ and neglected terms quadratic in $v_0=\hbar k_0/m$. The other gap $\delta_2(3T)-\delta_3(3T)=-\delta_1(3T)+\delta_2(3T)$ (to leading order) is calculated in a similar manner. If the electrons have a temperature $T_e$, we may insert the average electron velocity $v_0=\sqrt{k_B T_e/m}$. Comparing the gap with the thermal de Broglie wavelength $h/\sqrt{2\pi m k_B T_e}$ gives an upper limit on the electron temperature $T_e$ of
\be
\frac{h d \pi^{5/2}}{64D_4 v_r^2 T^3k_B}=T_e.
\ee
For the design parameters of Tab. \ref{electronparameters}, this amounts to 15\,mK, doable in a dilution refrigerator.

\subsubsection{Decoherence from axial damping}

Axial motion in the Penning trap is damped with a decay rate of $\gamma$, proportional to a loss resistance $R$, but independent of $\om_z$. The resistance $R$ is usually made as large as technically feasible, as this facilitates electron detection. For $R\sim 10^5\,\Omega$, we obtain $\gamma \sim 2\pi \times 10$\,Hz for a normal-sized trap. By choosing $R$ as low as possible, we can get $\gamma$ to, let's say, $2\pi \times 1\,\mu$Hz.

The decay rate for the $s^{\rm th}$ harmonic oscillator state is $s\gamma$ and limits the coherence time of the interferometer. Since $s \simeq \om_r/\om_z$ (the electrons are not in a pure HO quantum state), the coherence time is limited to $\tau \approx \om_z/(\om_r\gamma)$, and for reasonable interference contrast, we need $4T< \tau$. Let's say we are working at $4T=\tau$. The phase $\Phi$ of the interferometer is thus limited to $\frac 2\pi \om_r\frac{\om_z}{\om_r\gamma}=\frac 2\pi \frac{\om_z}{\gamma}=\frac 2\pi Q$, where $Q=\om_z/\gamma$ is the quality factor of the axial motion. Since $T=\tau/4=\om_z/(4\om_r\gamma)$ and $\om_z=\pi/(2T)$, we obtain $T=\frac{\pi}{8T\om_r\gamma}$, we get the following combination of parameters that lead to optimum phase $\Phi_{\rm opt}$:
\be
T=\sqrt{\pi/(8\om_r\gamma)}, \quad
\om_z=\frac{\pi}{2\sqrt{\pi/(8\om_r\gamma)}}=2\sqrt{\pi\om_r\gamma}, \quad
\Phi_{\rm opt}= \frac 4\pi \frac{\sqrt{\pi\om_r\gamma}}{\gamma}=4\sqrt{\frac{\om_r}{\pi\gamma}}
\ee
For $\om_r=2\pi\times 1.2\,$GHz, $\gamma=2\pi\times 1\,\mu$Hz, we obtain $T=2.8$\,ms, $\om_r=2\pi\times 122\,$Hz, and $\Phi_{\rm opt}=7.8\times 10^7$.

\subsubsection{Example}
Table \ref{electronparameters}  gives a numerical example. Obviously, many questions remain to be addressed, such as how to cool electrons. Given that we can use only one electron at a time, how can we repeat the experiment rapidly so as to obtain good statistics? This will probably require nondestructive detection of the electron. The purpose of these sections is not to answer all these questions, but to present the phase calculation, showing that high-precision electron interferometry is, in principle, possible, and what the challenges are.

\begin{table}
\caption{\label{electronparameters} Numerical example for an electron interferometer}
\begin{tabular}{ccc}
\hline
Parameter & & value \\ \hline
Laser wavelength & $\lambda$ & 1064\,nm \\
Effective wavenumber & $k$ & $1.2\times 10^7/$m \\
Recoil frequency & $\om_r$ & $2\pi\times 1.2$\,GHz \\
Recoil temperature & & 57\,mK \\
Axial frequency & $\om_z$ & $2\pi\times 10$\,kHz \\
Axial amplitude & $A$ & 2.1\,cm \\
Typical trap size & $d$ & 10\,cm \\
Axial loss resistance & $R$ & 10\,k$\Omega$ \\
Axial damping ($\kappa\sim 1$ characterizes trap geometry) & $\gamma=(e\kappa/2d)^2 R/m$ & $2\pi\times 0.7$\,mHz \\
Pulse separation times & $T=\pi/(2\om_z)$ & $25\,\mu$s \\
Phase & $\Phi_0=8\om_r T/\pi$ & $5.1 \times 10^5$\,rad \\ \hline
\end{tabular}
\end{table}

\section{Interferometry in space}

\subsection{Concept}
\begin{figure}
\centering
\epsfig{file=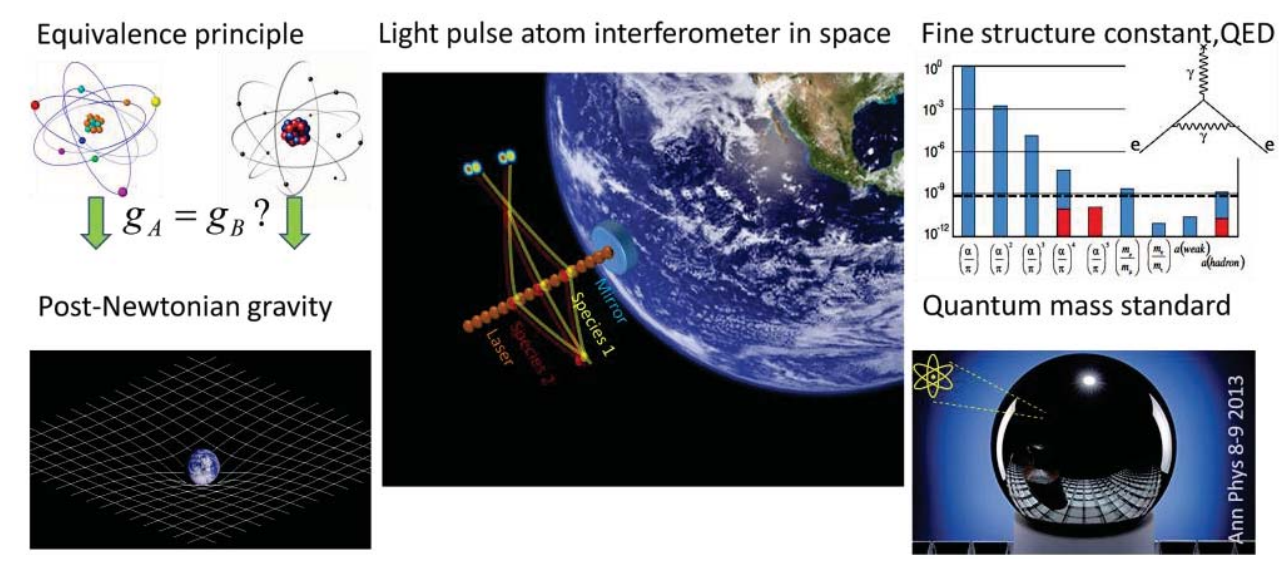,width=0.6\textwidth}
\epsfig{file=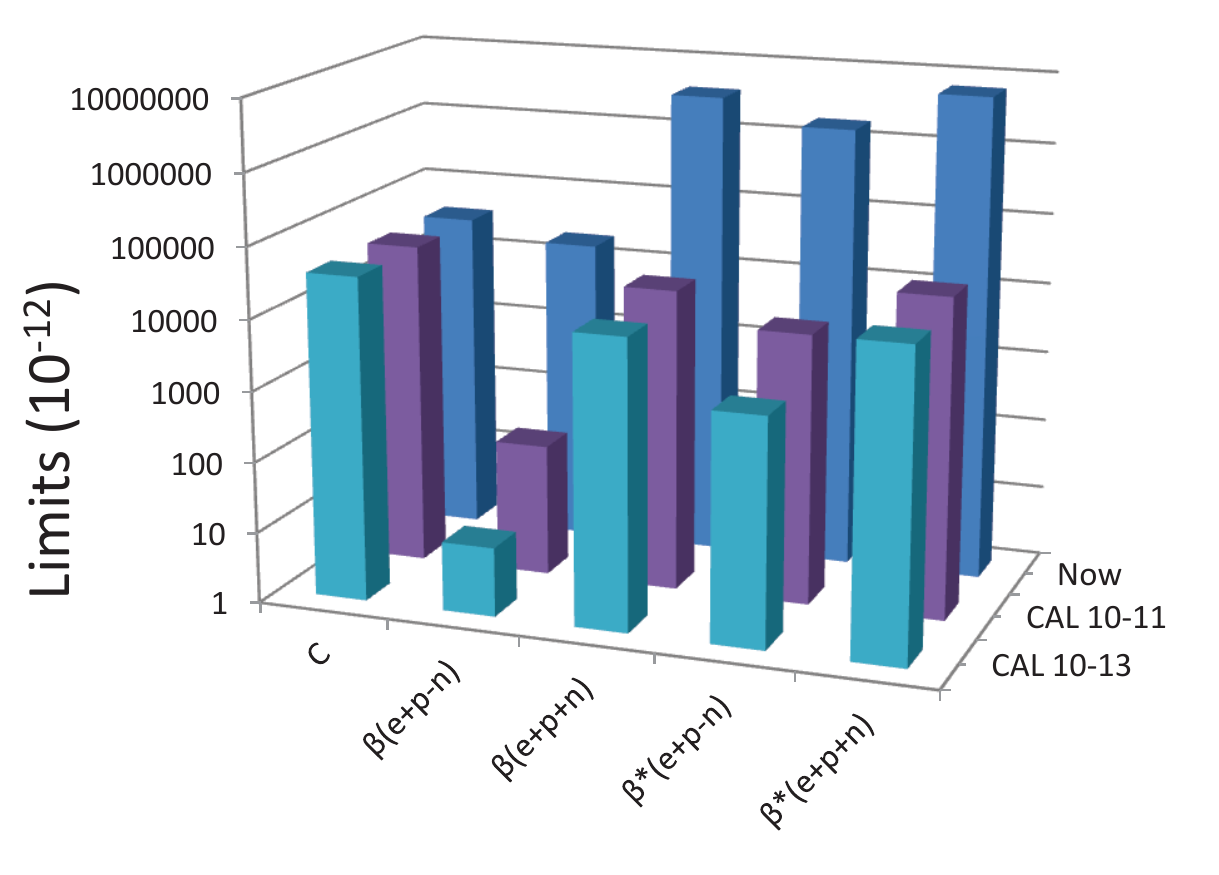,width=0.35\textwidth}
\caption{\label{spaceconcept} Configuration of a dual-species atom interferometer and its possible science goals.}
\end{figure}

Operation in space offers a number of potential advantages. Experiments can use long interrogation times due to absence of free fall of the atoms relative to the apparatus, the possibility to suppress systematic effects by inverting the experiment (``putting the Earth on the other side of the experiment"). Tests of fundamental phyiscs benefit from, e.g., the possibility to explore larger modulations of the gravitational potential and the velocity of the experiment relative to an inertial frame. Within 20 years after the invention of light-pulse atom interferometers,  the technology to needed to do so is finally within reach. Fig. \ref{spaceconcept} shows a dual-species atom interferometer in orbit and its possible science goals.

One possible platform for performing first steps is NASA's Cold Atom Lab (CAL). Considerable science can be achieved without assuming any special hardware beyond the basic configuration described in current CAL documents concerning atom number, atom temperature, and shielding of external influences. This science includes a spaceborne atom interferometer; a precise dual-species atom interferometer, reaching $10^{-11}$ differential gravity resolution even with very basic hardware; a test of Einstein's equivalence principle (EEP) with 87Rb and 40/41K, which have high sensitivity to unbounded modes of EEP violation; a search for new physics arising due to space-time curvature; a measurement of atomic mass ratios with competitive accuracy; a measurement of the fine structure constant; and a quantum mass standard.

A thick (several millimeter radius) laser beam for Bragg diffraction would be important for precision measurements of high performance. Even with a thin beam, however, the potassium to rubidium mass ratio can be measured to parts in $10^{10}$, improving the knowledge of the potassium atom mass 50-fold and delivering a powerful first demonstration of the use of atom interferometers in mass measurement. These efforts will also lead to long coherence times and large coherent splitting of matter waves.

\subsection{Test of the equivalence principle}

\begin{table}
\centering
\caption{\label{CalEEP}. Current limits (parts in $10^9$) on EEP violations compared with limits after CAL makes a $10^{-11}$ or $10^{-13}$ measurement using 87Rb/40K.}
\begin{tabular}{cccccc} \hline
	& $C$	& $\beta^{e+p-n}$ & $\beta^{e+p+n}$ & $\bar \beta^{e+p-n}$ & $\bar \beta^{e+p+n}$ \\ \hline
Present	& $19\pm37$ & $-13\pm21$ & $2400\pm3900$ & $ 	1100\pm1800 $ & $	 -4100\pm6700 $ \\
$10^{-11}$ test at CAL	& $19\pm37	$ & $-0.002\pm0.072$ & $-8.9\pm18	 $ & $1.4\pm6.5	$ & $16\pm31$ \\
$10^{-13}$ test at CAL	& $19\pm37$ & $	-0.0022\pm 0.0094$ & $	 -9.0\pm12	$ & $1.4\pm1.6$ & $	16\pm21$ \\ \hline
\end{tabular}
\end{table}

Table \ref{CalEEP} and Fig. \ref{spaceconcept} compare the global limits on EEP violations available from present data (see section \ref{GlobalLimits}) with the limits that would result from modest-precision measurements in space. The improvement is very strong. By a differential measurement between 87Rb and a potassium isotope (40K or 41K), the spaceborne experiment will measure a combination of coefficients that is considerably different from any combination measured before. Thus, even a modestly sensitive result can improve the global bounds on EEP violation for classical matter by a factor of 200-300 for four of the five coefficients.

\subsubsection{Advantages of quantum tests of the EEP in space}
Spaceborne atom interferometers are sensitive to modes in which the EEP might fail that are inaccessible to terrestrial or classical experiments. Compared to classical tests, they can operate with spin-polarized matter and thus place bounds on EEP violating effects that might couple to the quantum properties of matter. Spin-dependent gravitational couplings have long been studied in the context of theories of gravity with nonvanishing torsion.  A quantum tests of the EEP will perform the first or most sensitive measurement on several $b_\mu, d_{\mu\nu}, g^\mu, f_{\mu\nu}, g_{\lambda\mu\nu}$, and $H_{\mu\nu}$ coefficients of the SME. In addition, they are sensitive to the phase of the matter-wave packet, whereas classical tests sense only the center of mass-motion. Schiff's conjecture implies that information gained from this phase is equivalent to the one measured in classical tests. After 50 years, however, the conjecture has neither been proved nor disproved. Quantum tests of the EEP are tests of the gravitational redshift for matter waves and currently the most sensitive probes by far for relativity violations outside of Schiff's conjecture \cite{redshift}.

Compared to terrestrial tests, these experiments are sensitive to effects that might arise at higher post-Newtonian order, PNO. Effects of gravity can be grouped by their suppression in powers of $1/c$. Newtonian gravity enters the metric proportional to $1/c^2$ and are thus labelled PNO(2). Higher-order effects such as frame-dragging or perihelion precession arise at PNO(3) and PNO(4), respectively. The large modulation of position and velocity provided by the Space Station's orbit helps detecting PNO(3) and PNO(4)-signals that might arise from the nonlinearity of general relativity. This enables a search for effects of higher-mass dimension operators of the SME, e.g., EEP violation that scales with the velocity of the system. Such effects have never been the subject of direct experimental study, and could be extremely large, yet unnoticed so far.

\subsection{Recoil measurements and mass standard}
By reprogramming, the spaceborne atom interferometer can measure the recoil frequency $\om_r=\hbar k^2/(2m)=\om_L^2/\om_C$ of the atoms. If $\om_L$ is known, this translates to a measurement of $\hbar/m$ or equivalently the Compton frequency $\om_C$ of the atom.

\subsubsection{Fine structure constant}
Combined with the Rydberg constant Ry and the isotopes' mass ratios $m/m_e$ with the electron, this measurement can be used to determine the fine structure constant $\alpha$, see section \ref{finestructureconstant}. The performance of CAL in such a measurement will depend heavily on its design. A very basic CAL having a thin optical lattice beam will not reach beyond a $10^{-8}$ accuracy. This can be improved by precise calibration of the lattice beam's wavefronts, and/or by using a thicker beam. An interferometer having a 2.5-mm beam waist beam could reach well beyond the ppb level and provide the most precise measurement of h/m and the fine structure constant.

Comparison of $\alpha$ as measured by atom interferometers and by the electron's gyromagnetic ratio $g$ can yield the most precise test of QED and limit a possible inner structure of the electron.  If $g$ doesn't deviate from the expected value by more than $\delta g$, the energy scale $m*$ of such a substructure must be $m*>m_e/(\delta g/2)^{1/2}$ in the chirally invariant model \cite{Brodsky1980} (other models lead to a linear scaling and thus a larger scale). Current data yields $m*>0.7\,$GeV, limited by measurements of $\alpha$. Data from the large electron-proton collider sets a limit of 10\,TeV; a space-borne measurement at $10^{-10}$ precision could reach 1.5\,TeV and a future one at $10^{-12}$ could reach 15\,TeV (assuming equal progress in the measurement and theory of g). Thus, paradoxically, some of the coldest atoms might yield some of the highest-energy bounds on elementary-particle substructure. For other new physics that might be learned from such measurements, see \cite{Paris,Terranova}.

\subsubsection{Absolute masses}
The kg is the last unit that is defined by an artifact. This has obvious disadvantages, such as errors caused by contamination or damage, and runs counter to the ideal of a unit based on Nature's laws. In 2011, the General Conference on Weights and Measures expressed its intent to revise the definition of the kilogram by assigning an exact value to the Planck constant $h$.  The kilogram would then be referenced to the second through the defined values of the Planck constant. Recently, we have realized that atom interferometers can measure atomic masses directly from this definition.  The link to macroscopic masses could be made on the ground by Avogadro spheres, silicon crystals of accurately measured atom number.  At the time of this writing, this would yields one of the most accurately calibrated macroscopic masses under the proposed CGPM-2011 redefinition.

Spaceborne atom interferometers could establish atoms as calibrated microscopic mass standards that can be used anywhere in the world, based on measuring their Compton frequencies with a precision of $10^{-11}$. They will enable absolute atomic mass measurements anywhere on Earth with unprecedented precision, greater than 1000 times more accurate than in the present SI.

A quantum realization of the unit of mass through atom interferometers and the Avogradro project complements the Watt balance, using exactly the same definition of the units. It is based on inertial mass and not gravitational mass, but is is independent of Earth's gravity, tides, earthquakes and magmatic currents. It does not require mechanically moving parts or standard resistors that are prone to drift, and is based on fundamental laws of quantum mechanics, rather than macroscopic quantum effects, for which first-principles theory doesn't exist and probably never will. It realizes high precision in the microscopic world, where it is most needed: Already now, microscopic masses can be compared to $10^{-11}$ precision or better. All these measurements would become absolute mass measurements.

\subsection{Design parameters}

We will discuss two scenarios, a ``conservative" one where we will not assume availability of any laser beams, wavelengths, and other equipment that is not available according to the CAL documentation at the time of this writing. The ``realistic" scenario assumes reasonable extrapolations from there. The two configurations are outlined in Tab. \ref{Calparam}. Table \ref{Calgravpar} and \ref{Calrecoilpar} list the most important parameters of the equivalence principle test and recoil; measurements, respectively, and the most important systematic effects.

\begin{table}
\centering
\caption{\label{Calparam}. Dimensions of the conservative and realistic scenario}
\begin{tabular}{lccp{4cm}}\hline
	& Conservative	& Realistic &	Remarks \\ \hline
Atom temperature	&100 pK	&100 pK	& \\
Atom number (Rb/K)	&$ 10^4/10^4$	& $2\times 10^5/10^5$ & \\ 	
Lattice wavelength	& 850 nm	& 850 or 676 nm	& 676 nm leads to same lattice depth for Rb and K \\
Lattice laser power	& 10-50 mW	& 0.1 W	& 2-3 recoil lattice depth \\
Lattice beam $1/e^2$ radius & 0.5 mm& 	3 mm	& Limited by separation of atoms from chip \\
Free expansion time	& 5 s	& 5 s	& CAL documents suggest up to 20 s\\
Tip-tilt mirror for lattice beam	& No &	Yes	& \\
RF reference stability &	$10^{-8}$ & $10^{-11}$ & 	For recoil measurement \\
Magnetic shielding factor & 	100	& 1000	& \\
Vibrations	& Not critical	& Not critical	& \\ \hline
\end{tabular}
\end{table}

\begin{table}
\centering
\caption{\label{Calgravpar} Equivalence principle tests in space using CAL. Note that the ``conservative" experiment does not assume a pulse separation time any longer than what is now routine in the lab. Taking advantage of microgravity allows the realistic scenario to gain much better performance. A EEP test at $10^{-11}$ sensitivity is easily compatible with the estimated systematic effects, leaving considerable room for tradeoffs with other projects at CAL.}
\begin{tabular}{p{7cm}cc} \hline
Parameter	& Conservative	& Realistic \\ \hline
Pulse separation time &	0.5 s	& 2 s\\
Momentum transfer & $2\hbar k$ & $2\hbar k$ \\
Gravity phase [rad]	& $1.8\times 10^7$ & $3\times 10^8$ \\
Differential resolution $\delta g/g$ (1 day) & $1\times 10^{-11}$ & $	 1.5\times 10^{-13}$ \\
Magnetic field systematic &	$3\times 10^{-11}$ & $1.5\times 10^{-13}$ \\
Gravity gradient influence due to initial cloud mismatch as caused by magnetic fields	& $1.4\times 10^{-11} g$ & $1.4\times 10^{-13}g$ \\ \hline
\end{tabular}
\end{table}

\begin{table}
\centering
\caption{\label{Calrecoilpar} Recoil measurement at CAL. These experiments require the realistic scenario to surpass the precision attained  in terrestrial experiments, but even the conservative scenario will lead to a 50-fold improvement in the Potassium mass.}
\begin{tabular}{p{6cm}cc} \hline
Parameter	& Conservative	& Realistic \\ \hline
Pulse separation time & 0.5 s	& 0.5 s\\
Momentum transfer	& $2\hbar k$ & $4\hbar k$ \\
Recoil phase [rad]	$4\times 10^4$ & $3\times 10^5$ \\
Differential resolution (1 day) &	$5\times 10^{-9}$ & $1.5\times 10^{-10} $ \\
Magnetic field systematic 	& $3\times 10^{-10}$ & $1.5\times 10^{-11}$ \\
Gravity gradient (known to $10^{-3}$) & $(150\pm0.15)\times 10^{-9}$ & $(1500\pm1.5)\times 10^{-10}$ \\
Beam splitter phase	& $1.4\times 10^{-10}$ & $1.4\times 10^{-10}$ \\
Guoy phase, wavefront curvature (characterized to 1\%) &	$(3\pm0.03) \times 10^{-7}$ & $	 (8\pm0.1) \times 10^{-9}$ \\
RF reference uncertainty & $10^{-8}$ & $10^{-11}$ \\
Total systematic error + noise & $1.1\times  10^{-8}$ & $3\times 10^{-10}$ \\ \hline
\end{tabular}
\end{table}
			
The experiments will use delta-kick ``cooling" to produce samples with residual kinetic energy below 100 pK and free expansion times greater than five seconds. Use of Bragg diffraction will reduce magnetic field sensitivity, allowing operation with modest magnetic shielding. Vibrations will be canceled by dual-species differential measurement in the case of gravity experiments, and by simultaneous conjugate interferometers  for recoil measurements.

As interferometer geometries, standard Mach-Zehnder and a diamond-shaped interferometers are suitable, the latter having the advantage of canceling the signal due to the gravity gradient, provided the two species are at the same place at the time of the initial beam splitter.

\subsection{Inversion of the setup}
Higher-precision experiments that are possible in a dedicated mission will most likely be limited by the gravity gradient, which causes a parasitic signal if the overlap of the two species is not perfect. It takes an overlap accuracy of nanometers to reach a precision in the $10^{-15}$ range in an EEP tests. In space, however, it is possible to invert the setup on a gimbal, thus ``putting the Earth on the other side of the experiment." This will help to greatly suppress systematic effects.

\section{Summary and outlook}
The snapshot of atom interferometry presented in this paper is by no means complete. However, we hope to have shown that the field is interesting from both the point of view of fundamental physics and applications. On the fundamental side, it inspired taking a new look at de Broglie's view that matter-wave packets are like oscillators (``clocks") ticking at an incredibly high frequency. We showed that this concept is powerful enough to derive all equations of motion of quantum mechanics, by contrast to what has been thought before. We hope that the concept might prove fruitful in the further development of quantum theory, e.g., for directly obtaining a theory of fermions from the Nambu-Goto action \cite{strings} in string theory. It might reveal new effects, such as gravitational Aharonov-Bohm effects described by the Dirac equation in curved space-time.

Might relativistic effects be observable in electron interferometers? Such effects should scale like $(n \om_L/\om_C)^2$, where $n$ is the number of photon momenta transferred by the laser and $\om_C/(2\pi)\sim 1.1\times 10^{20}\,$Hz is the electron's Compton frequency. For a laser frequency $\om_L$ corresponding to a wavelength of 266\,nm, this amounts to $4\times 10^{-10}$; measurement is not completely out of the question. Such relativistic effects could include a dependence of the interferometer phase on the relative orientation of spin and momentum (``spin-orbit coupling"). Observation of spin-gravity coupling would be even more exciting. In an electron interferometer (similar to Fig. \ref{Penning}) with a pulse separation time of $T$, the phase induced by spin-gravity coupling will be $\sim (\om_L/\om_C)^2 n k g T^2\sim 2\times 10^{-7}$\,rad for the above parameters and $T=1\,$ms. The splitting of the electron trajectories in such a device would amount to 6\,meters; maybe the high velocities can be contained in the cyclotron motion in the trap.

In experiments, we have tested the equivalence principle and other properties of gravity, in particular the gravitational redshift and the isotropy of gravity. We obtained comprehensive bounds on equivalence-principle violations in the standard model extension (SME). These limits are comprehensive: No experiment may evade them unless one assumes physics beyond the SME, such as violation of energy-momentum conservation or the existence of additional matter fields or forces. Nevertheless, in a universe in which we cannot explain the observed dominance of matter over antimatter, or account for 95\% of the observed mass-energy, it would be presumptuous to categorically rule out any possibility of equivalence principle violations not described by the SME.

Atom interferometry has already now achieved the accuracy needed to play a significant part in the measurement of microscopic masses in the future redefinition of the international system of units. While measurement of macroscopic masses will rarely be more precise than a few parts in $10^9$ (due, e.g., to contamination and outgassing), microscopic masses can already now be compared to parts in $10^{11}$. With better Compton-frequency mass standards, all these relative measurements will become absolute. Macroscopic standards can be derived from microscopic ones by Avogradro spheres, which at the time of this writing have an accuracy of 30\,ppb. They are expected to improve in the future.

Finally, we hope to have given a glimpse of prospects for interferometry with new types of particles, such as antimatter or charged (anti-)particles, and the new types of experiments enabled by them, such as equivalence principle tests with charged particles.

Many other ideas are currently pursued by researchers world-wide, e.g., gravitational wave detection, compact atom interferometers, navigation and geophysics, gravity gradient measurements, or single-atom interferometry, to name just a few. The field of atom interferometry has a bright future.

\acknowledgments

I would like to acknowledge the support of the National Science Foundation, the National Aeronautics and Space Agency, the David and Lucile Packard Foundation, the Alfred P. Sloan Foundation, and Lawrence Berkeley National Lab.

Many thanks to Paul Hamilton for his very careful and insightful reading of the manuscript. I have the privilege to work with a number of great scientists and am grateful to all of them, in particular all members of my group at Berkeley since 2009. I would like to highlight those involved directly with the work reported here: Justin Brown, Steven Chu, Brian Estey, Joel Fajans, Ori Ganor, Paul Hamilton, Mike Hohensee, Sabine Hossenfelder, Matt Jaffe, Pei-Chen Kuan, Alan Kostelecky, Shau-Yu Lan, Jay Tasson, Bob Wiringa, Jonathan Wurtele, Chenghui Yu, Nan Yu, Anton Zeilinger, and Andrei Zhmoginov. Thanks to Guglielmo Tino and Mark Kasevich for the invitation to the Summer School. I thoroughly enjoyed the science and friendship experienced there. Finally, thanks to my family for their patience.

\end{document}